\documentclass[aps,pre,preprint]{revtex4-1}

\usepackage{algorithm}
\usepackage{algpseudocode}
\usepackage{booktabs}
\usepackage{moreverb}
\usepackage{graphicx}
\usepackage{latexsym}
\usepackage{graphics}
\usepackage{epsfig}
\usepackage{amsmath,amssymb,amsfonts,theorem,color,bm}
\usepackage{multirow}
\usepackage[normalem]{ulem}

\begin{document}

\newcommand{\ra}[1]{\renewcommand{\arraystretch}{#1}}

\newcommand{\dfracc}{\displaystyle\frac}
\newcommand{\dint}{\displaystyle\int}
\newcommand{\dsum}{\displaystyle\sum}
\newcommand{\sen}{\mbox{$\rm sen$}}
\newcommand{\Beta}{{|\!\bf B}}
\newcommand{\ray}{\vec{\lambda}}
\newcommand{\RR}{\mbox{I}\!\mbox{R}}
\newcommand{\bA}{{\bf A}}
\newcommand{\bD}{{\bf D}}
\newcommand{\bG}{{\bf G}}
\newcommand{\bI}{{\bf I}}
\newcommand{\bJ}{{\bf J}}
\newcommand{\bH}{{\bf H}}
\newcommand{\bL}{{\bf L}}
\newcommand{\bM}{{\bf M}}

\bmdefine\bu{\bm u} \bmdefine\bU{\bm U} \bmdefine\ba{\bm a}
\bmdefine\bb{\bm b} \bmdefine\bn{\bm n} \bmdefine\bk{\bm k}
\bmdefine\bx{\bm x} \bmdefine\by{\bm y} \bmdefine\bz{\bm z}
\bmdefine\bq{\bm q} \bmdefine\bQ{\bm Q} \bmdefine\bv{\bm v}
\bmdefine\bV{\bm V} \bmdefine\bw{\bm w} \bmdefine\bW{\bm W}
\bmdefine\bX{\bm X}
\bmdefine\be{\bm e} \bmdefine\bE{\bm E} \bmdefine\br{\bm r}
\bmdefine\bo{\bm 0} \bmdefine\boo{\bm o} \bmdefine\bOO{\bm O}
\bmdefine\bSigma{\bm \Sigma}
\bmdefine\btau{\bm \tau} \bmdefine\bh{\bm h} \bmdefine\bg{\bm g}
\bmdefine\bp{\bm p} \bmdefine\bP{\bm P}
\bmdefine\bs{\bm s}\bmdefine\bS{\bm S}
\bmdefine\bt{\bm t}
\bmdefine\bT{\bm T}

\bmdefine\bcalG{\bm{\mathcal G}} \bmdefine\bcalI{\bm{\mathcal I}}
\bmdefine\bcalT{\bm{\mathcal T}} \bmdefine\bcalL{\bm{\mathcal L}}
\bmdefine\bcalM{\bm{\mathcal M}} \bmdefine\bcalN{\bm{\mathcal N}}
\bmdefine\bcalR{\bm{\mathcal R}} \bmdefine\bcalS{\bm{\mathcal S}}
\bmdefine\bcalA{\bm{\mathcal A}} \bmdefine\bcalB{\bm{\mathcal B}}
\bmdefine\bcalC{\bm{\mathcal C}} \bmdefine\bcalD{\bm{\mathcal D}}
\bmdefine\bcalQ{\bm{\mathcal Q}} \bmdefine\bcalH{\bm{\mathcal H}}
\bmdefine\bcalU{\bm{\mathcal U}} \bmdefine\bcalV{\bm{\mathcal V}}
\bmdefine\bcalD{\bm{\mathcal D}}

\bmdefine\bPhi{\bm{\Phi}} \bmdefine\bvarphi{\bm{\varphi}}
\bmdefine\bSigma{\bm{\Sigma}} \bmdefine\bOmega{\bm{\Omega}}
\bmdefine\bF{\bm{F}} \bmdefine\bR{\bm{R}} \bmdefine\bbf{\bm{f}}
\bmdefine\bnabla{\bm{\nabla}}

\newcommand\red{\color{red}}
\newcommand\blue{\color{blue}}
\newcommand\green{\color{green}}
\newcommand\magenta{\color{magenta}}

\bmdefine\ip{\bm \cdot} \bmdefine\dip{\bm :}
\newcommand\er{\mbox{\rm e}}
\newcommand\ir{\mbox{\rm i}}
\newcommand\smallir{\footnotesize\mbox{\rm i}}
\newcommand{\cierto}[1]{\fbox{#1}\label{#1}}

\def\rin{{\rm in}}
\def\at{{\rm at}}
\def\and{{\rm and}}
\def\for{{\rm for}}
\def\cc{{\rm c.c.}}
\def\HOT{{\rm HOT}}
\def\Rey{{\rm Re}}
\def\tif{\text{\quad it \quad}}
\def\tand{\text{\quad and \quad}}
\def\at{\text{\quad at \quad}}
\def\cc{\hbox{c.c.}}
\def\llsim{\,\stackrel{\ll}{\sim}\,}\def\ggsim{\,\stackrel{\gg}{\sim}\,}

\newcommand{\beqn}{\begin{equation}}
\newcommand{\eeqn}{\end{equation}}


\title{Uncovering spatio-temporal patterns in semiconductor superlattices
 by efficient data processing tools}

\author{F.\ Terragni}
\affiliation{G. Mill\'an Institute for Fluid Dynamics, Nanoscience and Industrial Mathematics, and Department of Mathematics, Universidad Carlos III de Madrid, 28911 Legan\'es, Spain}

\author{L.\ L.\ Bonilla}
\affiliation{G. Mill\'an Institute for Fluid Dynamics, Nanoscience and Industrial Mathematics, and Department of Mathematics, Universidad Carlos III de Madrid, 28911 Legan\'es, Spain}

\author{J.\ M.\ Vega$^{*}$}
\affiliation{E.T.S.I. Aeron\'autica y del Espacio, Universidad Polit\'ecnica de Madrid, 28040 Madrid, Spain \\
$^{*}$Corresponding author: josemanuel.vega@upm.es \\}

\date{\today}

\begin{abstract}
Time periodic patterns in a semiconductor superlattice,
relevant  to microwave generation, are
obtained upon numerical integration of a known set of  drift-diffusion
 equations.
The associated spatio-temporal transport mechanisms
are uncovered by applying (to the  computed data) two recent
data processing tools, known as the higher order dynamic
mode decomposition and the
spatio-temporal Koopman decomposition.
Outcomes include a clear identification of
the asymptotic self-sustained oscillations of the current
 density (isolated from the transient dynamics)
and an accurate description of the electric field
 traveling pulse in terms of its dispersion diagram.
In addition, a preliminary version of a
novel data-driven reduced order model is  constructed,
which allows for extremely fast
online simulations of the system response
 over a range of different configurations.
\end{abstract}

\maketitle


\section{Introduction \label{sec:intro}}
{\it Semiconductor superlattices}
are periodically layered  structures, which are
formed by epitaxial growth of layers  of two or more different
semiconductors with similar lattice parameters. These superlattices show
damped, high-frequency, spatially uniform
Bloch oscillations \cite{EsakiT70,Feldmann1992}. In principle, it should
be possible to find inhomogeneous Bloch oscillations  that
coexist with much slower self-sustained oscillations due to periodic
generation and motion of charge dipoles \cite{Bonillaetal2011}.
In experiments, the frequency range of the more robust self-sustained
oscillations in semiconductor superlattices may be as large as 100--200 GHz,
which makes them useful in, e.g., fast
oscillators and detectors \cite{Bonilla2005},
whereas similar self-sustained oscillations in quantum cascade laser
devices may reach ultra-high frequencies well in the THz
range \cite{Winge2018}, which
are needed in, e.g., nano-patterned antennas \cite{Leeetal2018}.
Superlattice-based devices exhibit very different spatio-temporal
 patterns and rich
nonlinear dynamics, including static high-field domains,
excitability due to collective charge dynamics, as well as
self-sustained periodic, quasi-periodic, and chaotic current
oscillations \cite{Bonilla2005,BT10,li13,li15,yin17}.
Under an external magnetic field, which is tilted with
respect to the growth direction, electron motion in the
superlattice miniband is two-dimensional (2D). Experiments show
a variety of high-frequency periodic and chaotic oscillations
of the current \cite{fro04,ale12}. It is interesting that
single electron motion in a miniband exhibits Hamiltonian
chaos and stochastic webs \cite{fro04,ale12,sos19},
whereas collective charge motion, taking scattering into
account through the Boltzmann-Bhatnagar-Gross-Krook
(BBGK) equation, displays dissipative chaos through
2D patterns \cite{bon17}.

In this work, we consider $n$-doped, strongly coupled one
miniband semiconductor superlattices \cite{but77,hof96,Bonillaetal2003,Bonilla2005,BT10}.
Under appropriate {\em dc} voltage bias and configuration
parameters, these superlattices exhibit self-sustained
oscillations of the current, which have been the subject
of extensive theoretical and experimental studies \cite{Bonilla2005}.
The relevant state variables in a {\em dc} voltage-biased
superlattice configuration are
the time-dependent scalar {\it current density} and
the distributed {\it electric field}.
Control parameters such as the width of the layers,
the doping density,
the voltage bias, and the conductivity at
the injecting contact (cathode)
determine the existence, shape, and frequency of the oscillations
of the current  density through the
superlattice \cite{Bonilla2005,BT10}.
The evolution of the state variables can
be computed using  one of the numerical methods presented in
\cite{Alvaroetal2013},
which simulate the electron transport in a
single miniband superlattice
by integrating either the governing BBGK
 kinetic equations
or a drift-diffusion approximation of  the latter.
In the spatially one-dimensional (1D) case,
numerical simulations show that,
for convenient values of the various involved
nondimensional parameters,
the electric field exhibits an interesting spatio-temporal pattern.
After a transient,
the dynamics become temporally periodic,
showing a {\it traveling solitary wave} for the electric field in
 the bulk (which can also be seen as
a charge dipole wave).
This wave goes from the cathode to the anode
(as in the Gunn effect of bulk semiconductors \cite{Gunn1965,kro72,bon97,shi98,BEH03,BT10})
and, when
it reaches the anode, another wave is created at the cathode,
in such a way that
the process is periodically repeated. Correspondingly,
the current through the superlattice displays self-sustained
oscillations. The current self-oscillations and the
spatio-temporal patterns of the electric field and charge density
are highly non-monochromatic, namely they show a {\it large number of
harmonics}. The intrinsic spectral properties and the amplitude
of these  phenomena depend on the specific
superlattice configuration and are
important to fully characterize the underlying electron transport.
Thus, extracting the frequency spectrum with accuracy,
analyzing its variation in terms of the main involved parameters,
and describing the evolution of the state variables
through the structure can provide interesting insights
into these devices and their setup.

In this paper,
numerically computed data for the current density
and the electric field  during current self-oscillations
will be treated using two recently introduced {\em data processing tools:}
the {\it higher order dynamic mode decomposition}  (HODMD)
 \cite{LeClaincheVegaSIADS17}
and the  {\it spatio-temporal Koopman decomposition} (STKD)
\cite{LeClaincheVegaJNLS18}.   These decompositions
 isolate the
dynamical features behind the given data, quantitatively
uncovering the dominant
components of complex nonlinear signals. For the superlattice
 self-oscillations, we use data extracted from the transient dynamics that decay to the periodic attractor to isolate and characterize the latter.
Specifically,  the HODMD and STKD methods extract
the frequencies and the growth rates of spatio-temporal patterns. Applications to other systems can be found in \cite{VegaLC2020}.

The HODMD method deals with
spatio-temporal data associated with dynamics exhibiting time-dependent
exponential growth or decay and oscillations. Firstly, we define a time-dependent vector state variable  $\bq$ comprising the total current density and the electric field at a number of grid points. Then, we compute snapshots of $\bq$
at temporally equispaced values of $t$ in a limited timespan.
The HODMD outcome is a discrete Fourier-like expansion of the
form
\beqn
\bq_k\equiv\bq(t_k)\simeq\sum_{n=1}^Na_n\bu_n\er^{(\delta_n+\smallir\,\omega_n)\, t_k},
\label{a1}
\eeqn
 with $t_k = (k-1)\,\Delta t$, for $k = 1,\ldots,K$. Here,
  $a_n > 0$ are real {\it amplitudes}, $\bu_n$ are conveniently normalized
(generally complex) {\it modes}, and $\delta_n$ and $\omega_n$ are the associated
{\it growth rates} and {\it frequencies},
respectively.
It is important to note that replacing $t_k$ by $t$ in eq.(\ref{a1})
(which involves automatic time interpolation)
leads to the continuous expansion
\beqn
\bq(t)\simeq\sum_{n=1}^Na_n\bu_n\er^{(\delta_n+\smallir\,\omega_n)\, t},
\label{a2}
\eeqn
which gives an analytical representation of the underlying dynamics.
When all growth rates, $\delta_n$,
are zero (or conveniently small
in absolute value), then the resulting modes
{\red are called {\it permanent modes} and  the associated}
 dynamics correspond to an {\it attractor}.
 If all frequencies $\omega_n$ are commensurable, then the attractor
is {\it periodic}, while it is
 quasi-periodic if some of the involved frequencies are
 incommensurable (within a small threshold).
It is worth mentioning that, for chaotic attractors,
an infinite number of modes would be involved and the analysis by means of HODMD would be subtle \cite{LeClaincheVegaSIADS17}.
If, in addition to the permanent modes,
there is a second group of modes that exhibit clearly negative
growth rates, then
the expansion (\ref{a2}) corresponds to a
{\it transient behavior approaching an attractor}.
The latter
 can be identified by retaining in
the expansion only those (permanent) modes with $\delta_n=0$ (or suitably small).
Such calculation of the final attractor via HODMD extrapolation
accelerates the computation of  asymptotic dynamics
\cite{LeClaincheVegaPoF17}. Collecting only the decaying modes
(with  non-small negative $\delta_n$)
gives the strictly decaying approach to the attractor.
On the contrary, if some growth rates are strictly positive and the remaining ones
are small in absolute value or equal to zero, then
the expansion (\ref{a2}) yields an unstable behavior departing from
(unstable) permanent dynamics,
which is useful to identify {\it instabilities} \cite{LeClaincheetal2020}.
 However, this latter
 scenario will not arise in the analysis carried out in the present paper.

Besides uncovering spatio-temporal patterns,
the expansion (\ref{a2}) can be used to build a purely data-driven
{\it reduced order model} (ROM), able to
simulate {\it online} the operation of the underlying dynamical system.
This ROM is extremely fast  because it
only requires performing algebraic computations, being thus
appropriate for {\it optimization} \cite{Parketal}
and {\it real-time active control} \cite{Gaoetal93}.
 The superlattice ROM will be used to reconstruct the
 spatio-temporal pattern of current self-oscillations for
 a range of values of the injecting contact conductivity.
 Importantly, the ROM uses
numerical data for very few conductivity values, while
efficiently predicting the superlattice response for many
different conductivity values. Furthermore, the same
approach may be extended to predictions of the device response
in multi-parameter searches.

The STKD method works for
dynamics exhibiting exponential/oscillatory
behavior in both the temporal variable and one or more
distinguished spatial variables (called {\it longitudinal} variables).
Focusing on the simplest
{\it spatially 1D} case with one scalar
state variable,
the STKD method leads to the continuous expansion
\beqn
q(x,t)\simeq\sum_{m=1}^M\sum_{n=1}^Na_{mn} u_{mn}\,
\er^{(\nu_m-\smallir\,\kappa_m) x +(\delta_n+\smallir\,\omega_n)\, t}.
\label{a3}
\eeqn
Here, $\nu_m$ and $\delta_n$ are {\it spatial} and
{\it temporal} growth rates,
respectively, while
$\kappa_m$ and $\omega_n$ are
 {\it wavenumbers} and {\it frequencies}, respectively.
 The real \textit{amplitudes}  $a_{mn}>0$ and the generally complex
\textit{modes} $u_{mn}$ (with $|u_{mn}|=1$) depend on the two indices,
$m$ and $n$.
Note that,
if $\omega_n/\kappa_m=c$, then the whole pattern is a {\it pure
traveling wave}, with
{\it phase velocity} $c$. In this case,
the {\it dispersion diagram} of $\omega_n$ vs. $\kappa_m$ becomes
a straight line passing through the origin.
When the dispersion diagram consists of a family of
parallel, oblique straight lines, the pattern is a
{\it modulated  traveling wave}.
For superlattice self-oscillations, the STKD method
produces an accurate reconstruction of the spatio-temporal
pattern of the periodic attractor, but it
involves too many spatial and temporal modes. Thus, it
is computationally
too costly to yield a fast ROM.

The  computational
costs of the various methods and codes can be measured, and compared
among each other, in terms
of the required CPU times. These will be given below for representative
cases, taking into account that
all computations were performed
using standard (uncompiled) MATLAB in a desktop PC, with a
microprocessor Intel Core i7--6500U at 2.5GHz.

The remainder of this paper is structured as follows. The drift-diffusion
 model equations and
the numerical solver for  the considered semiconductor
superlattice are described in Section \ref{sec:Numerics}.
The main  results, in connection with
uncovering spatio-temporal patterns using the HODMD method
for a representative
superlattice configuration, are given in Section \ref{sec:Results}.
Section \ref{sec:atractorSTKD}
shows how the STKD method characterizes the evolution of the electric field. Section \ref{sec:DataDriven} contains a preliminary version of a novel HODMD-based data-driven ROM for the present system, while the main conclusions
of the paper are found in Section \ref{sec:conclusion}.
Appendices \ref{sec:HODMD} and \ref{sec:STKD} provide
concise descriptions of the HODMD and STKD methods, respectively.


\section{Model equations and numerical solver \label{sec:Numerics}}
In nondimensional  form,  the BBGK-Poisson kinetic equations for
the distribution function, $f(x,k,t)$, of a one miniband
 semiconductor superlattice are \cite{Alvaroetal2013}
\begin{subequations}\label{kin1}
	\begin{eqnarray}
	&& \hskip-5mm  \lambda \left({\partial f\over \partial t} +
	\gamma \sin(k)\frac{\partial f}{\partial x}\right) + F\frac{\partial f}{\partial k}
	= \frac{f^{FD}-f}{\tau_e} - \frac{\tau_e^2-1}{2\tau_e}\,[f -f(x,-k,t)], \label{kin1a} \\
	&& \hskip-5mm \frac{\partial F}{\partial x} = n-1, \label{kin1b} \\
	&& \hskip-5mm n = \frac{1}{2\pi} \int_{-\pi}^{\pi} f(x,k,t) dk =
	\frac{1}{2\pi} \int_{-\pi}^{\pi} f^{FD}(k;\mu(n)) dk, \label{kin1c} \\
	&& \hskip-5mm f^{FD}(k;\mu) = \alpha\, \ln\left[1+
	\exp {\left(\mu - {\cal E}(k)\right)}\right]. \label{kin1d}
	\end{eqnarray}
\end{subequations}
Here, $n(x,t)$ and $F(x,t)$ are the electron density and the
electric field inside the superlattice, respectively.
The superlattice has a single populated miniband with a tight-binding
 dispersion
relation ${\cal E}(k) = \beta(1-\cos k)$
 between the energy and the wavevector
$k$. $f^{FD}(k;\mu)$ is the local Fermi-Dirac equilibrium
distribution function and $\mu$ is the chemical potential,
which is a function of the electron density given by the
solution of eq.\eqref{kin1c}.  In addition,
$\lambda$, $\gamma$, $\tau_e$, $\alpha$, and $\beta$ are dimensionless
 parameters,
where $\lambda$ is the ratio of the mean time between collisions
to the characteristic electron transport time, which is small
for most superlattices. In \cite{Bonillaetal2003}, this fact
was exploited by using the Chapman-Enskog perturbation method
(in the limit as $\lambda\to 0$) to derive a \textit{drift-diffusion}
equation for $F(x,t)$ which, in nondimensional form,
is \cite{Alvaroetal2013}
\beqn
\dfrac{\partial F}{\partial t} + {\cal A}\,\dfrac{\partial F}{\partial x}
+ {\cal B}\,\dfrac{\partial^2 F}{\partial x^2} + {\cal C}\,J = {\cal D},
\label{c1}
\eeqn
for $0 < x < L$. Here, $J(t)$ is the (scalar) total current
density and $L$ is the
nondimensional length of the superlattice.
The coefficients $\cal A$, $\cal B$, $\cal C$, and $\cal D$
appearing in this equation depend on $F$, its first and second
spatial derivatives, and the superlattice physical properties.
Their specific form
can be found in \cite{Bonillaetal2003,Alvaroetal2013}
but it will not be needed for our data-driven analysis.
Equation \eqref{c1} is solved with a nondimensional Ohmic
boundary condition at the injecting contact (cathode)
and a zero-charge boundary condition at the receiving
contact (anode) \cite{Bonillaetal2003},  namely
\beqn
\dfrac{\partial F(0,t)}{\partial t} + \sigma F(0,t) = J(t),
\qquad \dfrac{\partial F(L,t)}{\partial x} = 0, \label{c7}
\eeqn
where $\sigma > 0$ is the
{\it contact conductivity}. Finally, the total current
 density is determined
by imposing the {\em dc} voltage bias condition,
\beqn
V_\text{bias} = \int_0^L F(x,t)\,dx, \label{c8}
\eeqn
where $V_\text{bias}$ is a given positive constant.

As explained in \cite{Bonillaetal2003}, numerical solutions of the model equations \eqref{c1}-\eqref{c8} provide accurate values of the oscillation frequencies and their dependence on the different parameters as observed in experiments \cite{hof96,sch98}.  Experimental observations in strongly coupled one miniband
semiconductor superlattices resolve oscillation frequencies
and Fourier spectra, but they do not provide {\em time resolved}
currents as functions of time  because the involved frequencies
are too high (GHz-THz range) \cite{Bonilla2005,hof96,sch98,fro04,ale12}.
Thus, we cannot directly  compare numerically obtained time-dependent
current densities or electron densities with experiments. This also
precludes comparison of experiments and predictions of driven chaos
(by dc and ac voltage) obtained by numerically simulating
eqs.~\eqref{c1}-\eqref{c8} or related models \cite{cao99}.
In any event, for the model equations and the set of parameter
values considered in the present work, with constant $V_\text{bias}$,
the  asymptotic dynamics are always
periodic. More complex oscillations
(either quasi-periodic or chaotic) may be reproduced by setting a voltage varying in time.
On the other hand, weakly coupled superlattices are described by mathematical models different from the equations we study in this paper. They present oscillations in the MHz range and permit detailed comparison of theory with time resolved current from experiments \cite{Bonilla2005}, which includes spontaneous \cite{li13,li15,MompoCarreteroBonilla2021} and driven \cite{bul95,luo98,bul99} chaos.

The numerical integration of eqs.(\ref{c1})--(\ref{c8})
is carried out by means
of a solver described in
\cite{Alvaroetal2013} (see also \cite{Carpioetal2001}
for an analysis of the numerical method).
Specifically, spatial discretization is performed via
centered finite differences
for the spatial derivatives and the composite
Simpson's rule to evaluate the right-hand side of  eq.(\ref{c8}),
while temporal integration is performed using an improved
 implicit Euler method.

The \textit{numerical solver} outlined above, whose
convergence order is approximately quadratic in space and linear in time,
will be used throughout the paper to both calculate the snapshots
needed by the data processing methods and
compute the reference, `exact' solutions to compare with the
generated approximations.
Concerning the computational cost,
considering the time interval $0 \leq t \leq 300$ (which will be repeatedly
used in this work) and
the standard desktop PC indicated at the end of
Section \ref{sec:intro}, each run requires $\sim 20$ CPU minutes.


\section{Spatio-temporal patterns in a semiconductor superlattice \label{sec:Results}}
In this section, we first present some representative results obtained via
direct numerical simulation. Then,
the HODMD method
 is applied to compute, isolate, and describe
  the periodic attractor and the transient dynamics decaying
to the attractor for both the current density and the electric field.
Finally, it is shown
that HODMD improves/outperforms a fast Fourier transform (FFT) calculation
of the  asymptotic dynamics.

Here, we consider the model equations described
in Section \ref{sec:Numerics},
eqs.(\ref{c1})--(\ref{c8}), with the following representative
parameter values
\beqn
L = 33.19, \quad \sigma = 0.37, \quad V_\text{bias} = 1.2 \,L. \label{d1}
\eeqn
The initial conditions will always be taken as
\beqn
F(x,0) = 1.2, \quad J(0) = 1.2 \,\sigma.\label{d3}
\eeqn
In order to obtain the needed data, the numerical solver mentioned
in Section \ref{sec:Numerics} is run using
\beqn
I = 481 \label{d3a}
\eeqn
equispaced grid points in the spatial domain and a time step
equal to $0.01$ in the time interval
\beqn
0 \leq t \leq 300.\label{d4}
\eeqn

With this selection of the discretization
parameters, the expected relative  \textit{root mean square} (RMS)
 accuracy
of the numerical simulations is $\sim10^{-3}$.


\subsection{The numerically computed pattern\label{sec:OrigPattern}}
\begin{figure}[h!]
\vskip-0.2cm
	\begin{center}
		\includegraphics[width=7cm,height=4cm]{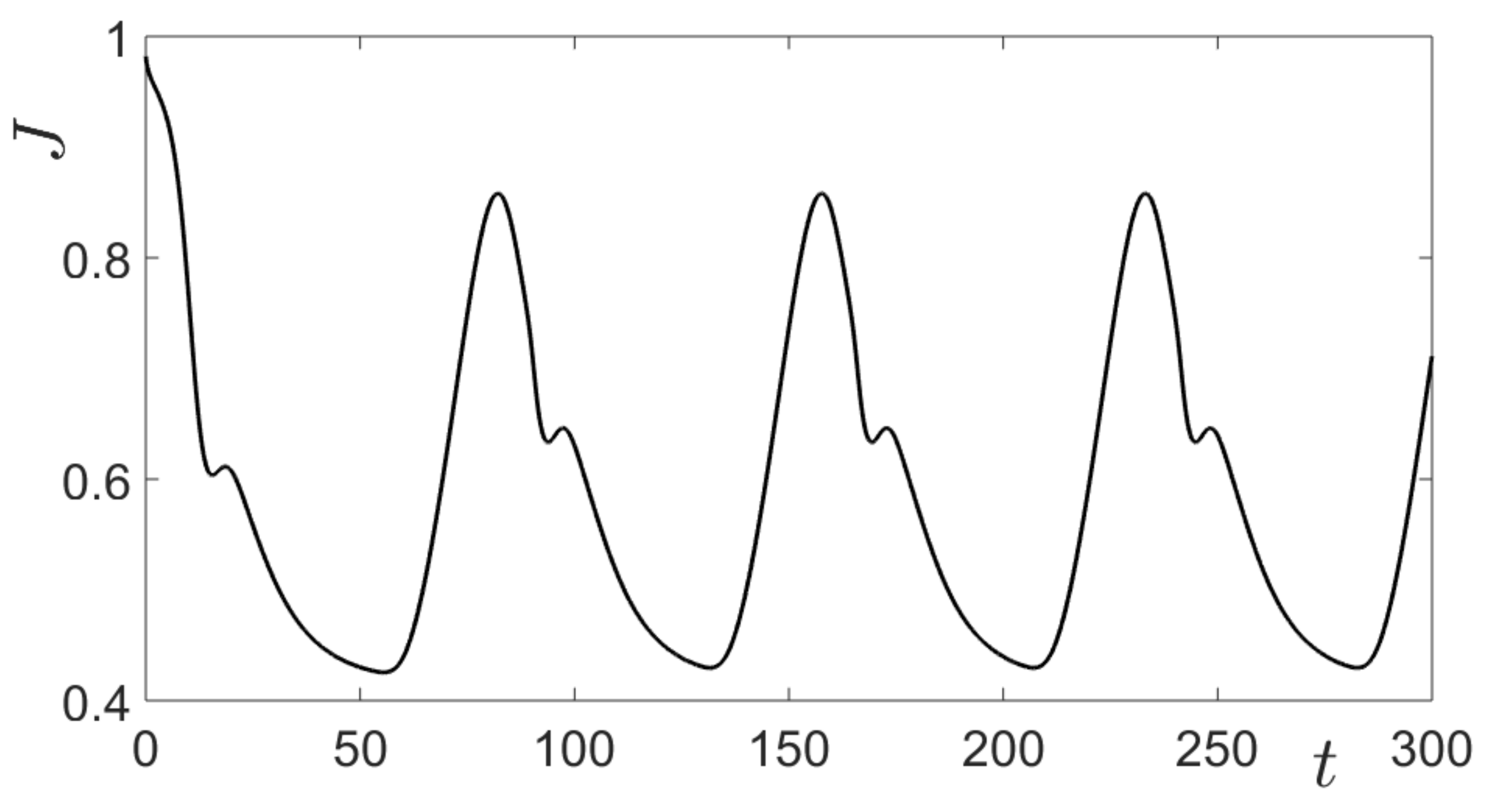}
\end{center}
\vskip-0.75cm
\caption{Temporal evolution of the current  density in the
	considered timespan. \label{fig:evolutionJ}}
\end{figure}

\begin{figure}[h!]
	\begin{center}
		\includegraphics[width=7cm,height=4cm]{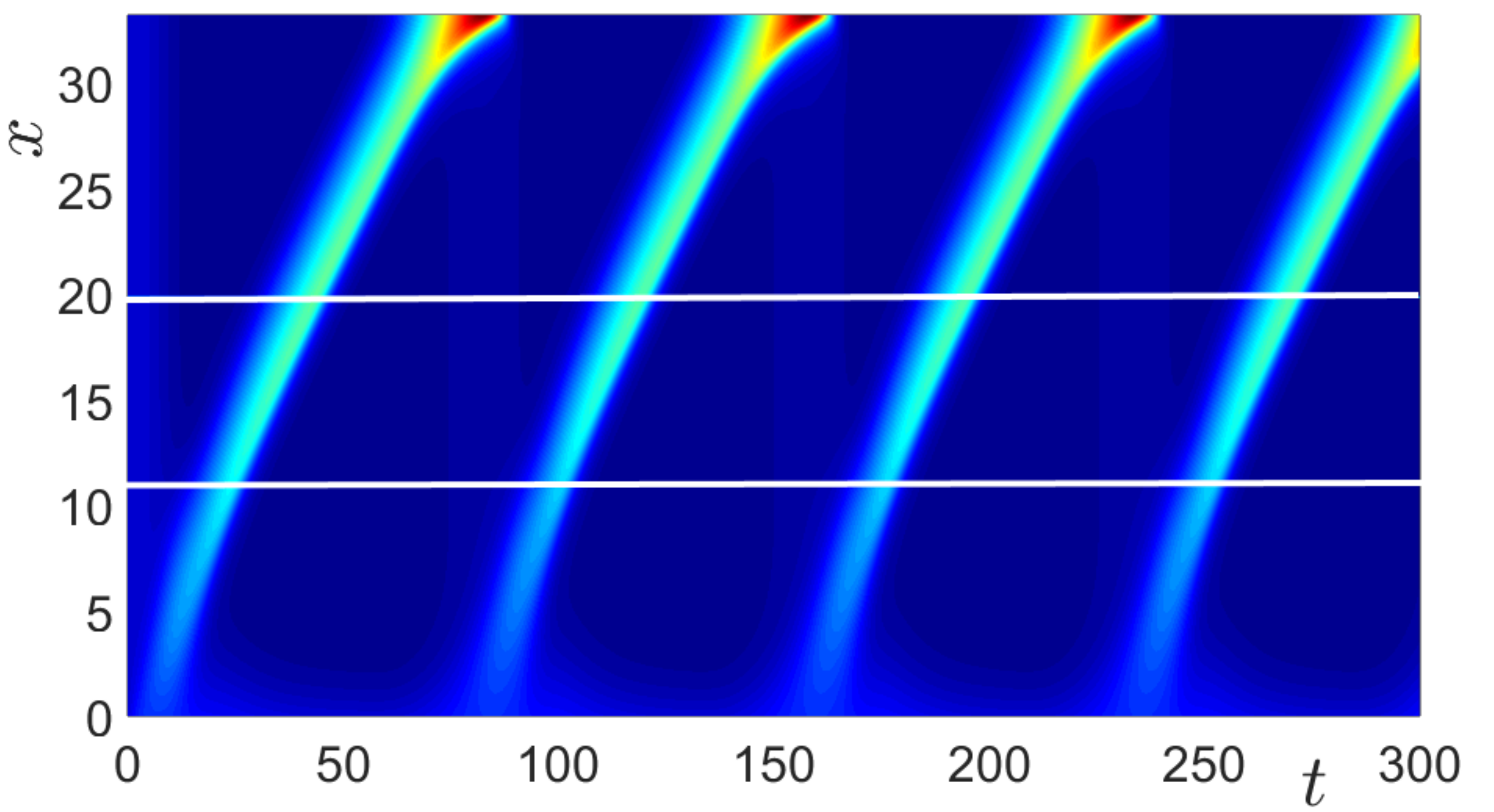}
	\end{center}
	\vskip-0.75cm
\caption{Spatio-temporal color diagram for the
evolution of the electric field in the
considered timespan. The intensity of $F$ varies from blue to
 red as $F$  increases.
The two horizontal white lines indicate the
values $x=\widetilde x_1$ and $x=\widetilde x_2$,
as defined in eq.(\ref{d9})\label{fig:STevolutionF}}
\end{figure}

Figure \ref{fig:evolutionJ} shows
how the current density in the time interval \eqref{d4} becomes time periodic after a transient. The transient duration, which depends on how far from the attractor
 the initial condition is, cannot be appreciated in the plot.
However, the highly non-monochromatic character of the attractor is
clearly  visible. For the same time interval, Fig.\ref{fig:STevolutionF}
displays the spatio-temporal density plot of the electric field $F(x,t)$.
As anticipated, the diagram shows a solitary wave that is created at
$x=0$ (cathode) and propagates  towards
$x=L$ (anode). As the solitary wave travels, its propagation velocity
decreases, while the
electric field peak  increases. Although not clearly appreciated
in the figure, the background electric field must decrease
along the  wave journey. This is because the voltage  bias, defined in
eq.(\ref{c8}), remains constant as time proceeds, while the area behind the
solitary wave increases.
On the other hand, this condition  is consistent with
the fact that, as seen in
Fig.\ref{fig:STevolutionF}, each time a solitary wave reaches the boundary
$x=L$, another wave is created at $x=0$.

\begin{figure}[h!]
\vskip-0.2cm
	\begin{center}
		\includegraphics[width=7cm,height=4cm]{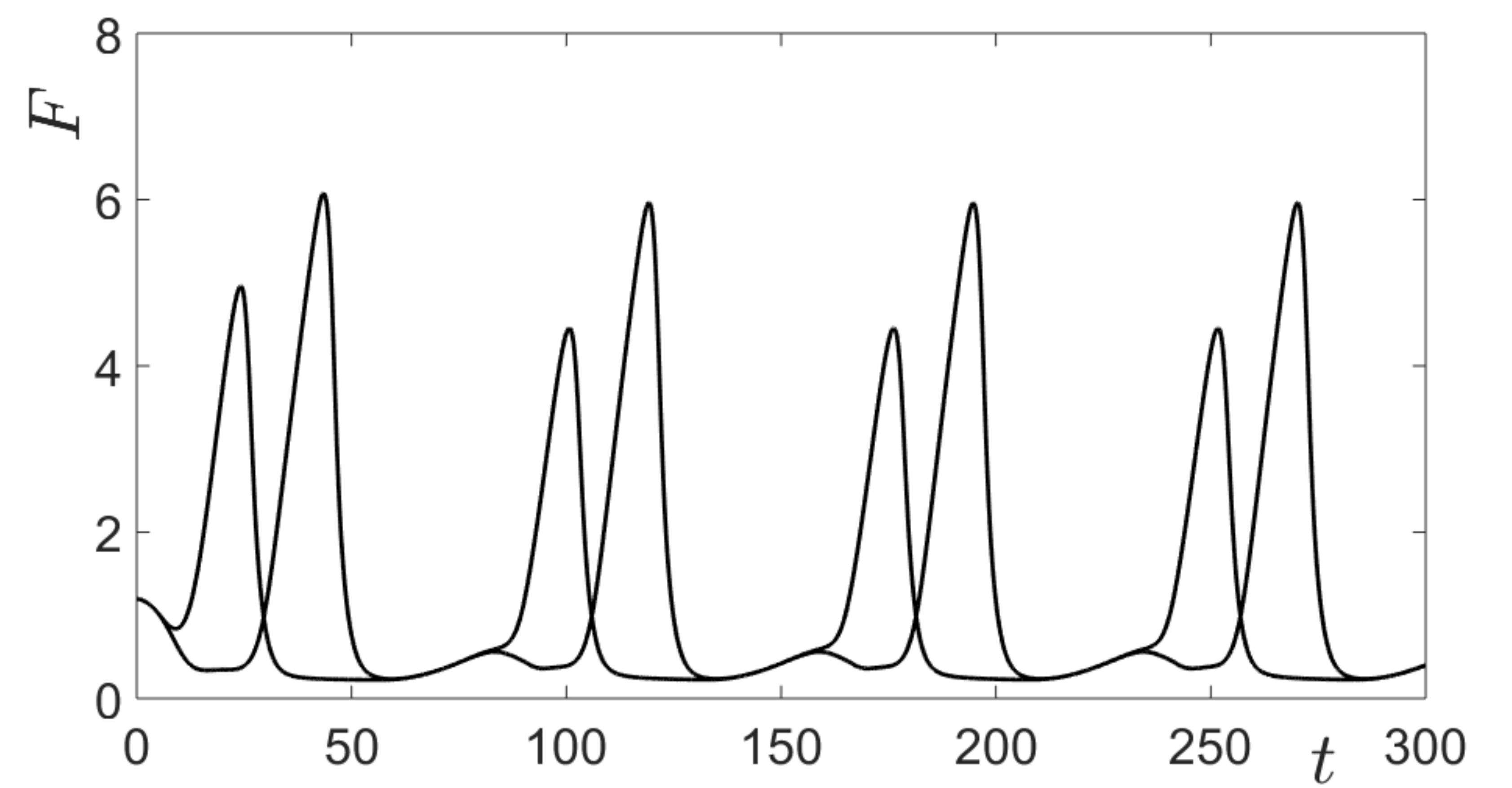}
	\end{center}
	\vskip-0.75cm
\caption{Plots of $F(\widetilde x_1,t)$ and $F(\widetilde x_2,t)$ vs. $t$,
with $\widetilde x_1$ and $\widetilde x_2$
as defined in eq.(\ref{d9}). Since $\widetilde x_1 < \widetilde x_2$,
the two curves can be distinguished
because, consistently with Fig.\ref{fig:STevolutionF},
$F(\widetilde x_2,t)$ shows higher peaks than
$F(\widetilde x_1,t)$. \label{fig:evolutionFx1x2}}
\end{figure}

However, illustrative as it is, the spatio-temporal diagram in
Fig.\ref{fig:STevolutionF}
does not have sufficient contrast
to quantitatively compare different approximations.
A more quantitative (though less complete in the spatial coordinate)
 account of the electric field
is given in  Fig.\ref{fig:evolutionFx1x2},
 which shows the evolution of $F$
at two representative points inside the spatial domain, namely
\beqn
x=\widetilde x_1=\frac{L}{3} \quad\text{and}\quad x=\widetilde x_2=\frac{3\,L}{5}.
\label{d9}
\eeqn
These points are the horizontal white lines in Fig.\ref{fig:STevolutionF}.
We shall use plots similar to Fig.\ref{fig:evolutionFx1x2} to compare
the electric field computed by direct numerical simulations
with its various approximations obtained via the HODMD method.


\subsection{Uncovering spatio-temporal patterns via HODMD\label{sec:Patterns}}
We now apply several variants of the HODMD method to the
superlattice dynamics discussed in the last subsection.
To this end, we consider $K=3000$ temporally
 equispaced snapshots at
\beqn
t_k=(k-1)\,\Delta\, t, \quad\text {for }\,k=1\ldots,K.
\label{d11}
\eeqn
Each snapshot contains the values of both the scalar current
density $J$ and the electric field $F$
at the
$I=481$ spatial grid points used in the numerical simulations.
In other words, for $k=1,\ldots,K$, the snapshot
$\bq_k$ is the $(I+1)$-vector given by
\beqn
\bq_k=[J(t_k),F(x_1,t_k),F(x_2,t_k),\ldots,F(x_I,t_k)]^\top. \label{d13}
\eeqn
 Applying HODMD to these snapshots,
we obtain the counterpart  of the discrete expansion
in eq.(\ref{a1}).
Replacing $t_k$ by $t$ in this expansion leads to its continuous counterpart,
given by  eq.(\ref{a2}), which invoking the structure
of the snapshots (\ref{d13}) yields the evolution
of both $J(t)$ and $F(x_i,t)$. As a result,  the current density and the
electric field are approximated as
\begin{alignat}{2}
&J(t)\,\simeq\sum_{n=1}^Na_n u^J_n\,
\er^{(\delta_n+\smallir\,\omega_n)\, t},\label{d15}\\
&F(x_i,t)\,\simeq\sum_{n=1}^Na_n u^F_n(x_i)\,
\er^{(\delta_n+\smallir\,\omega_n)\, t},\label{d16}
\end{alignat}
for $i=1,\ldots,I$.
Here, the amplitudes $a_n$ are common to both
expansions, as are the growth rates
$\delta_n$ and the frequencies $\omega_n$.
In fact, for the periodic attractors computed below,
the involved frequencies
will include the zero frequency, which is associated with the
 {\it temporal mean field},
 while the remaining ones will be {\it positive} and
 {\it negative harmonics} of a
{\it fundamental frequency} $\omega_1$. Namely,  they will be
of the form
\beqn
\omega_p=p\,\omega_1, \,\,\text{for}\,\, p=-P,-(P-1),\ldots,0,\ldots, P-1,P.
\label{d17}
\eeqn
This permits rewriting the expansions (\ref{d15})-(\ref{d16}) as
\begin{alignat}{2}
&J(t)\,\simeq\sum_{p=-P}^P a_p u^J_p\,
\er^{\smallir\,p\,\omega_1\, t},\label{d18}\\
&F(x_i,t)\,\simeq\sum_{p=-P}^P a_p u^F_p(x_i)\,
\er^{\smallir\,p\,\omega_1\, t},\label{d19}
\end{alignat}
where the growth rates have been set to zero because we are considering
attractors.

In the following, we will apply
the HODMD method to first identify both the final
periodic attractor
and the strict decay to the attractor in the
transient stage. Then, the attractor
will be computed by performing various applications of
\textit{extrapolated} HODMD,  using data extracted
in timespans of limited temporal length.
In these cases, with
the standard PC described at the end of Section \ref{sec:intro},
the required CPU time to compute
the expansions (\ref{d18})-(\ref{d19}) is $\sim$1 CPU minute.


\subsubsection*{Identifying the attractor and the transient
 decaying dynamics using snapshots in the whole timespan
 \label{sec:Attractor-Transient}}
\begin{figure}[h!]
	\begin{center}
(a)\includegraphics[width=7cm,height=4cm]{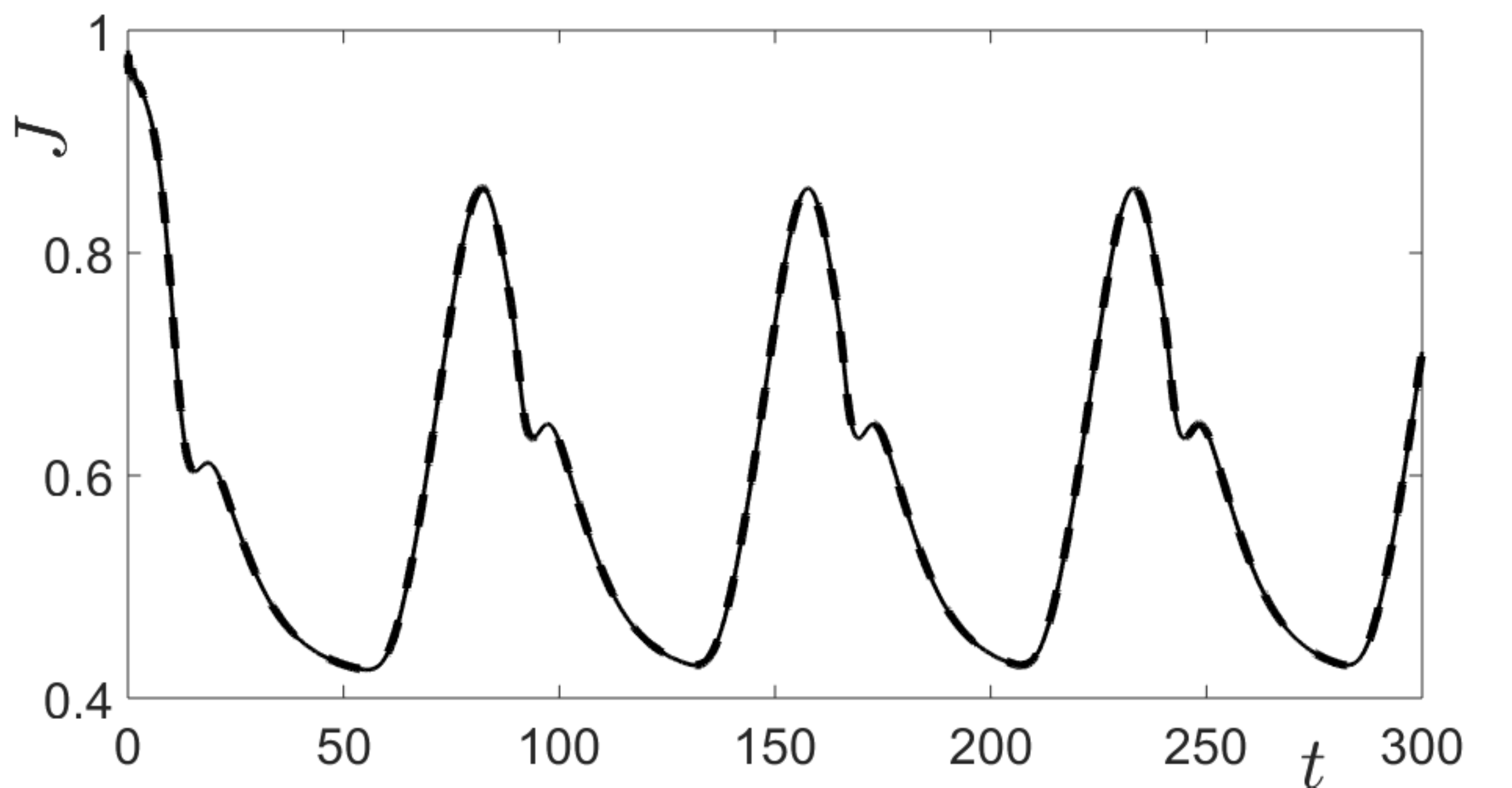}\hskip0.5cm
(b)\includegraphics[width=7cm,height=4cm]{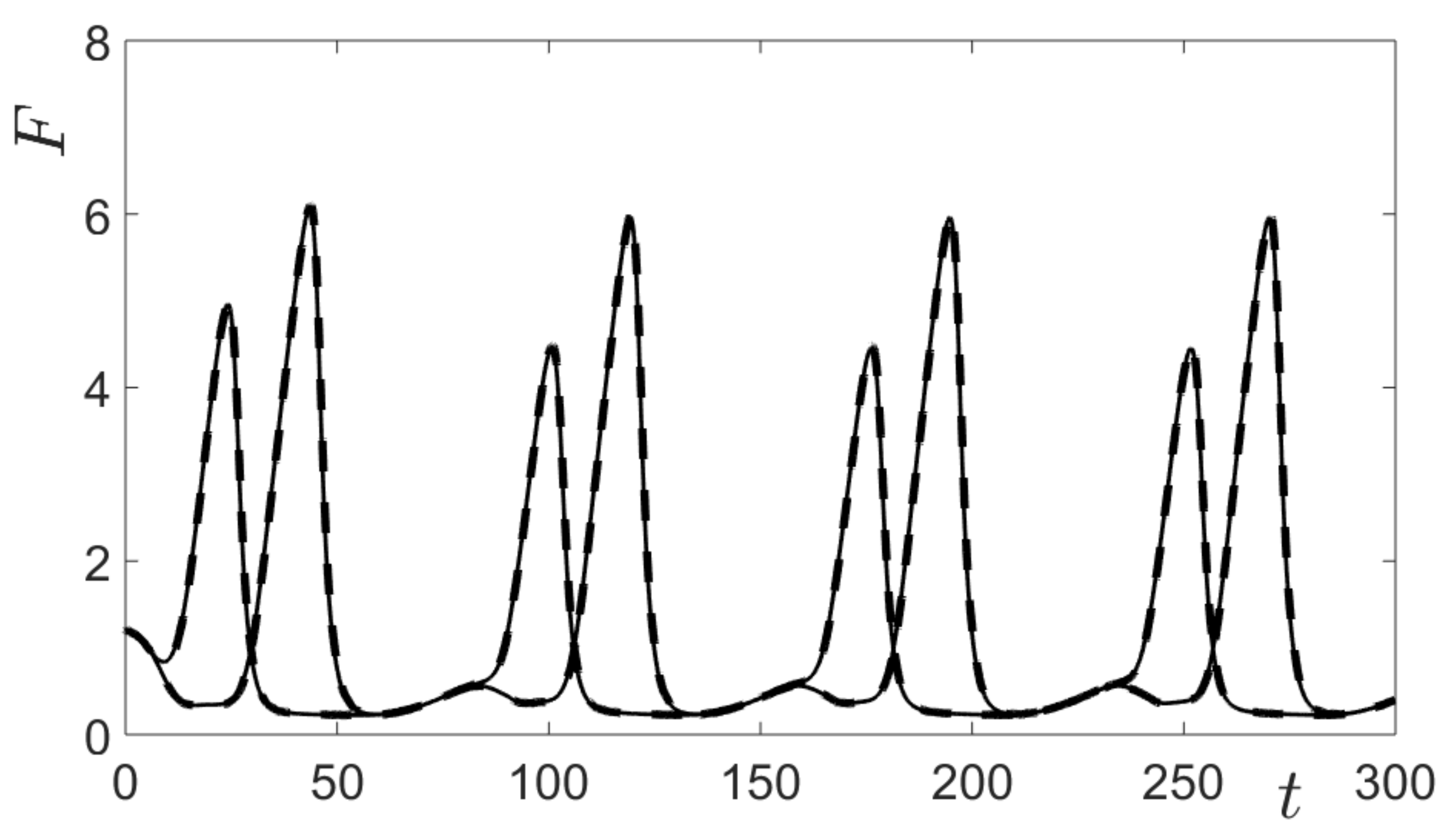}
	\end{center}
	\vskip-0.75cm
	\caption{Counterparts of Figs.\ref{fig:evolutionJ} (a)
and \ref{fig:evolutionFx1x2} (b) for the  HODMD method application
 with tunable parameters as in eq.(\ref{d20})
 and $K = 3000$ snapshots in $0 \leq t \leq300$.
The original data are plotted with thin solid lines,
while the  reconstructed data with
thick dashed lines. \label{fig:evolutionAllData}}
\end{figure}

To begin with, the HODMD method is applied to the entire set of $K=3000$ snapshots,
defined in eq.(\ref{d13}). After a slight calibration, the tunable parameters of the
HODMD method  (see Appendix \ref{sec:HODMD}), namely the dimension reduction threshold,
$\varepsilon_\text{SVD}$, the mode truncation threshold, $\varepsilon_\text{DMD}$,
and the index $d$ to apply the DMD-$d$ algorithm,  are selected as
\beqn
\varepsilon_\text{SVD}=10^{-7},\quad \varepsilon_\text{DMD}=10^{-4},
\quad\text{and }\,\,\,\,d=25.\label{d20}
\eeqn
Using these values, the DMD-25 algorithm reconstructs the given snapshots,
retaining $N = 113$ modes. To elucidate the performance of the HODMD method,
 we define the {\it relative root mean square} (RRMS) error
\beqn
\text{RRMS error}=\sqrt{\frac{\sum_ {k=1}^K\|\bq_k^\text{approx}-\bq_k\|^2_2}{\sum_ {k=1}^K\|\bq_k\|^2_2}},\label{d21}
\eeqn
where $\|\cdot\|_2$ denotes the usual Euclidean norm.
For the current density and the electric field, the RRMS errors are
$\sim9.4\cdot10^{-4}$ and $2.4\cdot10^{-3}$, respectively.
The approximations of both state variables are illustrated in
Fig.\ref{fig:evolutionAllData}.
It is interesting to point out
that using the DMD-1 algorithm (namely, standard DMD
\cite{SchmidH,Schmid2010}), with
the same values of
thresholds $\varepsilon_\text{SVD}$ and
$\varepsilon_\text{DMD}$ given in eq.(\ref{d20}),
the RRMS errors for $J$ and $F$
are much larger, namely $\sim0.034$ and 0.15, respectively.
The worse performance of the DMD-1 algorithm is due to the
fact that, to the approximation relevant here, the spatial
complexity (i.e., the rank of the set of modes), 64,
is clearly smaller than the spectral complexity
(i.e., the number $N$), 113.
The number of modes retained by the DMD-1 algorithm is 110, namely
it is slightly smaller than the spectral complexity
(see Appendix \ref{sec:HODMD} for further details on this issue).

\begin{figure}[h!]
	\begin{center}
(a)\includegraphics[width=7cm,height=5cm]{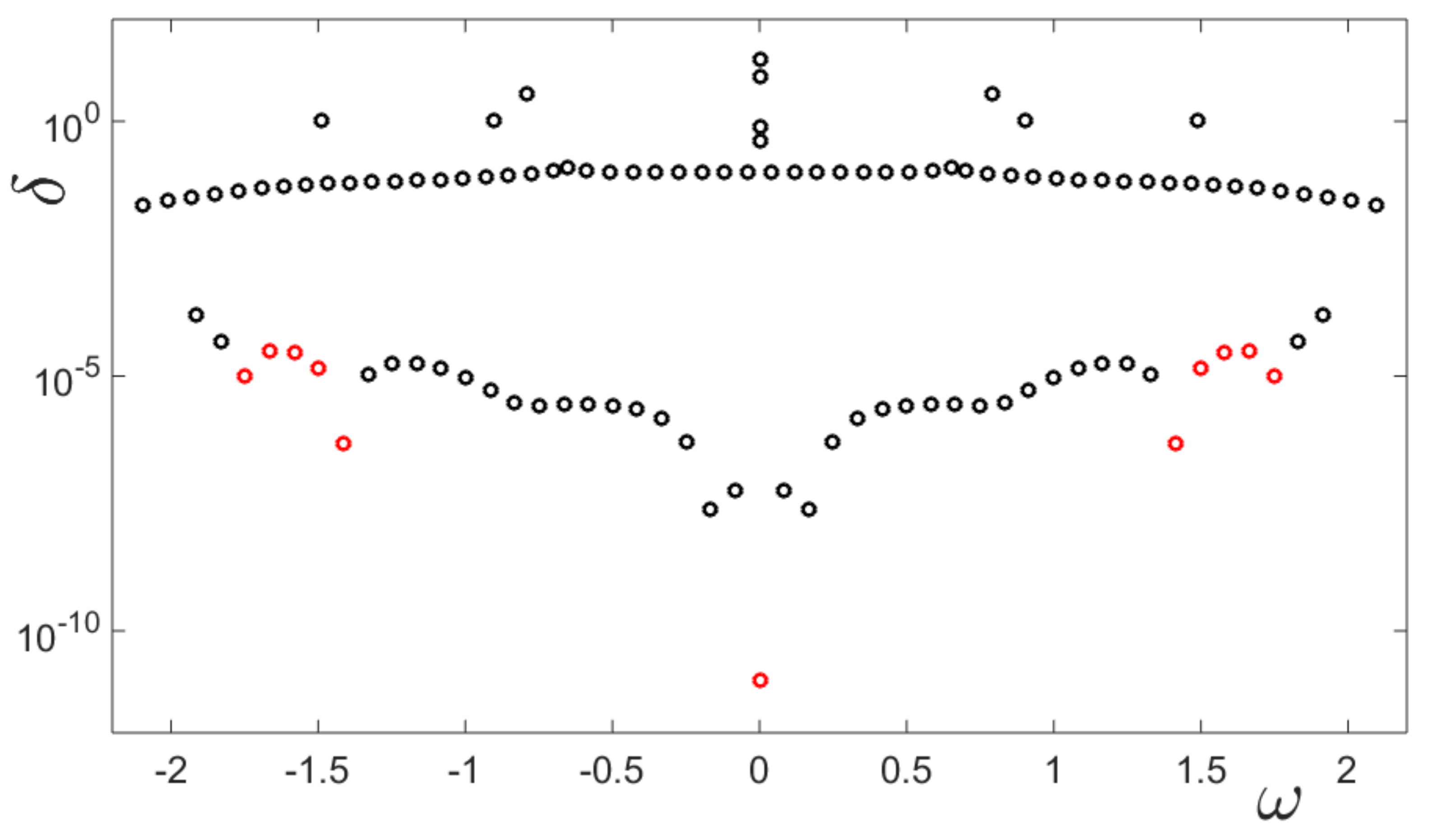}\hskip0.5cm
(b)\includegraphics[width=7cm,height=5cm]{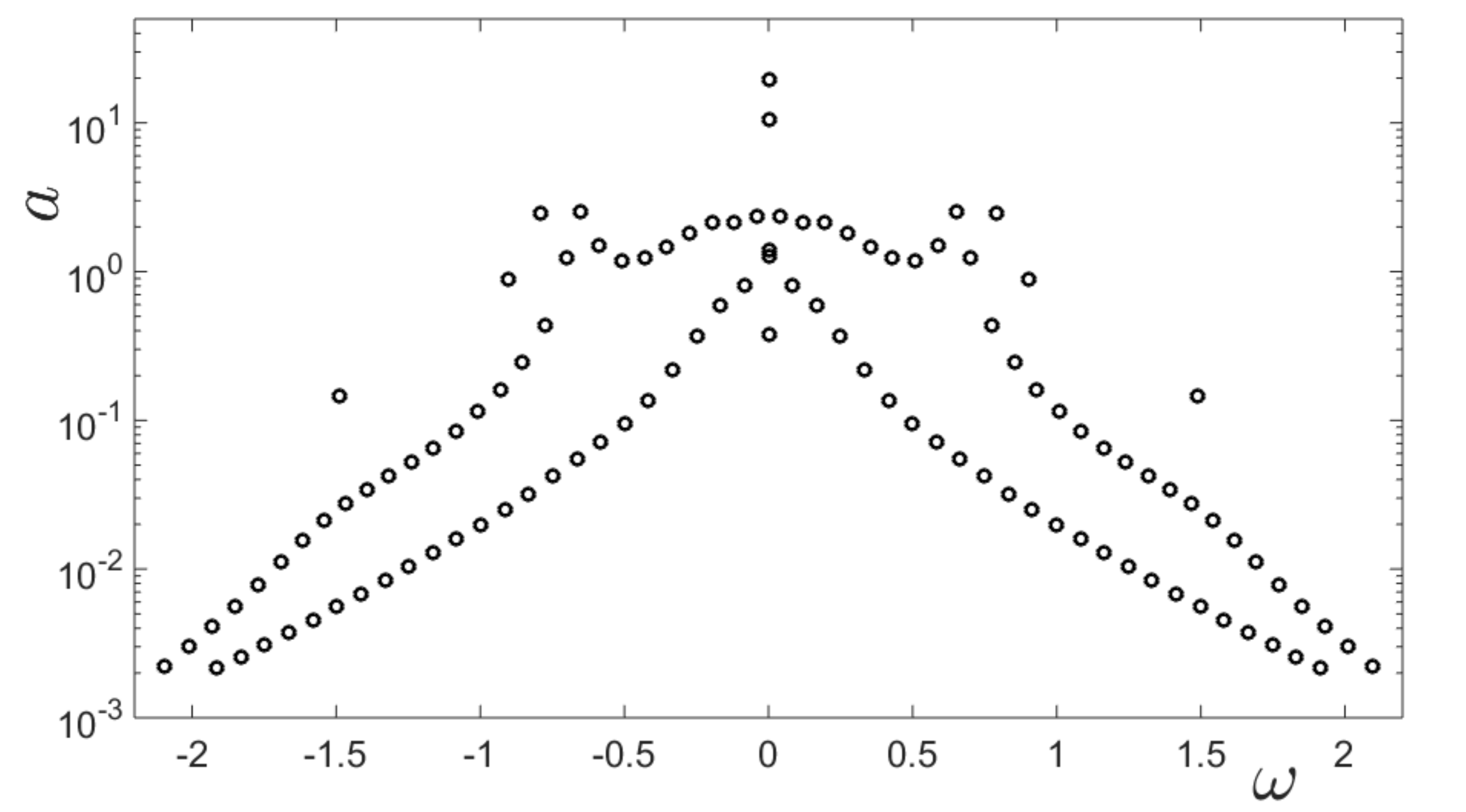}
	\end{center}
	\vskip-0.75cm
\caption{Growth rates (a)
and amplitudes (b) vs. the frequencies for the retained modes,
using the HODMD
method with tunable parameters as in eq.(\ref{d20})
and $K = 3000$ snapshots in $0\leq t\leq300$.
In plot (a), positive and negative growth rates are displayed
in red and black, respectively. \label{fig:diagramsAllData}}
\end{figure}

Inspection of the diagrams of the growth rates and amplitudes vs.
the frequencies
is enlightening. As can be seen in
Fig.\ref{fig:diagramsAllData},
both plots are symmetric around $\omega=0$, which is due
to the fact that the considered data are real. And, what is more important
in the present context, plot (a) shows
two distinguishable groups of growth rates, which are separated by a wide gap
between $\delta \sim -10^{-4}$ and $\delta \sim -10^{-2}$.
These two groups give rise to different spatio-temporal patterns, which can
be identified as anticipated in Section \ref{sec:intro}.
\begin{itemize}
	\item Since the 47 points in the lower
	group in Fig.\ref{fig:diagramsAllData}--(a)
	exhibit very small growth rates in absolute
	value (namely, smaller than $\sim10^{-4}$),
	this group is seemingly associated with the periodic {\it final attractor}.
	These
	(small but) non-zero growth
	rates are either positive or negative, but
	this is an artifact due to  numerical errors, both in the
	given data and in the HODMD computations.
	In fact, they are set to zero in the attractor reconstruction.
	\item The upper group
	of points in Fig.\ref{fig:diagramsAllData}--(a)
	exhibits significant, negative growth rates. Thus, it seems to be
	associated with the strictly {\it decaying approach to the attractor}.
\end{itemize}
These two groups are now analyzed to uncover the associated spatio-temporal
patterns.

\begin{figure}[h!]
	\begin{center}
(a)\includegraphics[width=7cm,height=5cm]{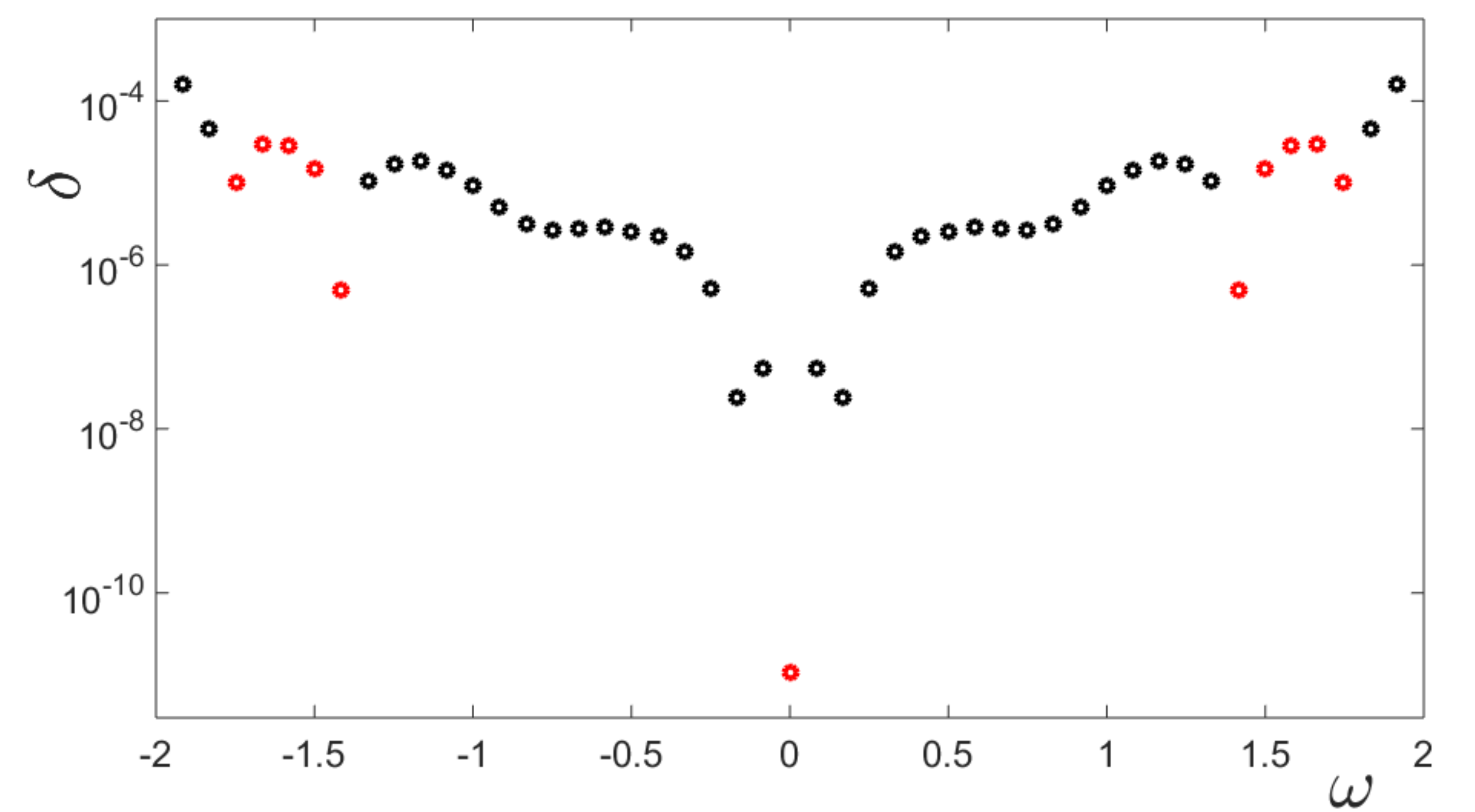}
		\hskip0.5cm (b)\includegraphics[width=7cm,height=5cm]{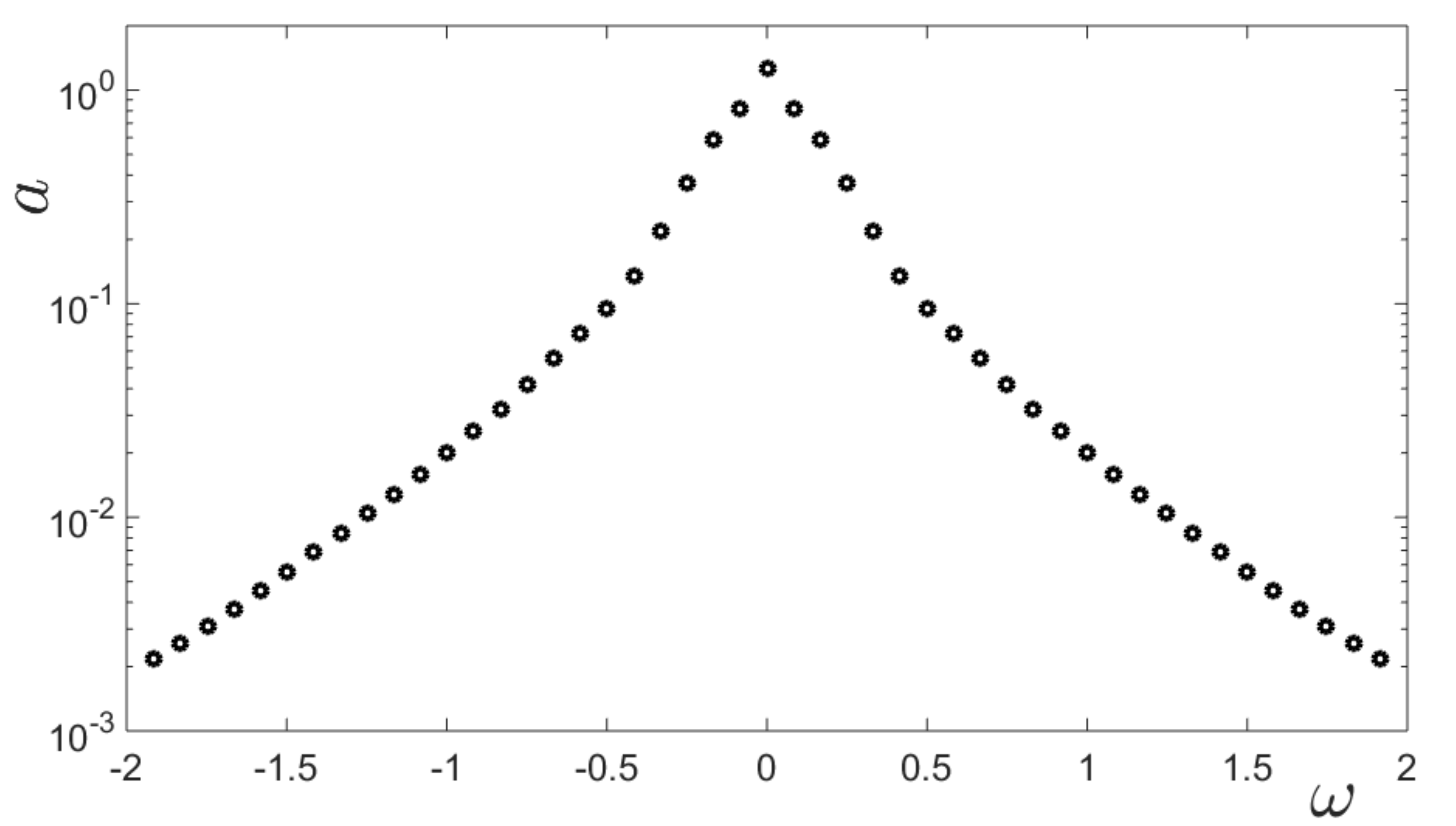}
	\end{center}
	\vskip-0.75cm
	\caption{Counterpart of Fig.\ref{fig:diagramsAllData}, considering
		only the lower group  of points in both plots.
		\label{fig:diagramsAllDataPeriodic}}
\end{figure}

For the 47 points in the lower group, the counterpart of
Fig.\ref{fig:diagramsAllData}
is given in Fig.\ref{fig:diagramsAllDataPeriodic}.
Note that the plot of the amplitudes vs. the frequencies shows
spectral decay, which is always seen in data resulting from
smooth periodic dynamics
\cite{LeClaincheVegaSIADS17}. Moreover, in addition to
the frequency $\omega=0$
(which corresponds to the temporal mean field), the
remaining 23 positive and 23 negative
frequencies are
positive and negative harmonics of the {\it fundamental frequency}
\beqn
\omega_1 = 8.3212\cdot10^{-2}.
\label{d22}
\eeqn
Namely, the various frequencies are of the form
$\pm p\,\omega_1$ for $p=0,1,\ldots,23$,
with four exact significant digits. In fact, the fundamental
frequency  $\omega_1$  is calculated as follows.
First, the 47 frequencies are  sorted in
increasing order. Then, the differences
between two consecutive frequencies are calculated and
seen to be approximately constant
along the sequence.  Finally, $\omega_1$ is
computed as the arithmetic mean of these differences.

The value of  $\omega_1$ in eq.(\ref{d22}) gives the period of the orbit
as
\beqn
T_1 = 2\pi /\omega_1 = 75.508.\label{d23}
\eeqn

\begin{figure}[h!]
	\begin{center}
(a)\includegraphics[width=7cm,height=4cm]{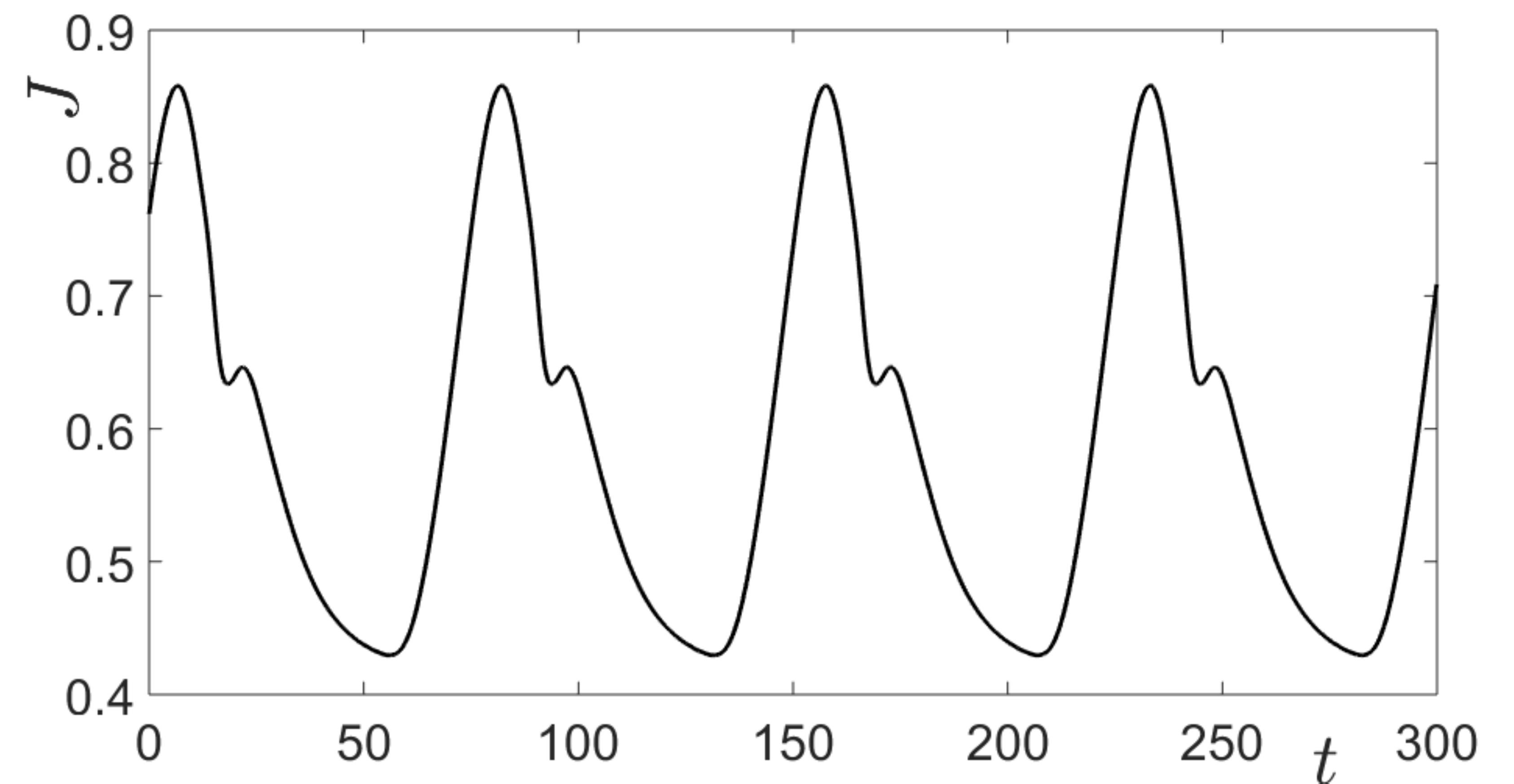}
\hskip0.5cm
(b)\includegraphics[width=7cm,height=4cm]{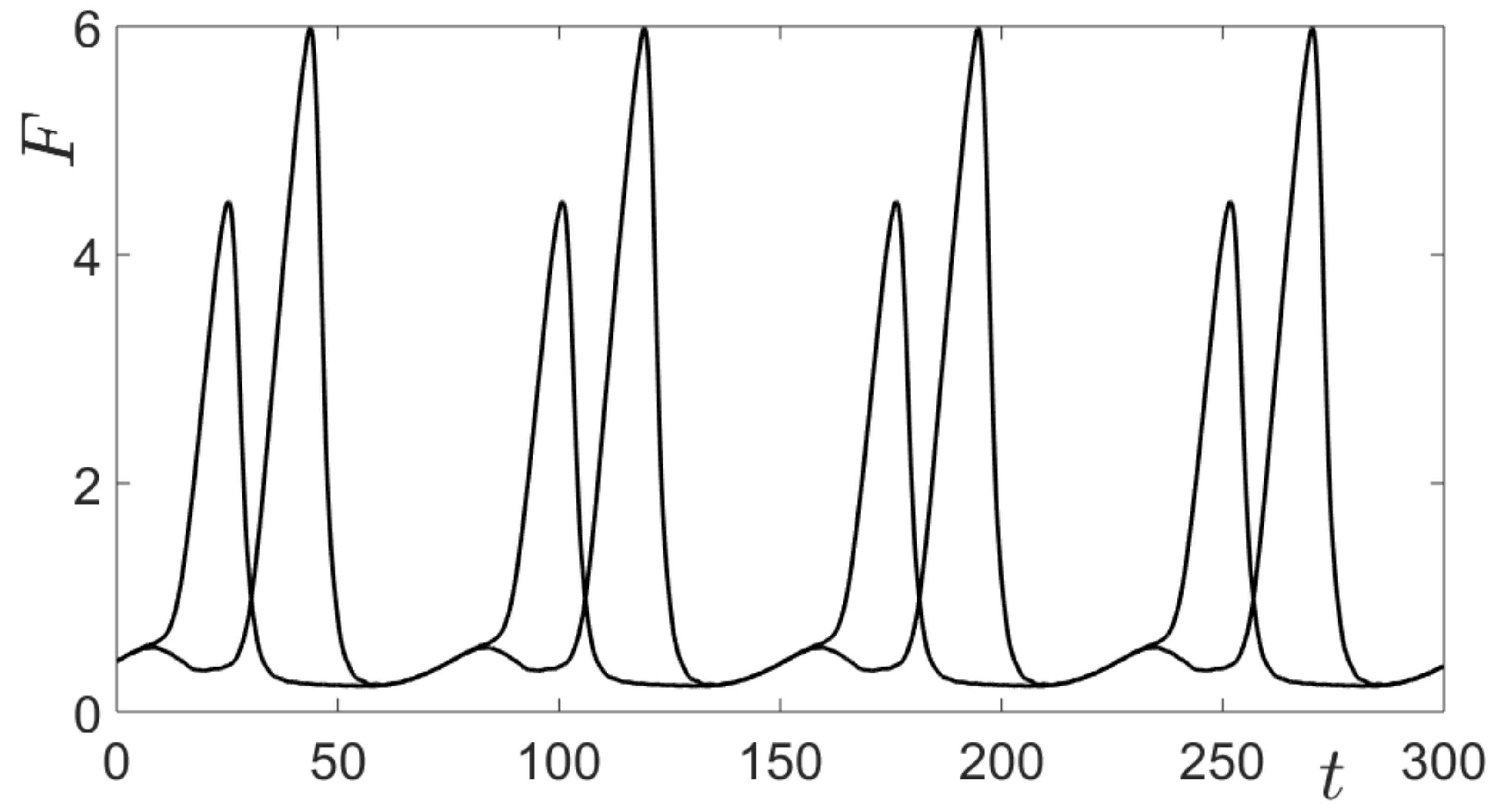}
	\end{center}
	\vskip-0.75cm
\caption{Counterpart of Fig.\ref{fig:evolutionAllData} for the periodic attractor. \label{fig:evolutionperiodic1}}
\end{figure}

On the other hand, as anticipated in Section \ref{sec:intro},
retaining only the 47 amplitudes, modes,
and frequencies appearing
in Fig.\ref{fig:diagramsAllDataPeriodic}, and setting
to zero the associated growth rates, the expansions
(\ref{d18})-(\ref{d19}) give the
evolution of the current  density and the electric field
for the periodic attractor.
It is interesting to note that, in the resulting
expansions, the spatial complexity
is 46, while the spectral complexity is 47.
These expansions
give (an approximation of) the periodic attractor in any
timespan.  Then, the attractor
is computed
in the whole time interval $0\leq t\leq300$ (although it
approximates the actual
dynamics only after the transient stage). In fact, this
computation of the asymptotic spatio-temporal pattern
will be compared with other approximations
obtained  below.
For illustration of the periodic attractor,
the current density and the electric field at $x=\widetilde x_1$
and $x=\widetilde x_2$
(see eq.(\ref{d9})) are given in Fig.\ref{fig:evolutionperiodic1}.

\begin{figure}[h!]
	\begin{center}
(a)\includegraphics[width=7cm,height=4cm]{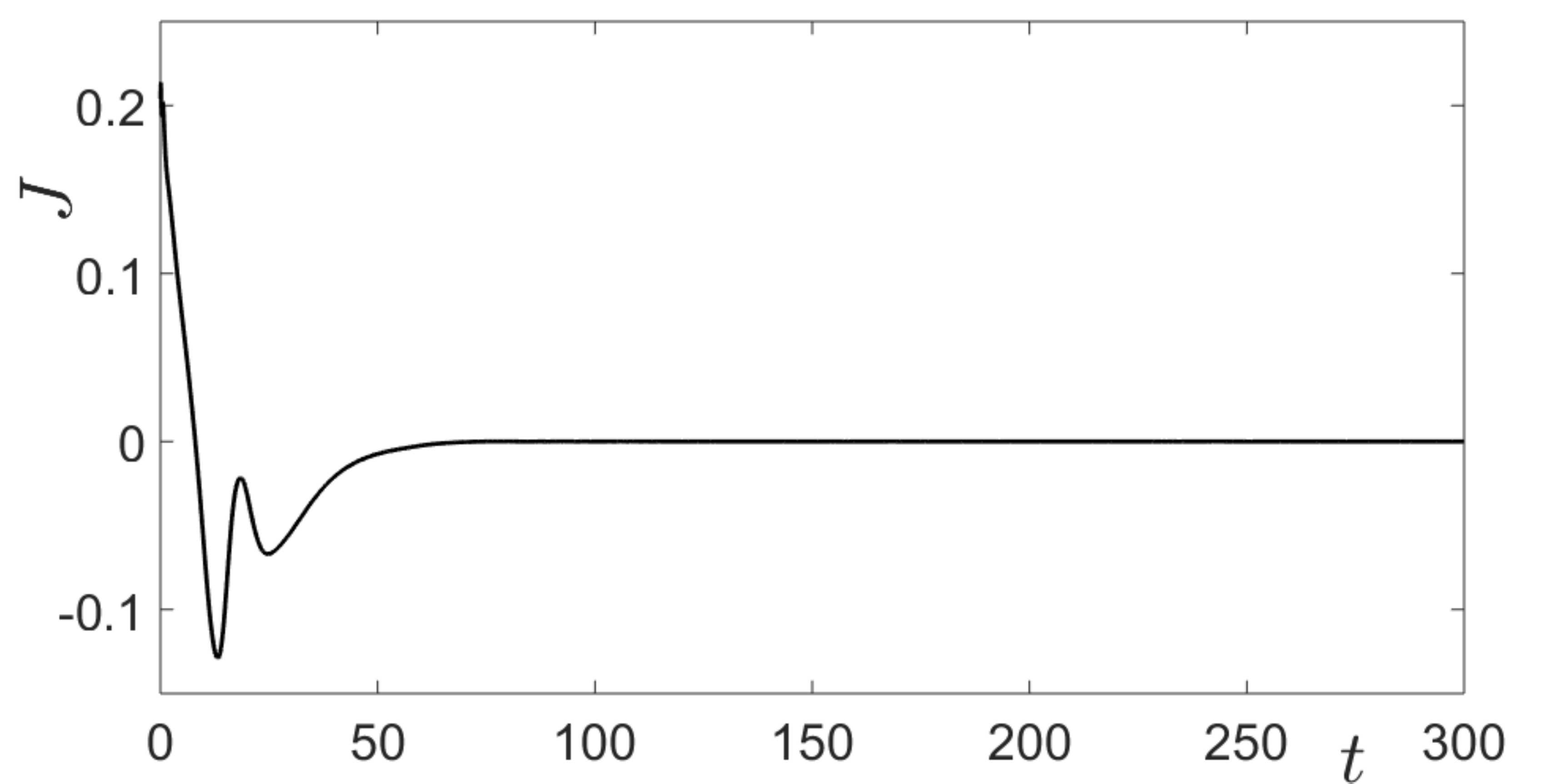}
\hskip0.75cm
(b)\includegraphics[width=7cm,height=4cm]{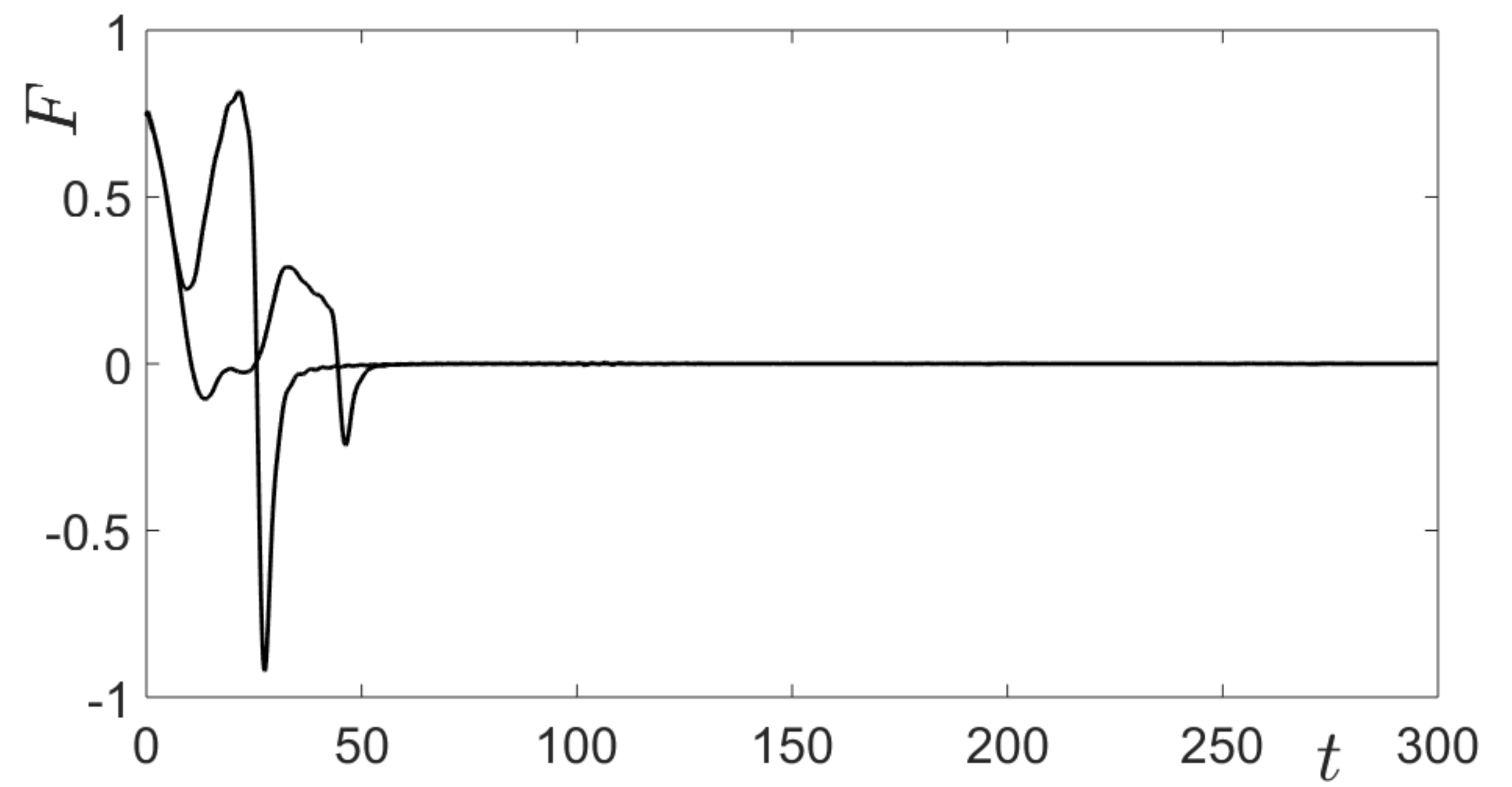}
	\end{center}
	\vskip-0.5cm
	\caption{Counterpart of Fig.\ref{fig:evolutionperiodic1} for the purely
		decaying dynamics. \label{fig:evolutiondecaying}}
\end{figure}

\begin{figure}[h!]
	\begin{center}
		\includegraphics[width=8cm,height=5cm]{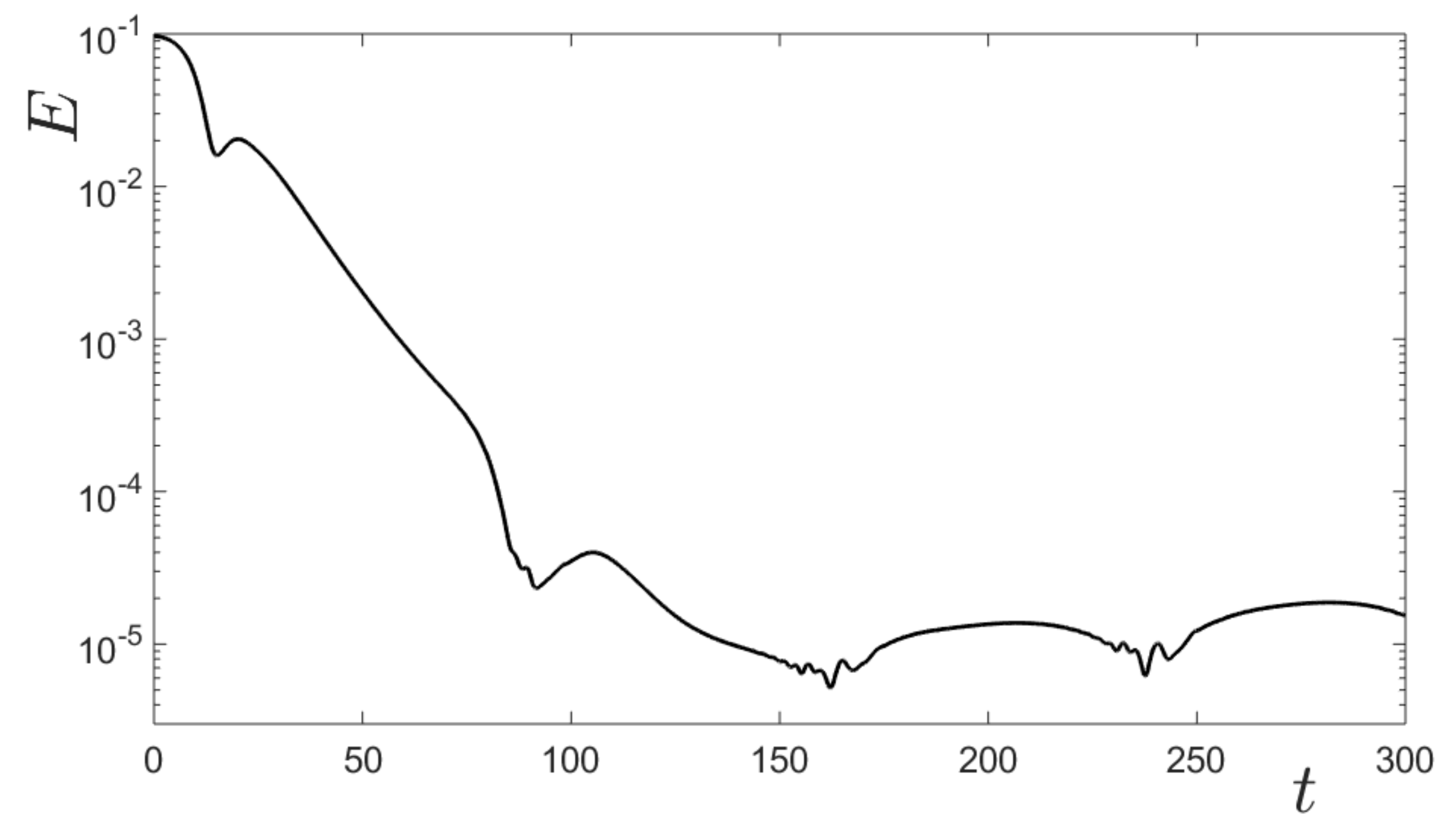}
	\end{center}
	\vskip-0.75cm
	\caption{Spatial RMS norm of the electric field for the strictly decaying
		dynamics. \label{fig:RMSdecayingF}}
\end{figure}

The spatio-temporal pattern associated with the upper group of $113-47=66$ points
in Fig.\ref{fig:diagramsAllData} is now analyzed.
In other words, the associated amplitudes, modes, growth rates,
and frequencies are used in the expansions (\ref{d15})-(\ref{d16}), which
permits reconstructing the transient dynamics
that obviously decay to zero as time goes to infinity.
Also, the purely decaying pattern is a good quantitative
description of the difference between
the whole dynamics and the periodic attractor, which
is seen in Fig.\ref{fig:evolutiondecaying}.
However, the temporal extent of the transient stage is not evident in this
figure to the naked eye. A good means to appreciate the
temporal convergence of the transient
to the attractor is to compute, for each value of $t$,
the spatial RMS norm of  the electric field for the
strictly decaying dynamics.
This RMS norm, denoted as $E$, is plotted vs.
 time in Fig.\ref{fig:RMSdecayingF},
which shows that such norm decreases (non-monotonously) until
$t\sim150$, where it saturates
due to computational errors. After this value of $t$,  it remains
approximately constant, except
for some oscillations. Thus, to the
approximation relevant here, the transient stage
can be assumed to be
\beqn
0 \leq t < 150.\label{d24}
\eeqn
In  this interval, the overall decrease of the strictly decaying
dynamics, as seen in the semi-logarithmic plot
in Fig.\ref{fig:RMSdecayingF}, is roughly a
straight line, whose negative slope is consistent with
the overall damping rate for the upper group of points
in  Fig.\ref{fig:diagramsAllData}--(a), which is
$-\delta\sim2\cdot10^{-2}$.


\subsubsection*{Computing the attractor via HODMD extrapolation\label{sec:Extrapolation}}
\begin{figure}[h!]
	\begin{center}
(a)\includegraphics[width=7cm,height=5cm]{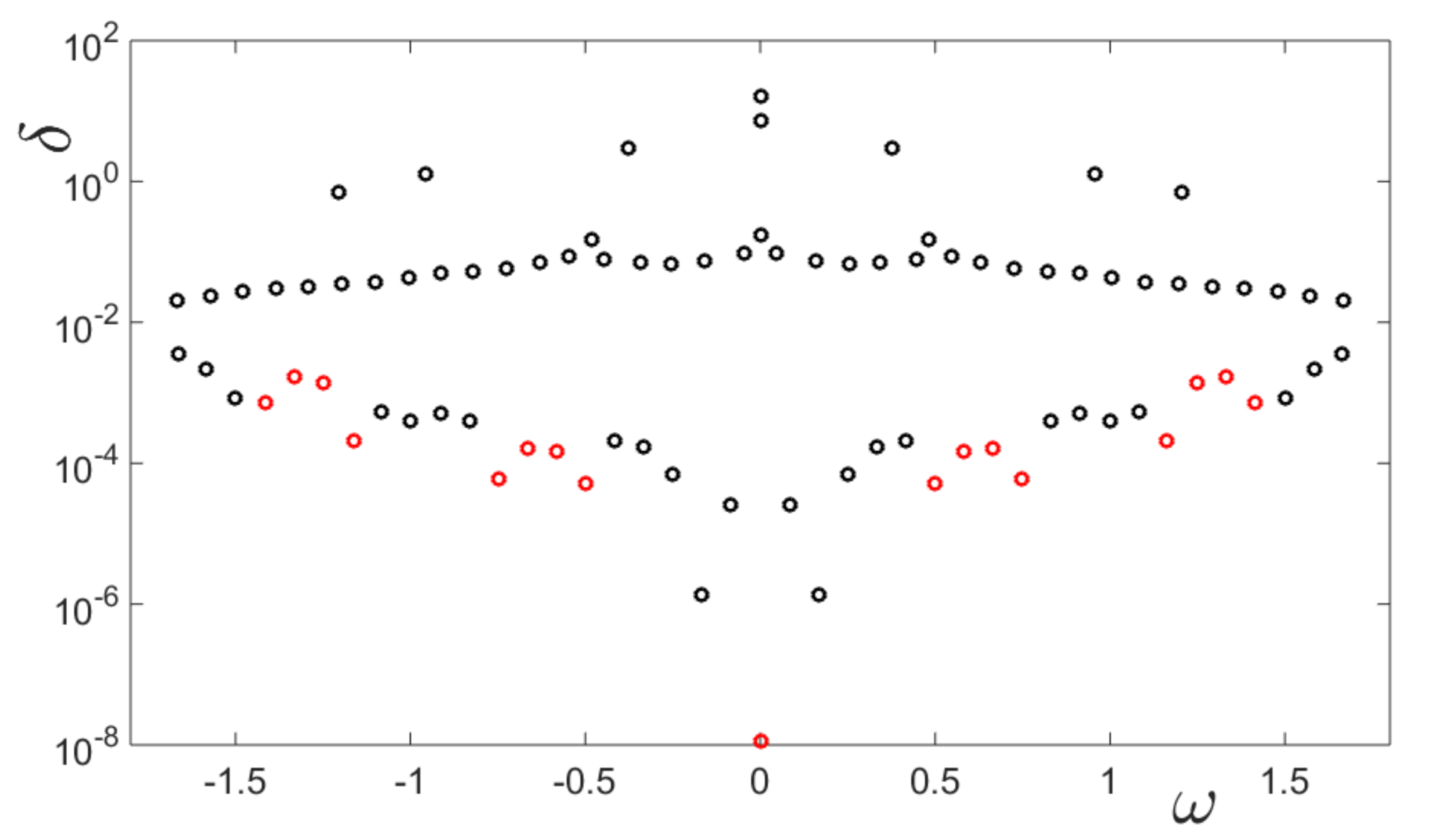}
\hskip0.5cm
(b)\includegraphics[width=7cm,height=5cm]{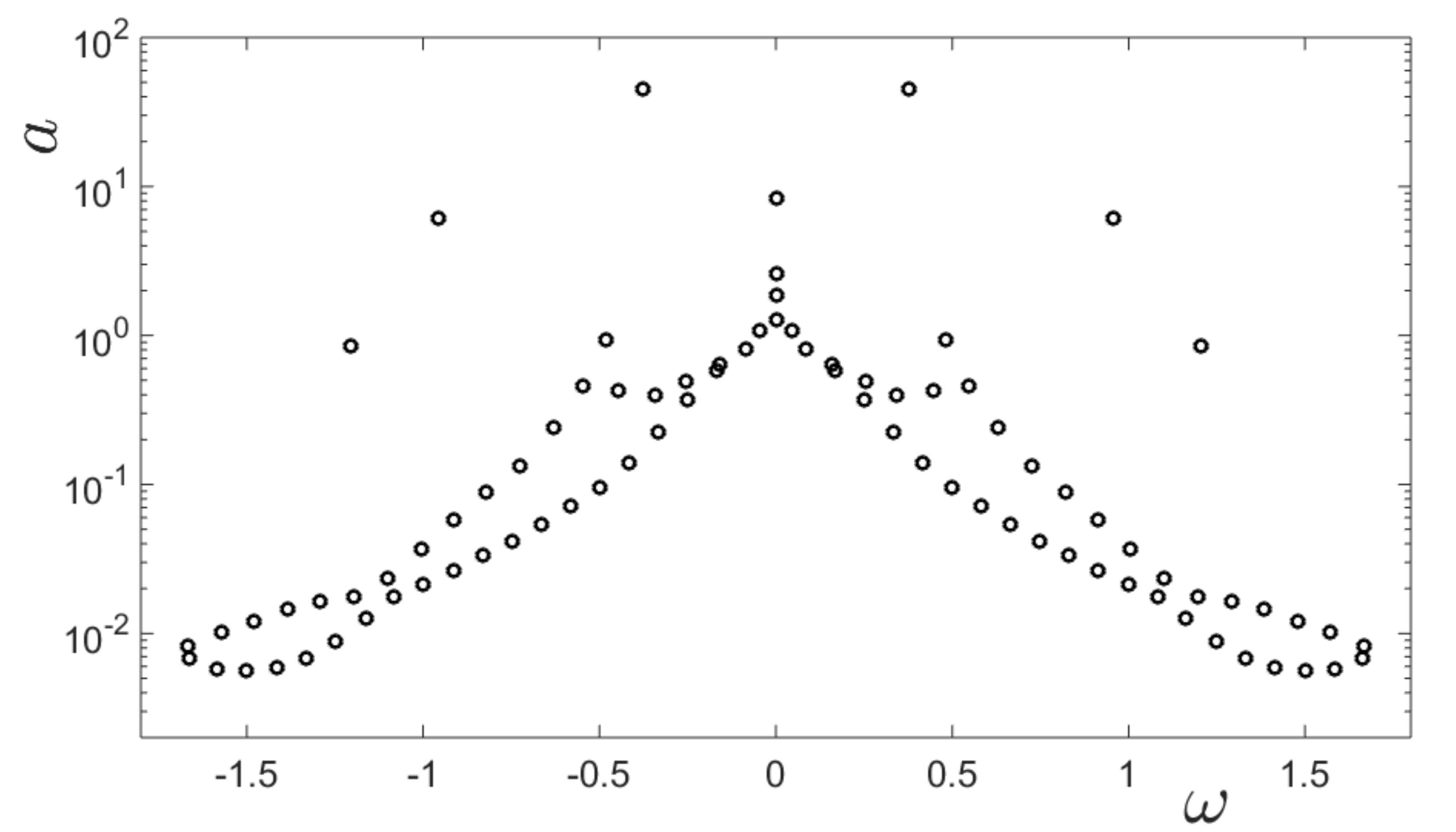}
	\end{center}
	\vskip-0.75cm
\caption{Counterpart of Fig.\ref{fig:diagramsAllData} for
		the application of HODMD to snapshots contained in the transient stage
		(\ref{d24}), with the tunable parameters given in eq.(\ref{d26}).
		\label{fig:diagramsTransient}}
\end{figure}

\begin{figure}[h!]
	\begin{center}
(a)\includegraphics[width=7cm,height=4cm]{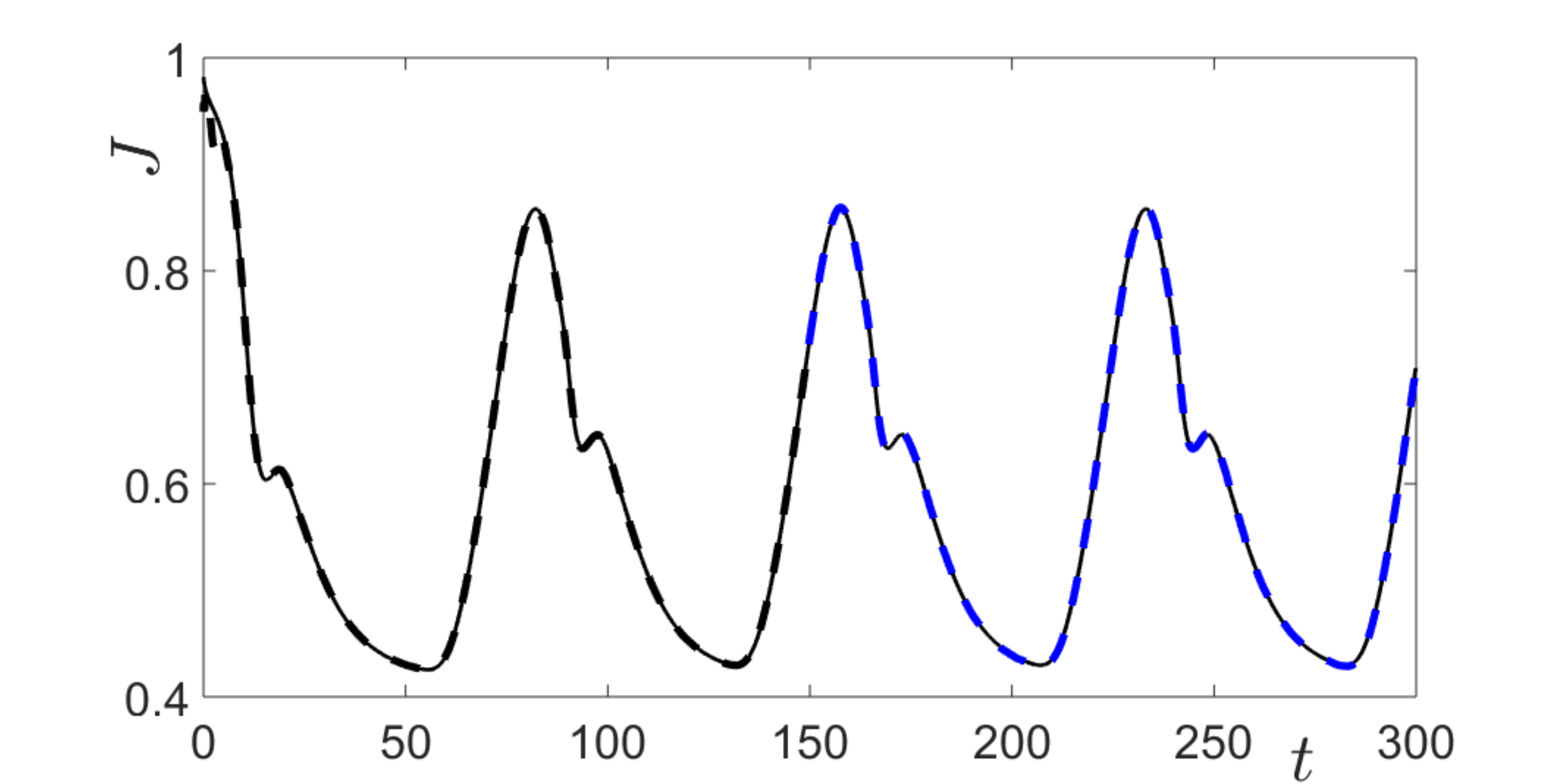} \hskip0.5cm
(b)\includegraphics[width=7cm,height=4cm]{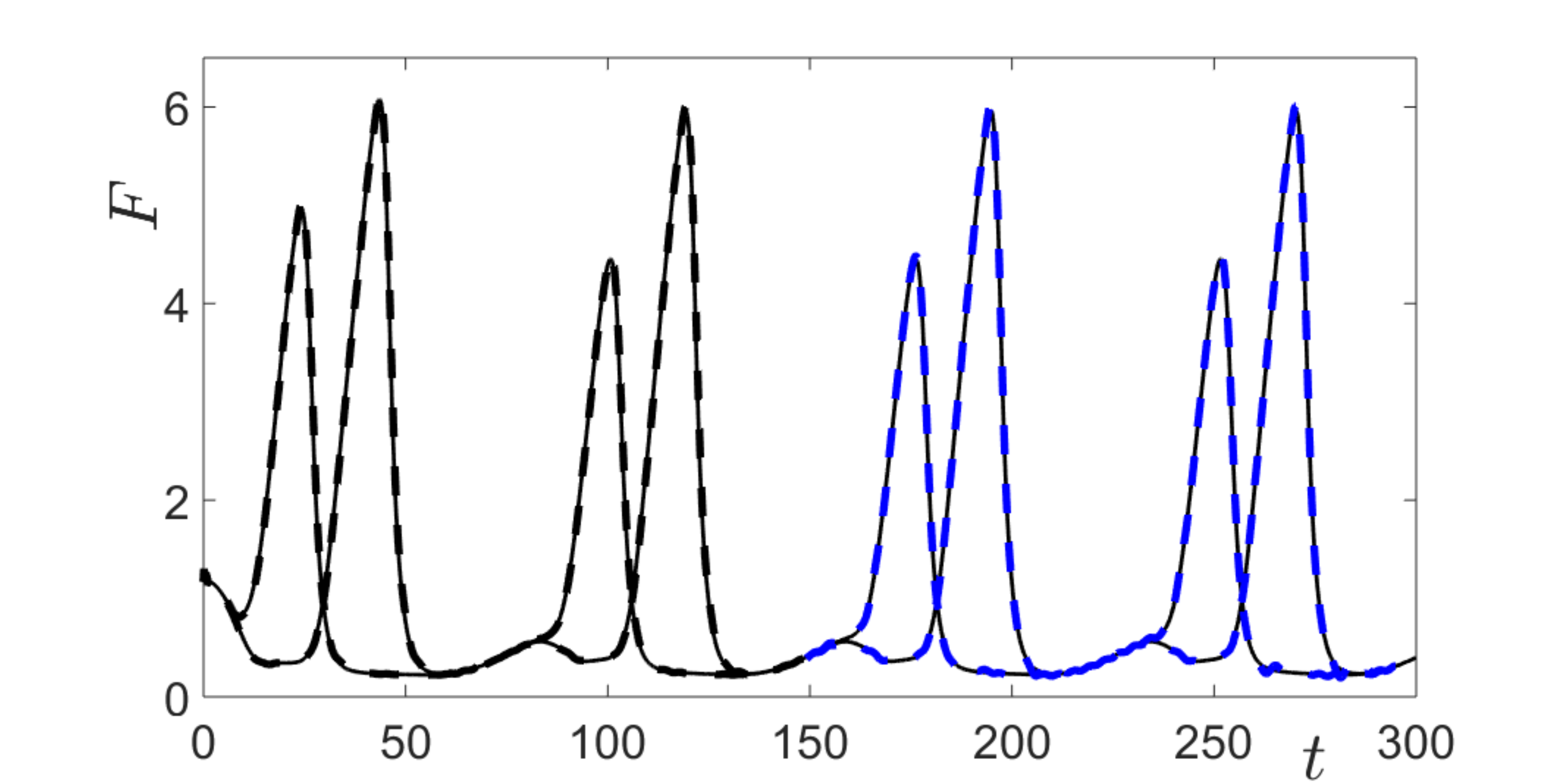}
	\end{center}
	\vskip-0.75cm
\caption{Counterpart of Fig.\ref{fig:evolutionAllData}
when applying extrapolated
HODMD using snapshots in the transient stage (\ref{d24}), with the tunable
parameters in eq.(\ref{d26}). In both plots,
the original data are shown
with thin solid lines, the reconstructed data in the
transient stage with thick dashed black lines,
and the extrapolated data with thick dashed blue lines.
\label{fig:evolutionTransient}}
\end{figure}

Now, let us consider the approach outlined in Section \ref{sec:intro}
to approximate the periodic attractor by HODMD \textit{extrapolation} from the
temporal interval (\ref{d24}), namely using  $K = 1500$ snapshots in the
transient stage. After some
slight calibration (controlling the RRMS
error, as defined in eq.(\ref{d21}),
of the HODMD reconstruction in the transient stage), the
HODMD tunable parameters are selected as
\beqn
\varepsilon_\text{SVD}=10^{-7} ,\quad \varepsilon_\text{DMD}=10^{-4},
\quad\text{and }\,\,\,\,\,d=45.\label{d26}
\eeqn
Using these values, the application of HODMD to the selected $K = 1500$
snapshots reconstructs  the latter with a RRMS
error $\sim5.91\cdot10^{-3}$ and $\sim6.95\cdot10^{-3}$ for the
current density and the electric field, respectively,
retaining $N = 88$ modes. The
associated growth rates, amplitudes,
and frequencies are displayed in Fig.\ref{fig:diagramsTransient}.
As seen in plot (a), the retained
 growth rates are organized in two groups, separated
by a gap around $\delta=-10^{-2}$. As in the previous case, this yields
two groups of modes. The upper group
seemingly corresponds to the purely decaying dynamics
and the lower group, which contains 41 points,
to the periodic attractor. The approximation
of the periodic attractor can be computed
beyond the transient stage
by extrapolation, using the
amplitudes, modes, and frequencies associated with the
lower group, and proceeding as anticipated
in Section \ref{sec:intro}.
It turns out that:
\begin{itemize}
	\item The obtained approximation of the fundamental frequency
	differs from  the value in eq.(\ref{d22})
	by a relative error $\sim5.11\cdot10^{-4}$.
	\item The approximation of the periodic attractor
	in the timespan $150\leq t\leq300$ is quite similar to
	its counterpart in Fig.\ref{fig:evolutionperiodic1}.
	Indeed, the RRMS difference between them
	is $\sim1.29\cdot10^{-3}$ for the
	current density and $\sim9.11\cdot10^{-3}$ for the electric field.
	Thus, HODMD extrapolation yields an accurate reconstruction of
	the attractor only using numerical data in the transient stage,
	which, in the present case, divides by two the CPU time of
	computing the snapshots.
\end{itemize}
These results are illustrated in Fig.\ref{fig:evolutionTransient},
which shows a reasonably good  approximation  in both the transient timespan and
beyond.

\begin{figure}[h!]
	\begin{center}
(a)\includegraphics[width=7cm,height=4cm]{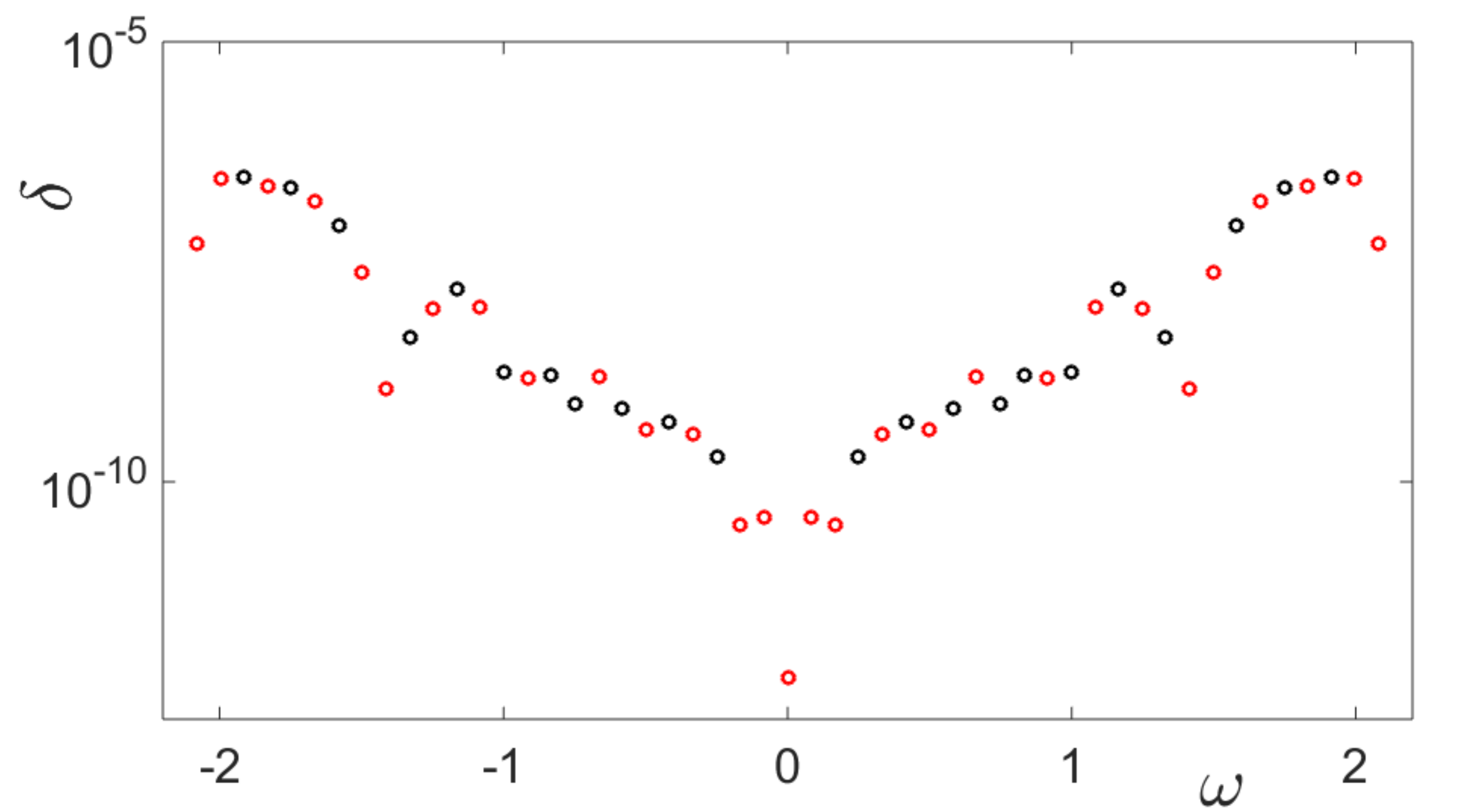}
\hskip0.5cm
(b)\includegraphics[width=7cm,height=4cm]{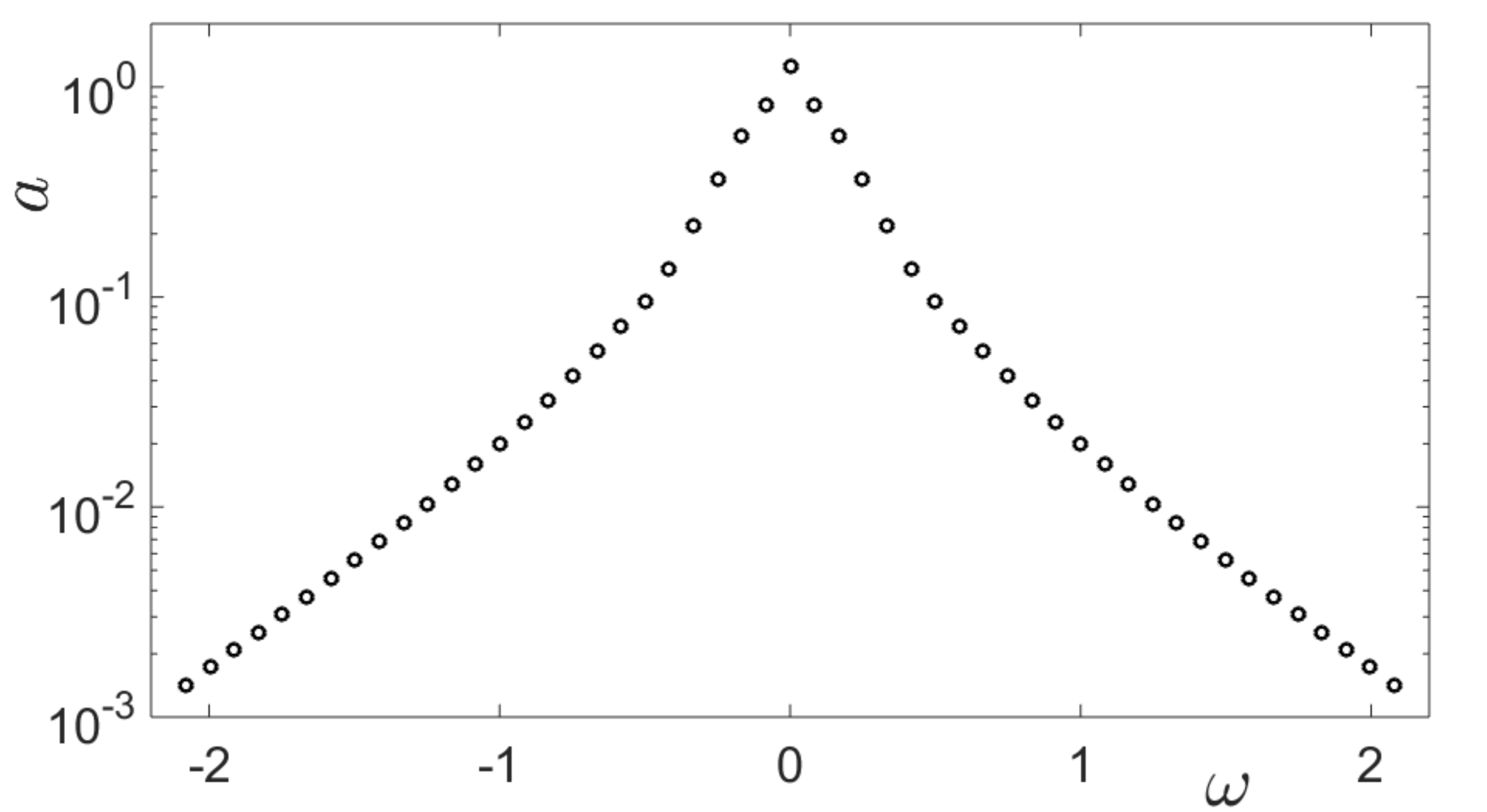}
	\end{center}
	\vskip-0.75cm
	\caption{Counterpart of Fig.\ref{fig:diagramsTransient} for
		the application of HODMD to snapshots contained in the timespan
		(\ref{d28}), with the tunable
		parameters given in eq.(\ref{d30}). \label{fig:diagramsExtrapPeriodic}}
\end{figure}

\begin{figure}[h!]
	\begin{center}
(a)\includegraphics[width=7cm,height=4cm]{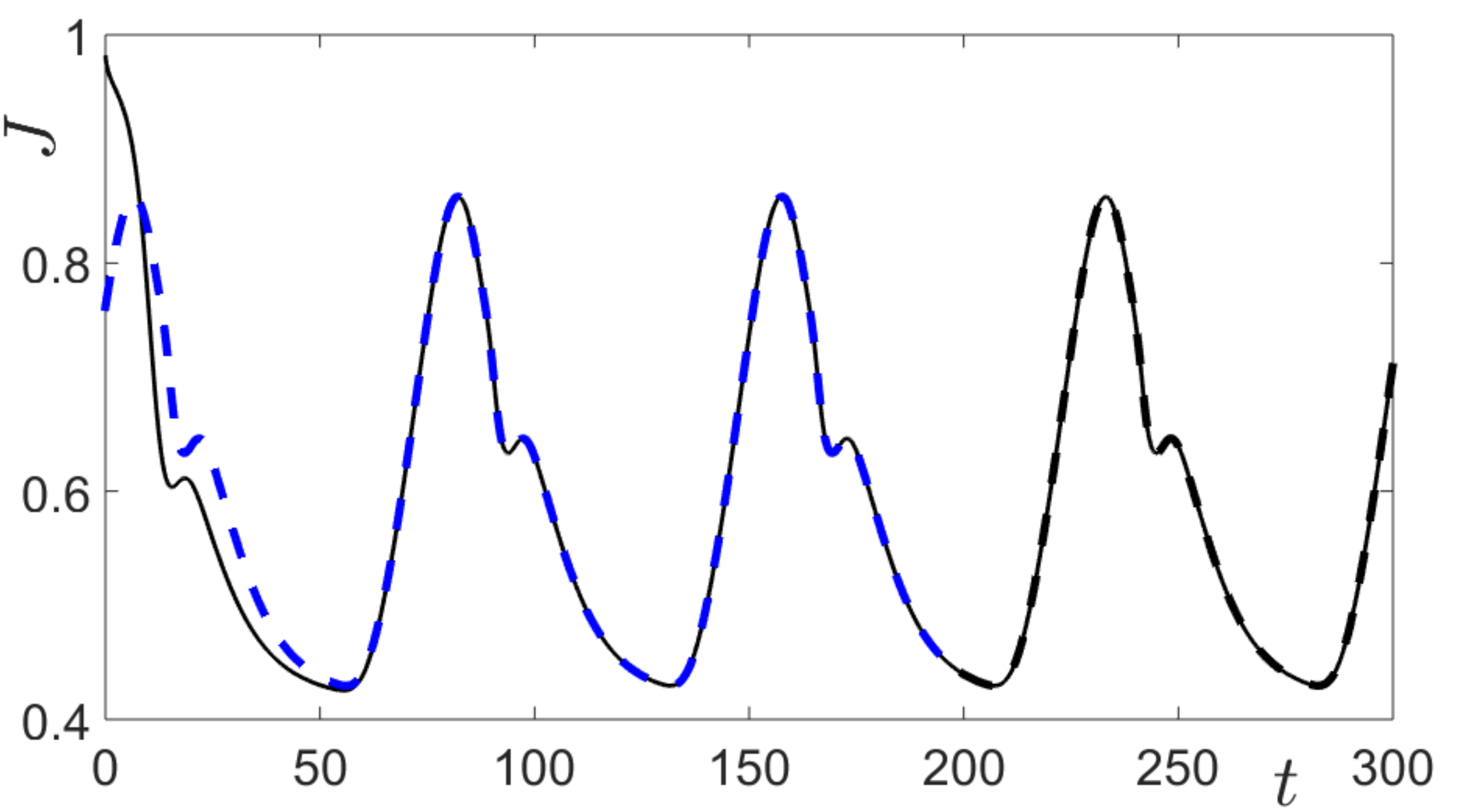}
\hskip0.5cm
(b)\includegraphics[width=7cm,height=4cm]{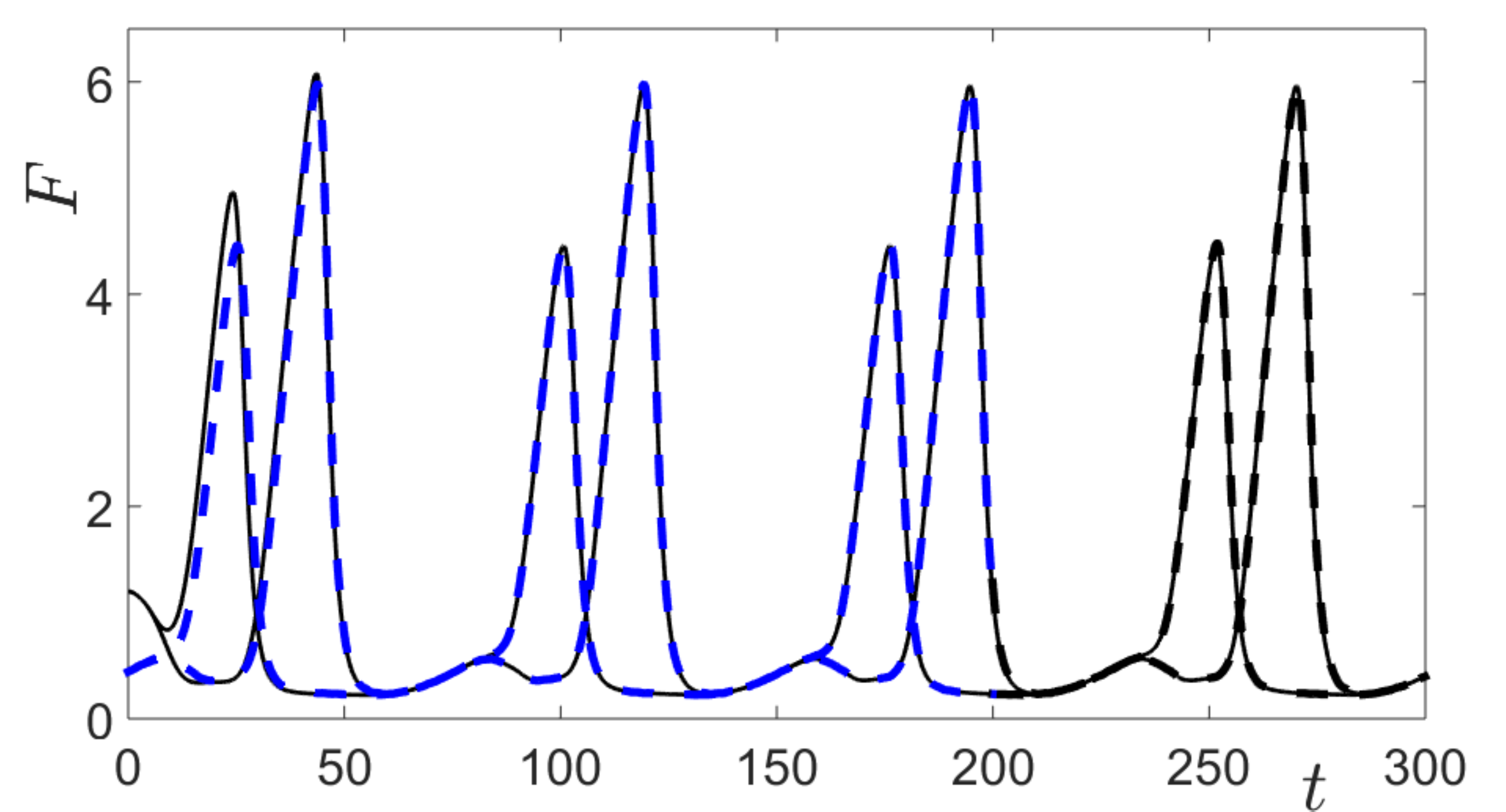}
\end{center}
	\vskip-0.75cm
	\caption{Counterpart of Fig.\ref{fig:evolutionTransient} when applying extrapolated
		HODMD using snapshots in the timespan (\ref{d28}), with the tunable
		parameters given in eq.(\ref{d30}). \label{fig:evolutionExtrapPeriodic}}
\end{figure}

As a second test case for HODMD temporal extrapolation, we reconstruct the attractor using
snapshots contained in the last third of the considered timespan, namely in
\beqn
200\leq t\leq300.\label{d28}
\eeqn
Note that this timespan is expected  to be in the region where the
dynamics are already very close to the periodic attractor.
The HODMD method is applied with the following
HODMD tunable parameter values
\beqn
\varepsilon_\text{SVD}=10^{-8} ,\quad \varepsilon_\text{DMD}=10^{-3},
\quad\text{and }\,\,\,\,\, d=20.\label{d30}
\eeqn
 Results show reconstructions of the current density and the
electric field, in the interval (\ref{d28}),  within a RRMS error,
as defined in eq.(\ref{d21}),
$\sim1.3\cdot10^{-4}$ and $\sim1.5\cdot10^{-3}$,
respectively, retaining $N = 51$ modes.
The diagrams giving the retained growth rates and amplitudes vs. the
frequencies are displayed in Fig.\ref{fig:diagramsExtrapPeriodic}.
Note that all growth rates are very small in absolute value, namely
smaller than $\sim10^{-6}$, while the amplitudes exhibit spectral decay.
The  frequencies include the zero frequency (related to the temporal mean field)
together with 25 positive and 25 negative harmonics
of a fundamental frequency. The latter
coincides  with the value in eq.(\ref{d22}) within a
relative error $\sim1.5\cdot10^{-4}$.
Also, the periodic attractor computed by HODMD backward
extrapolation differs from its counterpart in Fig.\ref{fig:evolutionperiodic1} by a
RRMS error $\sim1.5\cdot10^{-4}$ and
$\sim1.6\cdot10^{-3}$ for the current density and the
electric field, respectively.
 These good results are illustrated in
Fig.\ref{fig:evolutionExtrapPeriodic}.
As can be seen in this figure, the backward extrapolation separates from
the original data in the transient  stage since, in the present case,
it yields the periodic attractor by construction.


\subsection{Comparison with results obtained by the FFT \label{sec:FFT}}
In this subsection, the illustrated HODMD-based results
are compared with
those obtained using the MATLAB function `fft', which implements
a standard algorithm based on the
discrete Fourier transform (DFT). Recall that FFT applies
DFT to the given data, which means that the method
somewhat assumes that the dynamics are periodic, with a period equal
to the temporal interval where the data have been  computed. This spurious
period introduces {\it sideband artifacts}, consisting in erroneous
frequencies that roughly concentrate (with small amplitudes)
around the actual frequencies.

\begin{figure}[h!]
	\begin{center}
		\hskip-0.5cm (a)\includegraphics[width=7cm,height=6cm]{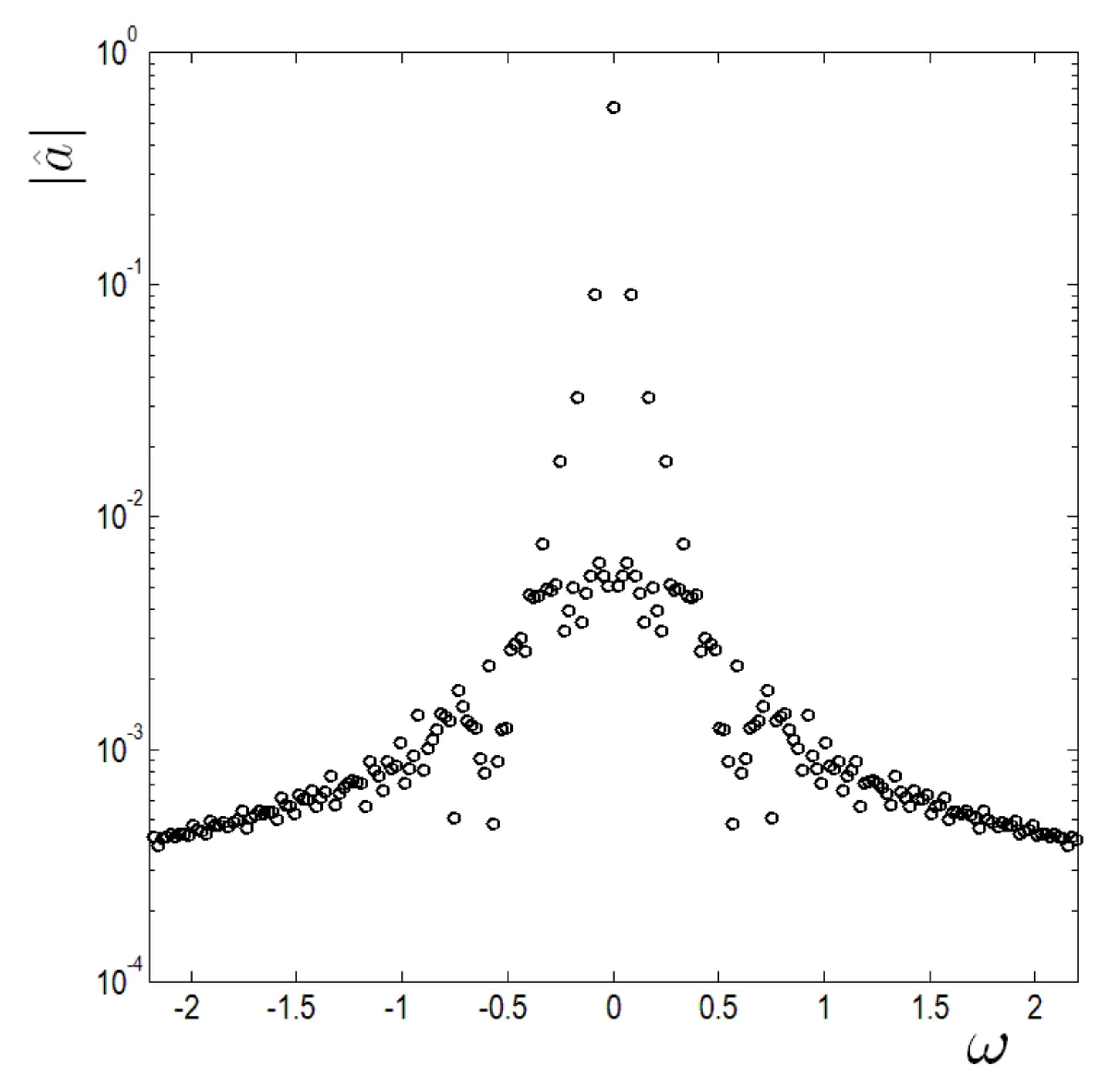}
\hskip0.5cm
(b)\includegraphics[width=7cm,height=6cm]{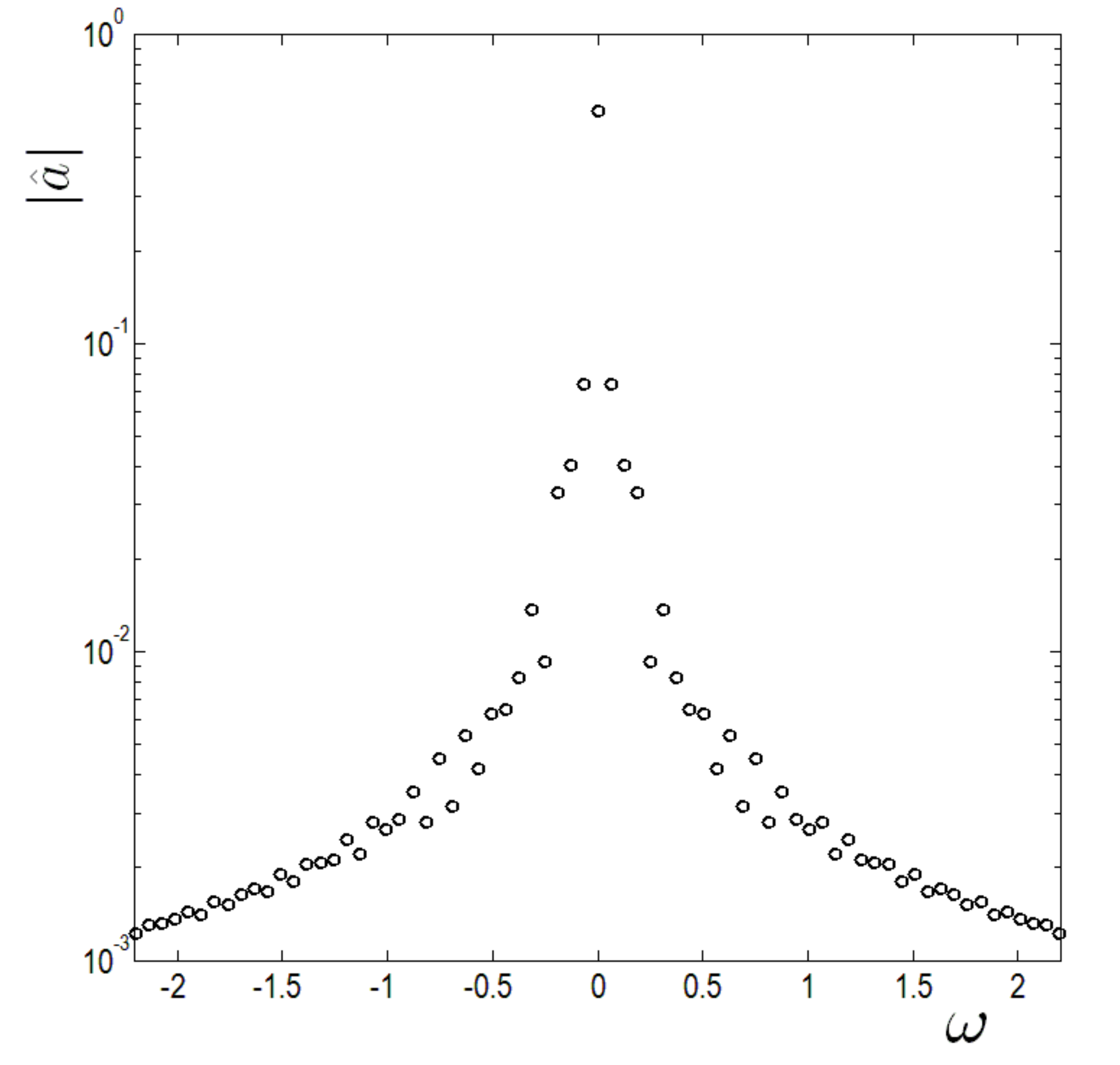}
	\end{center}
	\vskip-0.75cm
\caption{Fourier amplitudes vs. the frequencies obtained via FFT applied to data
of the current density in the temporal intervals $0\leq t\leq300$ (a)
and $200\leq t\leq300$ (b). \label{fig:fft}}
\end{figure}

For illustration, only data regarding the
current density will be used.
First, the FFT is applied to the values of the current density
computed by the numerical solver described in
Section \ref{sec:Numerics} at $K = 3000$ time instants, as in eq.(\ref{d11}),
collected in the timespan (\ref{d4}), $0\leq t\leq300$, and
sampled at distance $\Delta t = 0.1$.
Thus, the sampling frequency is the same as
the one implicit in the application of HODMD in the previous subsection.
Plot (a) in Fig.\ref{fig:fft}
shows the absolute
value of the FFT-calculated amplitudes vs. the frequencies.
As can be seen, the FFT captures
a large number of frequencies,
which include the mean field
frequency $\omega=0$, the fundamental frequency
$\omega_1 = 8.378 \cdot 10^{-2}$, and three harmonics of
it, all with reasonable accuracy.
The remaining frequencies computed by the FFT are all spurious
and include a few (positive and negative) subharmonics of the fundamental
frequency. This is obviously much worse
than the outcome of HODMD that, in addition to
 isolating the modes associated with the final periodic attractor
from those yielding the transient approach to the attractor,
was able to very accurately identify 23 harmonics of the fundamental frequency
in the periodic attractor, as
seen in Fig.\ref{fig:diagramsAllDataPeriodic}.
Instead, the FFT sideband artifacts produce a
large amount of additional spurious frequencies.
Indeed, with these spurious modes, the FFT `tries' to describe
the decaying behavior
due to the actual modes exhibiting strictly negative growth rates, which
cannot be computed by the FFT  itself.

As a second comparison between FFT and HODMD, the current density
data (already used for HODMD)
in the timespan $200 \leq t \leq 300$, as in eq.(\ref{d28}),
are treated via FFT, which identifies a large number of frequencies,
all of them spurious, as seen in Fig.\ref{fig:fft}--(b).
In other words, not even the fundamental frequency is captured
by the FFT in the present case, while HODMD was able to identify it well, together
with 25 harmonics, as seen in Fig.\ref{fig:diagramsExtrapPeriodic}.

Summarizing, the performance of HODMD is  superior to that of FFT
in connection with both reliability and efficiency.
In other words, HODMD identifies  more clearly a larger number of
the actual frequencies describing
the periodic attractor. Besides, HODMD is able to  separate the
modes describing the permanent dynamics from
their counterparts associated with the transient behavior.


\section{Analysis of the periodic attractor via STKD\label{sec:atractorSTKD}}
In this section, we further study the spatio-temporal structure
of the periodic attractor and the underlying traveling wave dynamics
of the electric field by using the STKD method. The periodic
attractor has been computed from its transient by means of the HODMD method
in Section \ref{sec:Results}. As anticipated in Section \ref{sec:intro},
the STKD method  yields an expansion of the form
\beqn
F(x,t)\simeq\sum_{(m,n)\in R_\text{ST}}\,a_{mn}u^F_{mn}\,
\er^{(\nu_m-\smallir\,\kappa_m)\, x + \smallir\,\omega_n\, t},\label{d40}
\eeqn
where $R_\text{ST}$ is the range of retained indices through
spatio-temporal truncation; see eq.(\ref{ap50})
in Appendix \ref{sec:STKD}.
Note that, since we shall describe an attractor, all temporal growth rates have
been set to zero. The spatial growth rates  $\nu_m$, instead,
must be retained because the wave of the electric field is far from being periodic in $x$.

\begin{figure}[h!]
	\begin{center}
(a)\includegraphics[width=7cm,height=4cm]{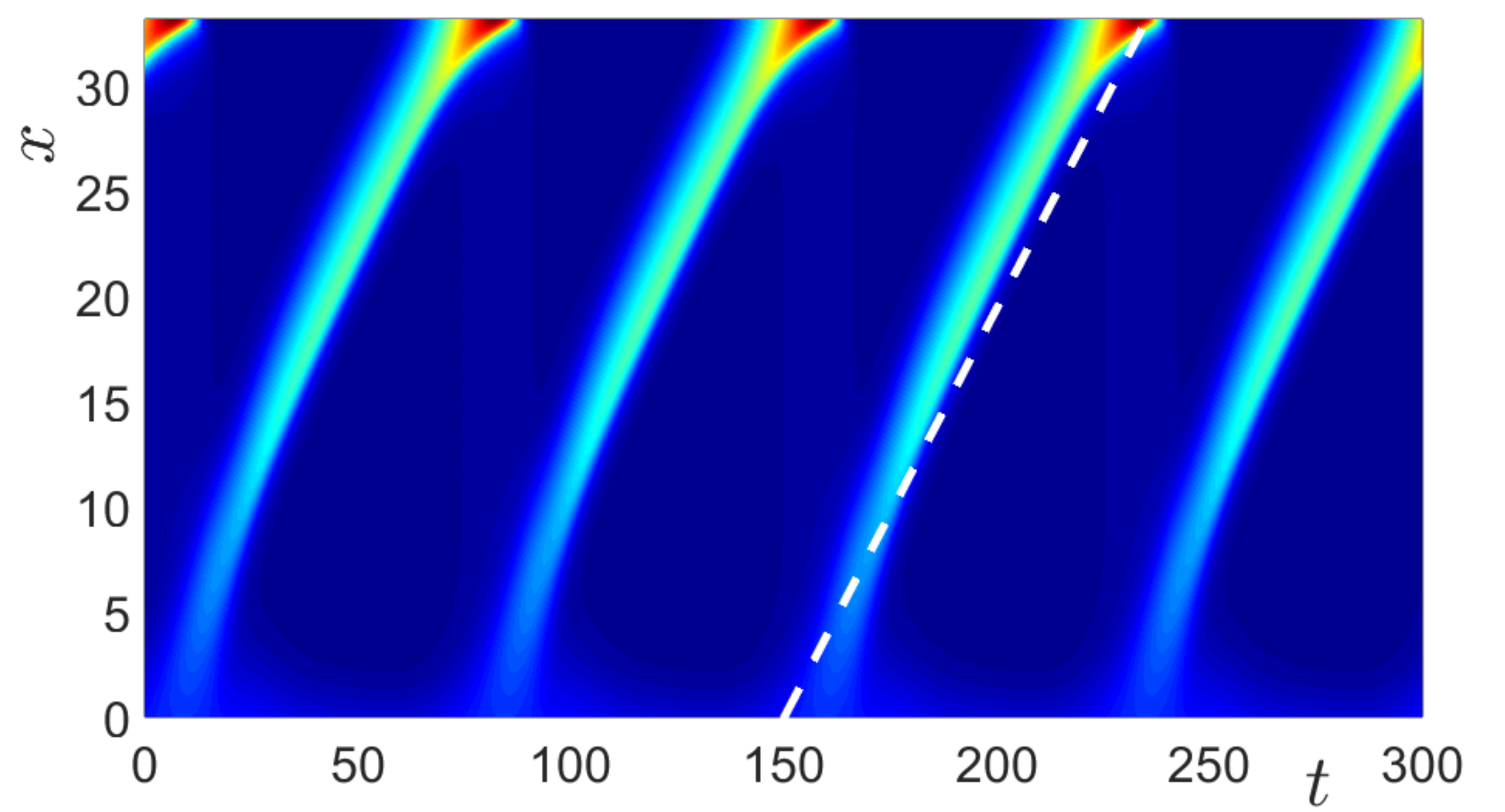}
\hskip0.5cm
(b)\includegraphics[width=7cm,height=4cm]{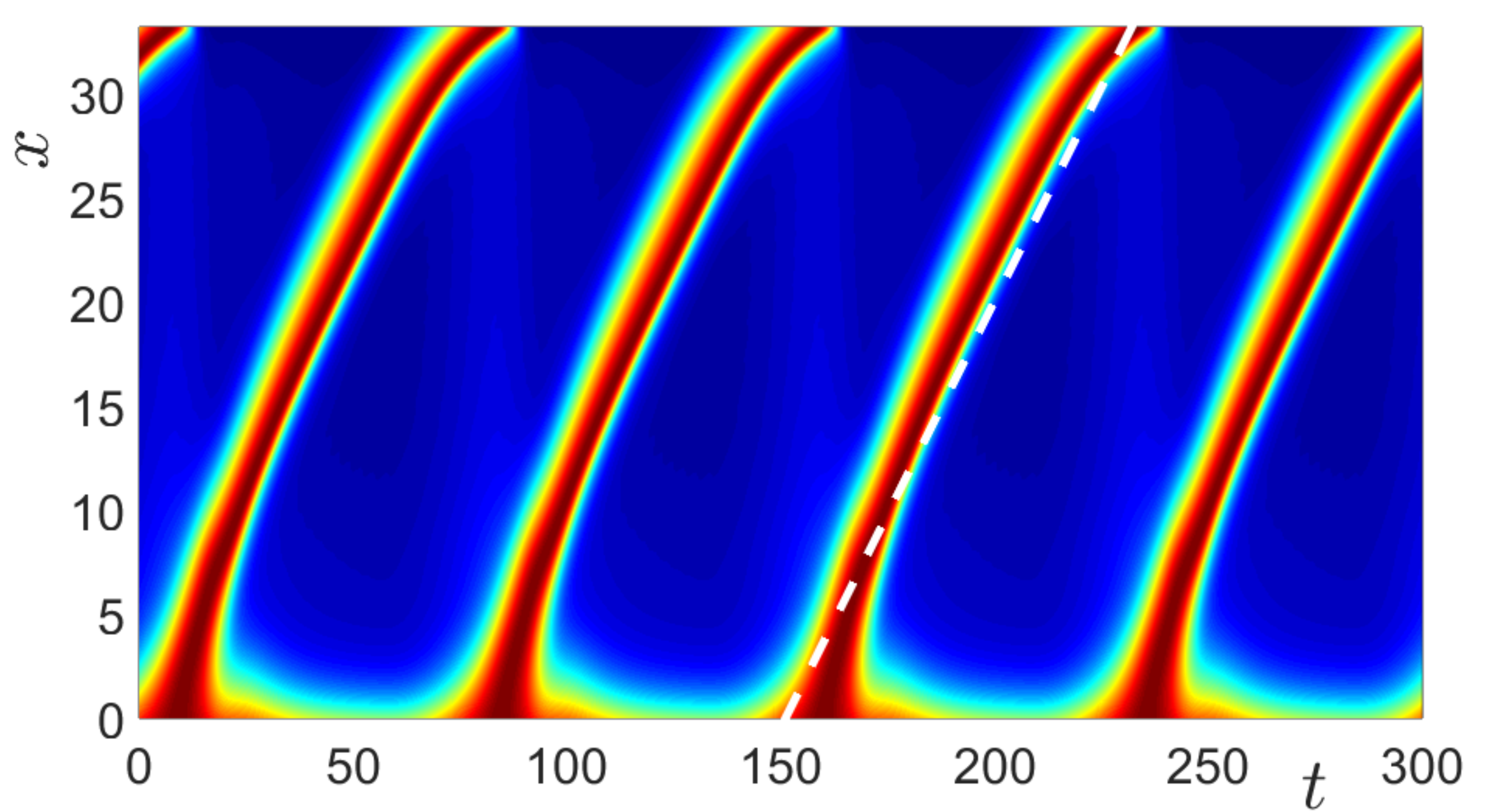}
	\end{center}
	\vskip-0.75cm
\caption{Counterpart of Fig.\ref{fig:STevolutionF} for the electric field in
the periodic attractor (a)
and the scaled electric field defined in eq.(\ref{d42})
(b). In both plots, the
dashed white line indicates the trajectory of
a particle moving with the overall propagation
velocity $c$ defined in eq.(\ref{d41}). Thus, the slope
of this line coincides with $c$.
		\label{fig:STevolutionF-Attractor}}
\end{figure}

\begin{figure}[h!]
	\begin{center}
		\includegraphics[width=7cm,height=4.5cm]{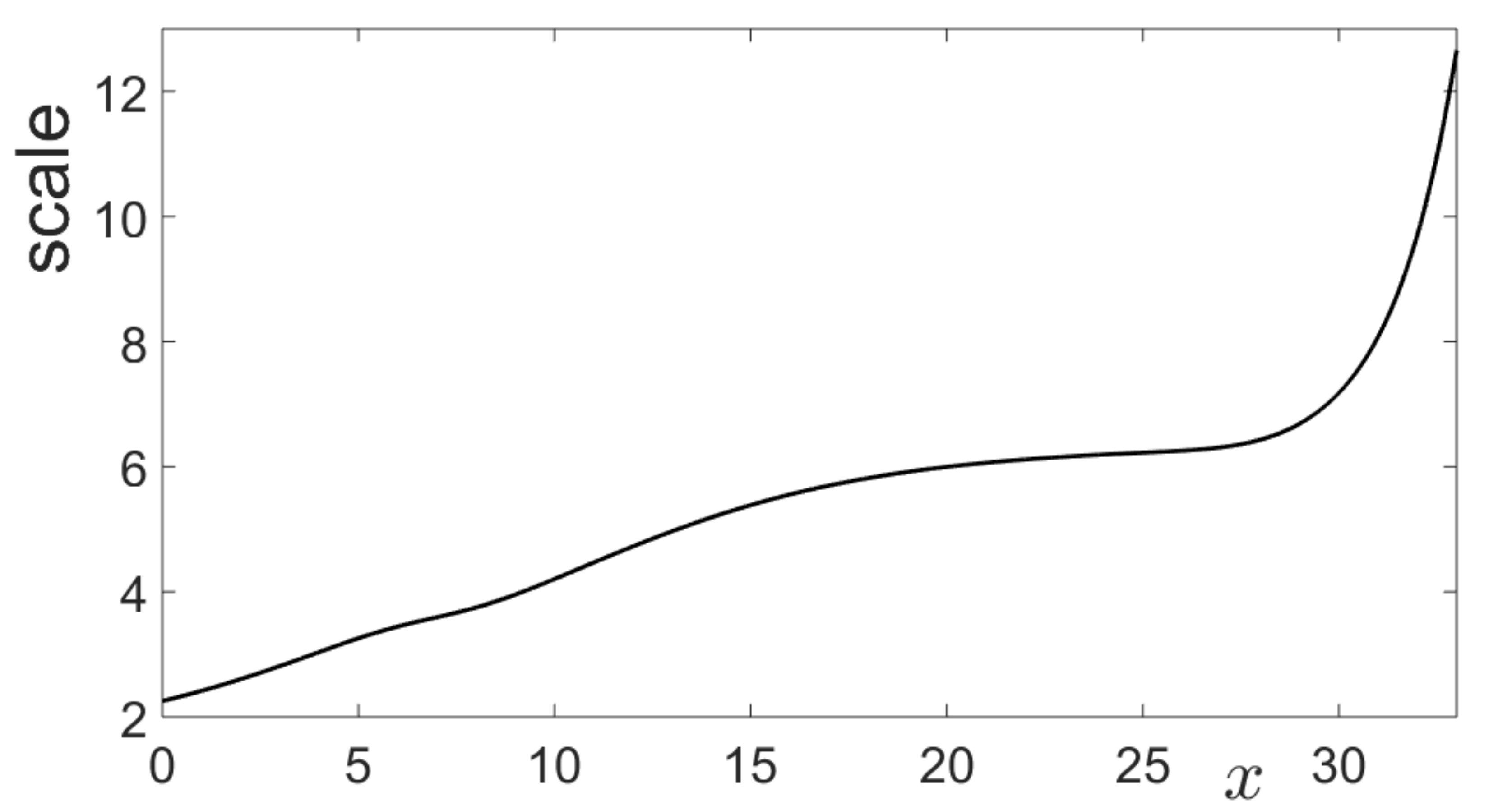}
	\end{center}
	\vskip-0.75cm
	\caption{The scale defined in eq.(\ref{d44}). \label{fig:scale}}
\end{figure}

The average velocity $c$ of the solitary wave of $F$
can be estimated noticing
that the latter moves the length $L$ in one period. Thus
\beqn
c\simeq L/T_1\simeq0.44,\label{d41}
\eeqn
where the numerical values of $L$ and $T_1$ are given in
eqs.(\ref{d1}) and (\ref{d23}), respectively.
The spatio-temporal diagram
for the electric field in the periodic
attractor that will be considered is shown in
Fig.\ref{fig:STevolutionF-Attractor}--(a).
Note that this plot does not coincide with its
counterpart in Fig.\ref{fig:STevolutionF},
where the whole evolution, including the transient stage,
was  shown.
This diagram reveals two difficulties in connection with a
STKD description
through an expansion of the form (\ref{d40}):
\begin{itemize}
	\item The (solitary) traveling wave is quite concentrated in both space and
	time. This means that the expansion (\ref{d40})
	must be broadband  for both the retained
	wavenumbers and frequencies, namely a fairly large number of relevant
	spatio-temporal modes is to be expected.
	\item The height of the wave significantly increases during its journey.
	This means that, in principle,
	very large, positive values of the spatial
	growth rate are to be  expected.
\end{itemize}
The first difficulty cannot be overcome because it
is inherent  in the very nature of the solitary traveling wave.
The second one, instead, can be alleviated
by appropriately scaling the electric
field as
\beqn
F^\text{scaled}(x,t)= F(x,t)\,/\,\text{scale}(x),\label{d42}
\eeqn
where, as suggested by Fig.\ref{fig:STevolutionF-Attractor}--(a),
the scale is defined as
\beqn
\text{scale}(x)=\max_t F(x,t). \label{d44}
\eeqn
For illustration, this scale is plotted in Fig.\ref{fig:scale},
which shows that it significantly varies from $\sim2$ at $x=0$ to $\sim12$ at
$x=L$. Using such scale, the spatio-temporal diagram of $F^\text{scaled}$
is given in Fig.\ref{fig:STevolutionF-Attractor}--(b), which shows a clearer
traveling wave structure with $F^\text{scaled}$ having a roughly constant
intensity.

\begin{figure}[h!]
\vskip-0.2cm
	\begin{center}
		\includegraphics[width=9cm,height=5cm]{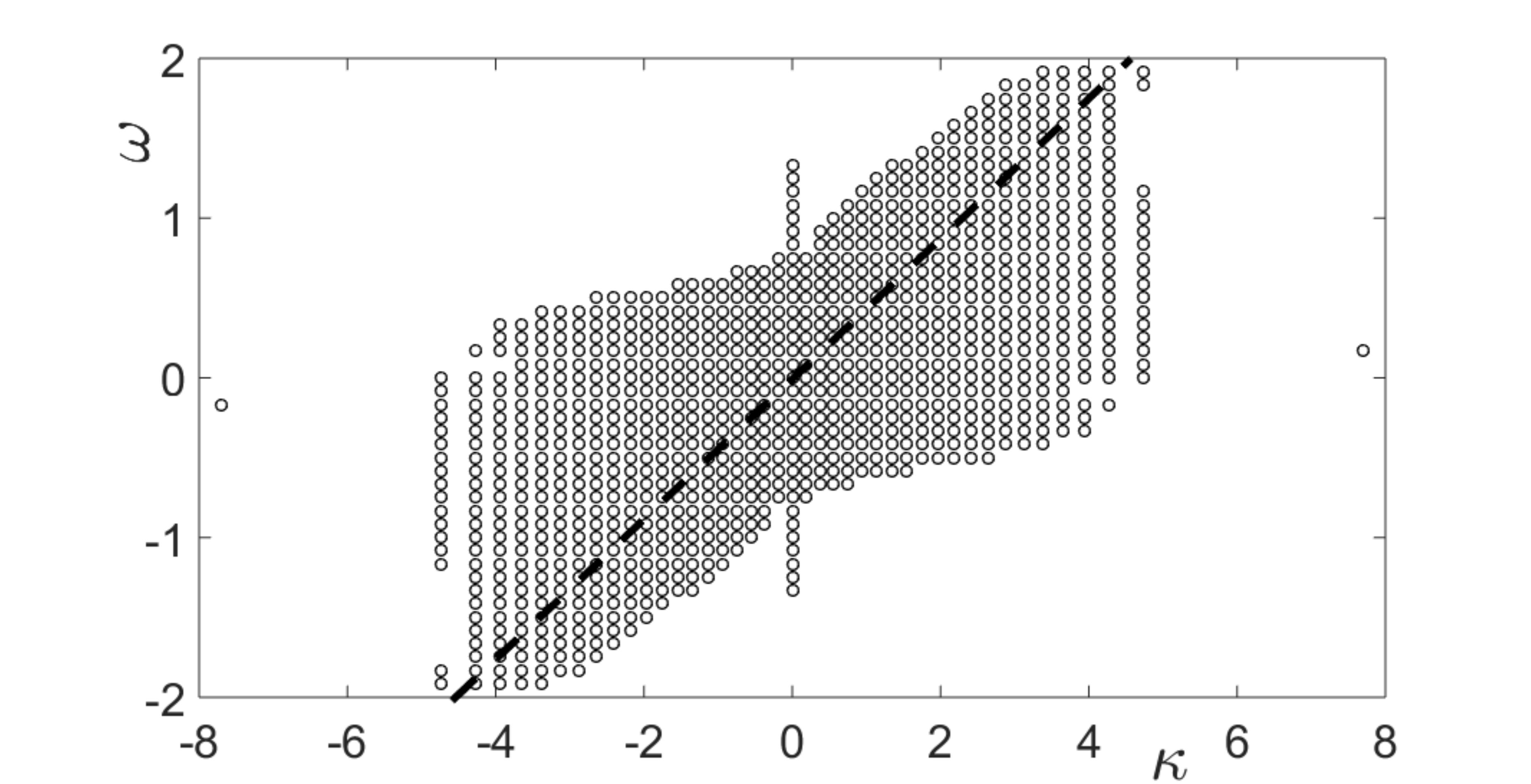}
	\end{center}
	\vskip-0.5cm
	\caption{Dispersion diagram resulting from the application of the STKD method to the
		scaled electric field in eq.(\ref{d42}) in the periodic
		attractor, using the tunable parameters in eq.(\ref{d50}).
		\label{fig:DispDiagramScaleAttractor}}
\end{figure}

Now, the STKD method is applied to $F^\text{scaled}$ with the following
tunable STKD parameter values (see Appendix \ref{sec:STKD})
\beqn
\varepsilon_\text{SVD}=5\cdot10^{-5},\quad
\varepsilon_\text{DMD}=10^{-4},\quad d^x=1,\quad d^t=25. \label{d50}
\eeqn
The reconstructed pattern shows a reasonable  RRMS error,
as defined in eq.(\ref{d21}), namely $\sim 2.75\cdot10^{-2}$
retaining 1038 spatio-temporal modes, which is a very large number,
as expected. The temporal growth rates are all zero because
the analyzed data correspond to the attractor.
The (positive and negative) spatial growth rates, instead, are $\sim 10^{-2}$
in absolute value.
This was to be expected
because, as anticipated, the pattern is far from being periodic in the spatial direction.
The associated \textit{dispersion diagram} is plotted
in Fig.\ref{fig:DispDiagramScaleAttractor}.
As can be seen,  most relevant points
in this diagram
are contained in a rhomboid, whose largest diagonal is indicated by
a thick dashed line.
The slope of this diagonal is $\omega/\kappa\simeq c$, where $c$
is the overall propagation velocity, given in eq.(\ref{d41}).

Summarizing, even in this very demanding situation, the STKD
method, with an appropriately large number of spatio-temporal modes, is able
to describe reasonably well
the present solitary traveling wave, estimating its overall propagation velocity.
The CPU time required by the application of the STKD method is $\sim 7$ CPU seconds,
using the standard PC described at the end of Section \ref{sec:intro}.
However, the analysis in this section shows that, in principle,
the STKD method is not appropriate to construct data-driven ROMs for the present problem.
This is due to the large number of spatio-temporal modes that should be retained,
which would increase the CPU time required by the online operation of the ROM.
Thus, in the next section, the proposed data-driven ROM will be constructed using
the HODMD method, which involves a moderate number of temporal modes.


\section{A preliminary version of a HODMD-based data-driven ROM \label{sec:DataDriven}}
In the previous two sections, we have used data obtained upon
numerical integration of the superlattice equations to
reconstruct and analyze the structure of  the underlying spatio-temporal pattern
 in different ways.
Let us now use the analysis in Section \ref{sec:Results} to {\it construct offline}
a data-driven ROM, able to
{\it simulate online} the  superlattice periodic attractor for continuous
values of one or more parameters in a given parameter range.
Here, for simplicity, we  restrict the discussion to
only one parameter, namely
the {\it contact conductivity}, $\sigma > 0$. The extension
of the  proposed data-driven ROM
to cases in which more than one parameter is involved
is straightforward, as is the extension to other related dynamical systems.
However, this is well beyond the scope of the present paper. \\

Let us consider the parameter range
\beqn
0.3\leq\sigma\leq0.7 \label{f0}
\eeqn
and the following {\it basic values} of the contact conductivity
\beqn
\sigma_1=0.3,\quad\sigma_2=0.4,\quad\sigma_3=0.5,\quad\sigma_4=0.6,
\quad\sigma_5=0.7. \label{f1}
\eeqn
The remaining parameters are kept fixed to the
values already used in Section \ref{sec:Results}.
The algorithm of the proposed, novel data-driven ROM follows three steps.
\begin{enumerate}
\item For each basic value of $\sigma$ in eq.(\ref{f1}), the periodic attractor is computed running the numerical solver
in the time interval $0 \leq t \leq 300$.
\item For each basic value of $\sigma$ in eq.(\ref{f1}),
the HODMD method is applied to snapshots in the timespan
$200 \leq t \leq 300$ using the following parameter values
\beqn
\varepsilon_\text{SVD}=10^{-7}, \,\,
\varepsilon_\text{DMD}=5\cdot10^{-4}, \,\, d=35, \label{f4}
\eeqn
which have been chosen after a slight calibration.
Since the number of computed modes
varies from 55 to 57, depending on the case,
55 modes are retained in all cases for consistency.
In this way, the RRMS error in
the reconstruction of the periodic orbit for the current
density and the electric field
is $\sim10^{-4}$ and $\sim10^{-3}$, respectively,
while all growth rates are smaller than
$10^{-4}$ in absolute value.
For each case, the involved fundamental
frequency, amplitudes, and modes are stored.
\item For any {\it new value} of $\sigma$ in the range (\ref{f0}),
the associated fundamental frequency, amplitudes, and modes
are calculated upon appropriate interpolation (see below),
using the fundamental frequencies, amplitudes, and modes
stored for the basic values of $\sigma$ in eq.(\ref{f1}).
This allows to reconstruct the \textit{new attractor}
using the HODMD expansions (\ref{d18})-(\ref{d19}).
\end{enumerate}
Note that steps 1--2 of the algorithm correspond to the \textit{offline} stage
of the ROM, namely the preprocess needed at the outset to construct all ingredients, while
step 3 implements the \textit{online} operation of the ROM, yielding the desired output.
We also recall that in the
expansions (\ref{d18})-(\ref{d19}), which are reported here for convenience,
\begin{alignat}{2}
& J(t) \,\simeq \sum_{p=-P}^P a_p u^J_p\,
\er^{\smallir\,p\,\omega_1\, t}, \label{f5} \\
&F(x_i,t)  \,\simeq \sum_{p=-P}^P a_p u^F_p(x_i)\,
\er^{\smallir\,p\,\omega_1\, t}, \label{f6}
\end{alignat}
the amplitudes $a_p$, the fundamental frequency $\omega_1$, and the number $P$ of positive and negative
harmonics are  common. Since data are real,
the various terms must be conformed in
complex conjugate pairs. Namely, expansions (\ref{f5})-(\ref{f6}) are invariant
under the transformations
\beqn
p\to-p, \quad u_p^J\to\overline u_p^J, \quad u_p^F\to\overline u_p^F, \label{f7}
\eeqn
where the overline denotes the complex conjugate.  In addition,
since the considered snapshots are as in eq.(\ref{d13}),
expansions (\ref{f5})-(\ref{f6}) can be recast in vector form, similar
to eq.(\ref{a2}), where
the vectors appearing in its left
and right-hand sides are given by
\beqn
\bq(t)=\left[\begin{array}{c}J(t)\\
F(x_1,t)\\F(x_2,t)\\  \vdots \\ F(x_I,t)\end{array} \right]
\quad\text{and}\quad
\bu_p = \left[\begin{array}{c}u^J_p\\
u_p^F(x_1)\\ u_p^F(x_2)\\  \vdots \\ u_p^F(x_I)\end{array} \right], \label{f8}
\eeqn
for $p = -P,\ldots,P$. \\

\begin{figure}[h!]
\begin{center}
\hskip-0.1cm (a)\includegraphics[width=7cm,height=4cm]{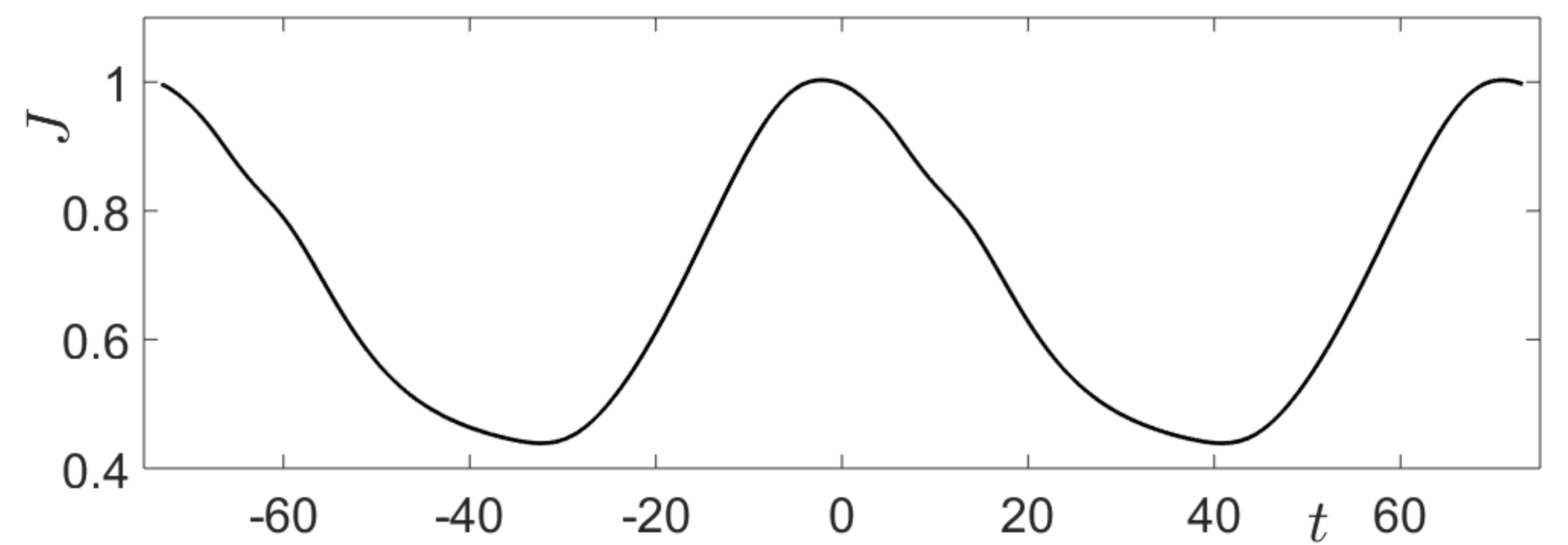}
\hskip0.5cm
(b)\includegraphics[width=7cm,height=4.1cm]{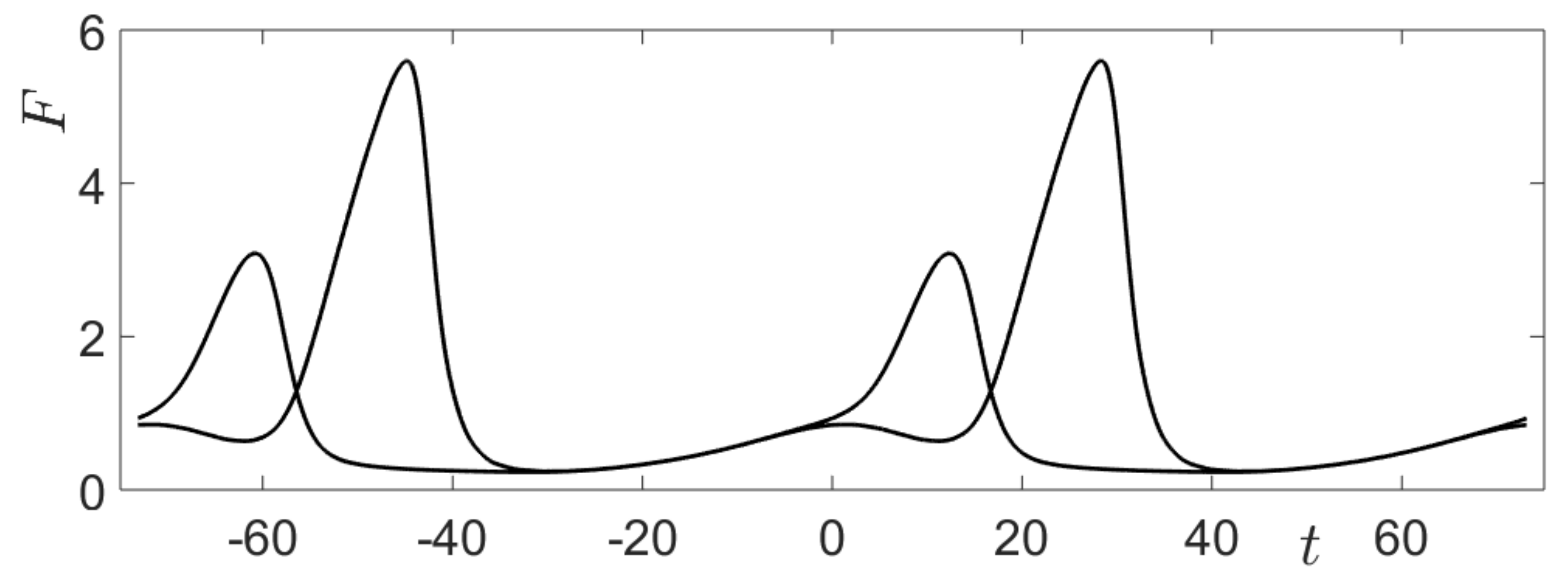}
\end{center}
\vskip-0.75cm
\caption{Temporal evolution of the current density (a) and the electric
field at $x=\widetilde x_1$ and  $x =\widetilde x_2$ as defined in eq.(\ref{d9}) (b),
for the synchronized HODMD reconstruction of the periodic attractor
corresponding to $\sigma = \sigma_5 = 0.7$, in the time interval (\ref{f15}). \label{fig:PeriodicEvolution-sigma07}}
\end{figure}

\begin{figure}[h!]
		\begin{center}
\hskip-0.1cm (a)\includegraphics[width=7cm,height=4cm]{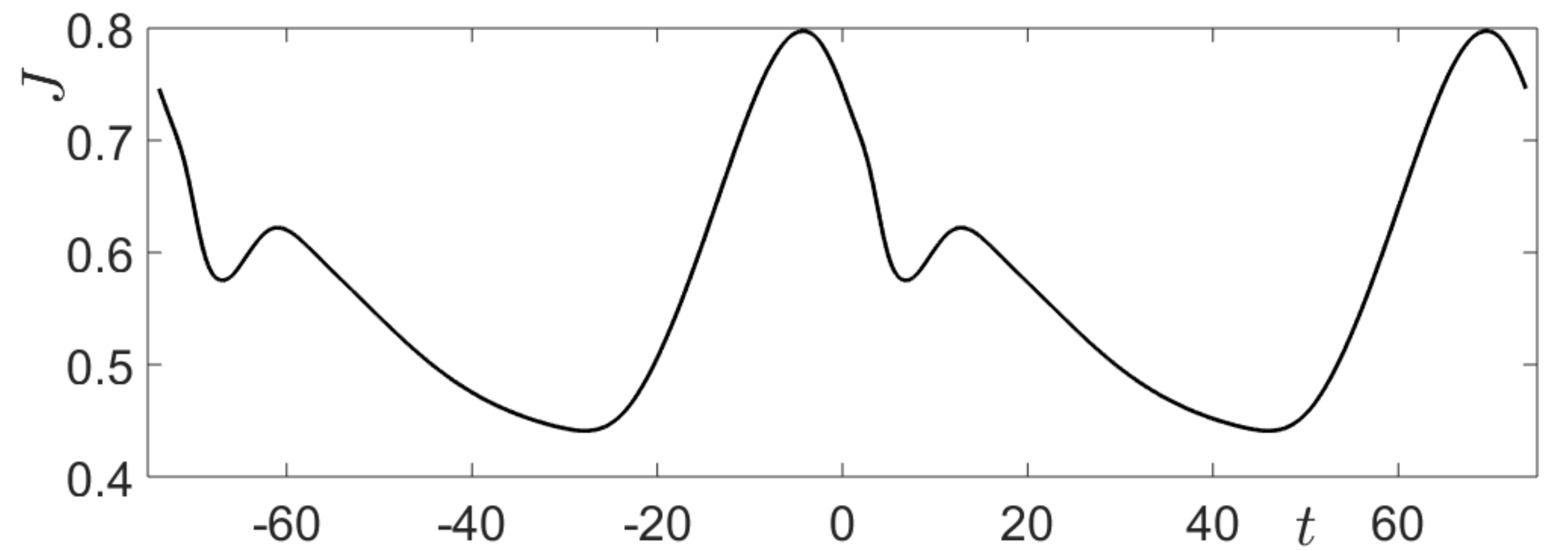}
\hskip0.5cm		(b)\includegraphics[width=7cm,height=4.1cm]{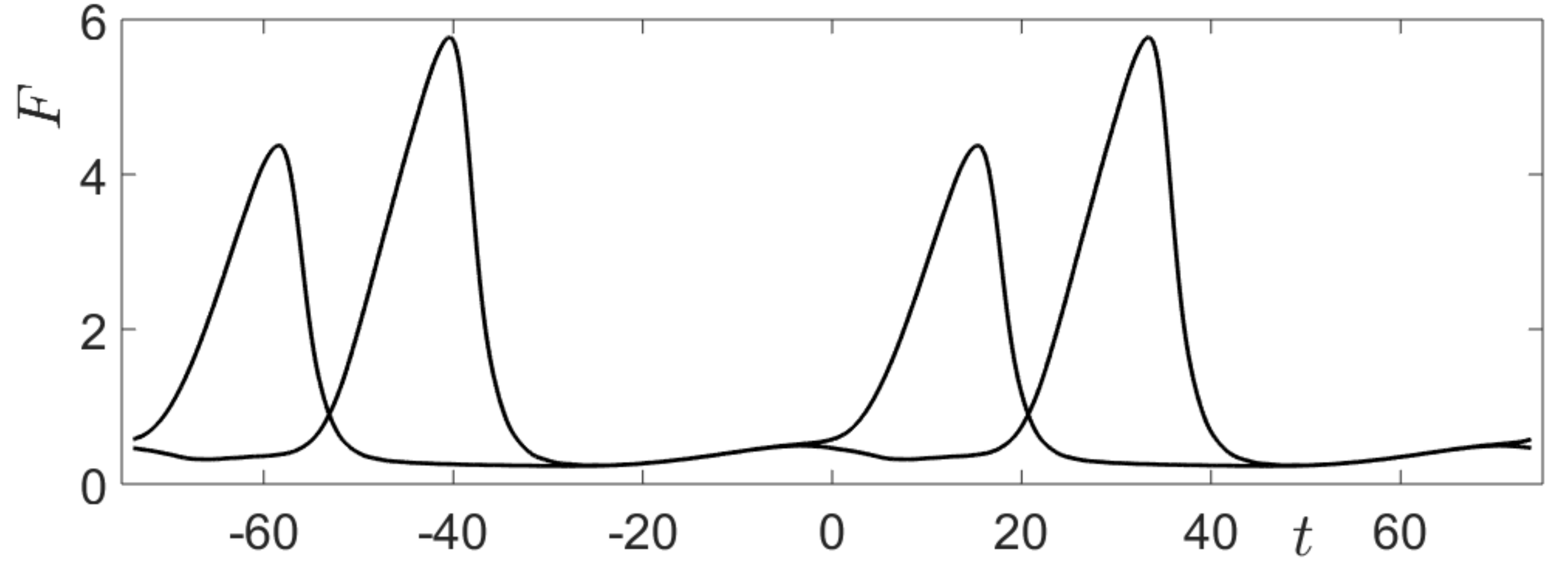}
		\end{center}
		\vskip-0.75cm
\caption{Counterpart of Fig.\ref{fig:PeriodicEvolution-sigma07}
for the periodic attractor corresponding to $\sigma = \sigma_1 = 0.3$. \label{fig:PeriodicEvolution-sigma03}}
\end{figure}

Before discussing how step 3 of the proposed algorithm is performed,
an important aspect must be pointed out.
Indeed, efficient interpolation of the computed modes requires that these be appropriately
synchronized. Note that \textit{synchronization}
was not necessary in Section \ref{sec:Results} since
all data came from a single run of the numerical solver,
namely a unique periodic attractor (from prescribed initial condition
and parameter values)
was computed and compared to its approximations in the same timespan.
If, instead, the numerical solver were run several times,
using different initial conditions and/or comparing the
outcomes in distinct
timespans, then the resulting asymptotic states
would not be automatically
synchronized, namely they would show
{\it time shifts} among each other.
Likewise, in the present case, the modes
for the various basic values of $\sigma$ come  from
different simulations of the numerical solver
and need to be synchronized too.
Synchronization means {\it shifting}
the time variable appearing  in eqs.(\ref{f5})-(\ref{f6})
as
\beqn
t \to t + t_\text{shift}, \label{f9}
\eeqn
for an appropriate value  of
$t_\text{shift}$, in order to make the
expansions for all considered values of the parameter
consistent among each other. Note that applying the action (\ref{f9})
also requires redefining the  modes as
\beqn
u_p^J\to u_p^J\,\er^{-\smallir\,p\,\omega_1 t_\text{shift}},
\quad u_p^F\to u_p^F\,\er^{-\smallir\,p\,\omega_1 t_\text{shift}}.\label{f11}
\eeqn
 The time shift can be calculated in various ways.
In the present method, it is computed by synchronizing the fundamental
mode for the current density in eq.(\ref{f5}), $u_1^J$,
which is done  defining
\beqn
t_\text{shift}=\frac{\log(u_1^J/|u_1^J|)}{\ir\,\omega_1}.\label{f13}
\eeqn
 Hence, with this selection of the time shift and
after performing the transformations
(\ref{f9})-(\ref{f11}),
$u_1^J$ becomes real and the associated
monochromatic oscillation is
\beqn
a_1\left[u^J_1\,\er^{\smallir\,\omega_1 t}+\overline u^J_1
\,\er^{-\smallir\,\omega_1 t}\right] =a_1\,|u_1^J|\,\cos(\omega_1\,t).\label{f14}
\eeqn
For convenience,
the synchronized HODMD reconstructions of the
periodic attractors  will be compared in the timespan
\beqn
-T_1 \leq t \leq T_1\label{f15}
\eeqn
in all simulations below,
where the period $T_1$ of the orbit is defined in terms of
the fundamental frequency as in eq.(\ref{d23}).
In order to illustrate the synchronization effect, we
consider the largest basic value of $\sigma$ defined in
eq.(\ref{f1}), namely $\sigma = \sigma_5 = 0.7$.
For this case, the fundamental frequency and period are
\beqn
\omega_1=0.085904 \quad \text{and} \quad T_1=73.142,
\label{f25}
\eeqn
both with six exact significant digits, while the time shift is
\beqn
t_\text{shift}= -22.696. \label{f27}
\eeqn
Using these values, the synchronized reconstruction of the current density
and the electric field in the time interval (\ref{f15}) is plotted in Fig.\ref{fig:PeriodicEvolution-sigma07}.
For the smallest basic value of $\sigma$ defined in
eq.(\ref{f1}), namely $\sigma=\sigma_1=0.3$, the fundamental frequency
is $\omega_1=0.085155$ and the associated period is
$T_1=73.786$, both with six exact significant digits. Note that these values
are fairly close to their counterparts  for $\sigma_5$.
Instead, the time shift is now
$t_\text{shift} = -33.417$, which is not close to its counterpart
for $\sigma_5$. Also, the synchronized reconstruction
of the current density
and the electric field in the time interval (\ref{f15}), plotted in Fig.\ref{fig:PeriodicEvolution-sigma03},
is quite different from its counterpart for $\sigma_5$,
especially for the current density,
as comparison with Fig.\ref{fig:PeriodicEvolution-sigma07}
shows. Thus, we may guess that the fundamental
frequency is fairly constant for the
basic values of $\sigma$ in eq.(\ref{f1}), but the
 periodic orbit varies significantly. \\

Let us now discuss how the proposed HODMD-based data-driven ROM
is able to efficiently simulate
the system response for any \textit{new} value of $\sigma$ in the range (\ref{f0}).
In other words, let us clarify how \textit{interpolation}
 is performed in step 3 of the introduced algorithm.
The new value of the fundamental frequency
is computed upon spline interpolation
using the values of $\omega_1$ for the basic values of $\sigma$ in eq.(\ref{f1}).
Interpolation of the spatial modes could be
performed point-by-point by, e.g., splines.
However, this strategy would be computationally  inefficient.
A more efficient interpolation procedure is developed as follows.
To begin with,  the retained modes defined in
eq.(\ref{f8}) are first synchronized and
then scaled with the mode amplitudes. Namely,  all synchronized
modes are scaled as
\beqn
\bu^*_p=a_p \bu_p,\quad\text{with }\,\,p=-P,\ldots,P, \label{f30}
\eeqn
for all basic values of $\sigma$ in eq.(\ref{f1}).
The resulting modes could be organized
in a third-order tensor, whose components are given by
\beqn
T_{ipq} = u^*_{ip,q}, \label{f32}
\eeqn
where $i = 1,\ldots,I+1$ labels
the mode component, $p = -P,\ldots,P$ stands for the mode index,
 and $q = 1,\ldots,Q$ indicates the basic value of
$\sigma$. Note that, in the present case,
the tensor $\bT$ will have dimension equal to
$(I+1)\times(2P+1)\times Q=482\times55\times 5$.
This tensor could be treated
using the HOSVD tensor decomposition \cite{KoldaB09}
combined with 1D interpolation, as explained
in \cite{LorenteVV2008}.
However, for simplicity,  here we unfold
the tensor $\bT$ into a matrix $\bA$
by collecting together the first
two indices appearing in eq.(\ref{f32}), $i$ and $p$,
into a single index, denoted as $l$.
The  matrix $\bA$ is then dimension-reduced using
truncated \textit{singular value decomposition} (SVD)
\cite{GolubvL96}, requiring that the
RRMS error of the reconstruction be smaller
than a threshold. This
yields the approximated matrix with  components
\beqn
A^\text{approx}(l,q)=\sum_{l'=1}^{L'}s_{l'}U_{ll'}V_{ql'}, \label{f34}
\eeqn
where $s_{l'}$ are the retained singular values,
while $U_{\cdot l'}$ and $V_{\cdot l'}$
are the corresponding left and right SVD modes, respectively.
Note that, for $q=1,\ldots,Q$, the two sides of this equation account for
the $Q = 5$ basic values of $\sigma$ in eq.(\ref{f1}).
Thus, eq.(\ref{f34}) can be rewritten
as
\beqn
A^\text{approx}_{l}(\sigma_q)=\sum_{l'=1}^{L'}s_{l'}U_{ll'}V_{l'}(\sigma_q). \label{f36}
\eeqn
Now, for each $l'=1,\ldots,L'$, 1D spline interpolation can be
applied to $V_{l'}(\sigma_q)$ to compute  this quantity for the new
value of $\sigma$, say $\sigma^\text{new}$.
 Hence, from eq.(\ref{f36}) we get
\beqn
A^\text{approx}_{l}(\sigma^\text{new})=
\sum_{l'=1}^{L'}s_{l'}U_{ll'}V_{l'}(\sigma^\text{new}). \label{f38}
\eeqn
More details about combining SVD with interpolation can be found in
 \cite{BuiTanh03,BuiTanhDW04}.
Finally, splitting the index $l$ into the original indices,
$i$ and $p$, which were collected together
in the matrix $\bA$, and invoking eq.(\ref{f30}) give the
scaled, synchronized modes for the new value of $\sigma$, namely
\beqn
\bu_p^{*\text{new}},\quad\text{for }\,\sigma=\sigma^\text{new}.
\label{f40}
\eeqn
Using these modes, the amplitudes and the original normalized modes
(exhibiting unit RMS norm) for $\sigma=\sigma^\text{new}$
are obtained from eq.(\ref{f40}) as
\beqn
a_p^\text{new}=\frac{\|\bu_p^{*\text{new}}\|_2}{\sqrt{I+1}},\quad
\bu_p^\text{new} =
\frac{\bu_p^{*\text{new}}}{a_p^\text{new}},\quad\text{for }\sigma=
\sigma^\text{new}, \label{f42}
\eeqn
and the associated synchronized HODMD reconstructions
of the current density and the electric field are
readily obtained via the expansions (\ref{f5})-(\ref{f6}). \\

\begin{figure}[h!]
		\begin{center}
			\hskip-0.1cm
(a)\includegraphics[width=7cm,height=4cm]{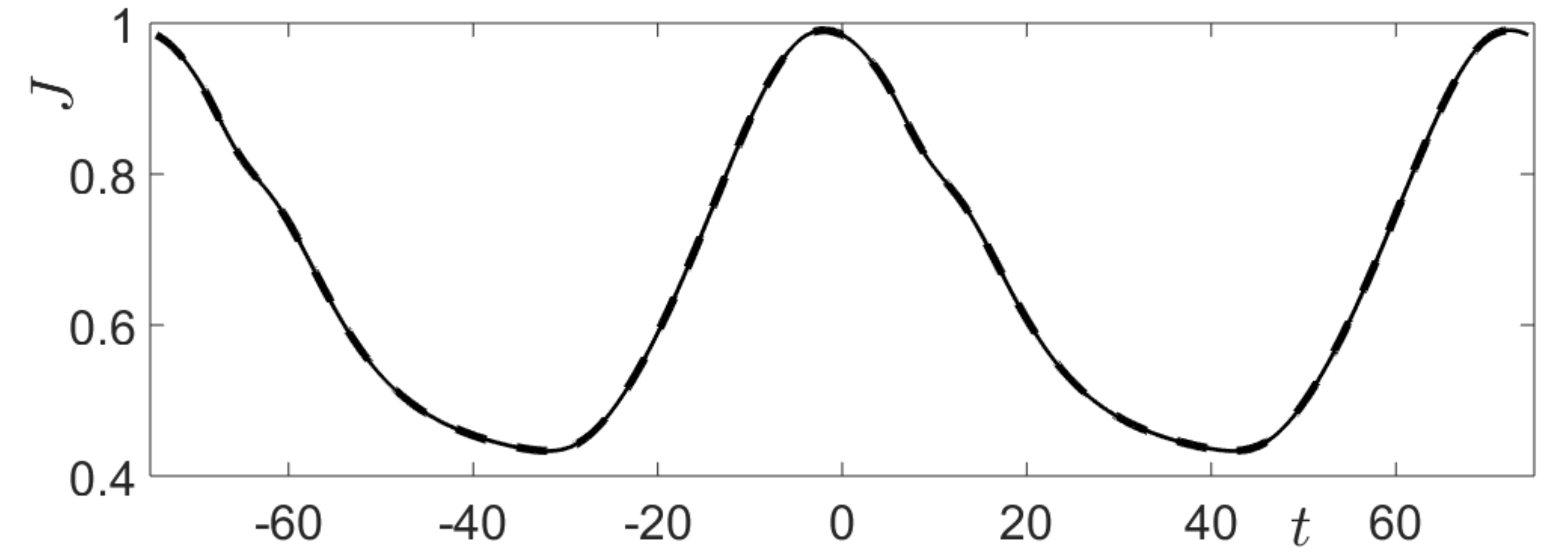}
\hskip0.5cm		(b)\includegraphics[width=7cm,height=4cm]{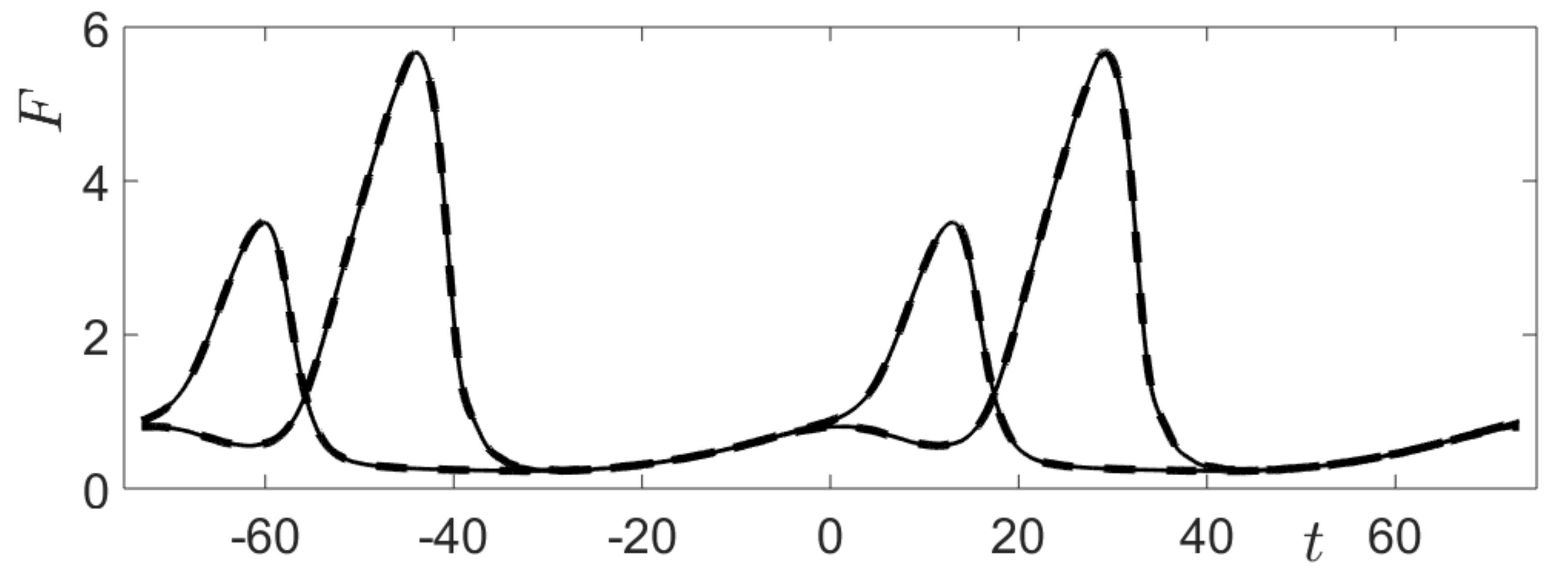}
		\end{center}
		\vskip-0.75cm
		\caption{Synchronized attractor for $\sigma^\text{new}=0.65$,
			in the time interval (\ref{f15}), for the current density (a)
			and the electric field at $x=\widetilde x_1$ and $x = \widetilde x_2$
			as defined in eq.(\ref{d9})
			(b). In these plots, the `exact' evolution is
			displayed in thin solid lines, while the
			outcome of the HODMD-based data-driven ROM
			is depicted with thick dashed lines. \label{fig:PeriodicEvolutionROM-sigma065}}
\end{figure}

\begin{figure}[h!]
\begin{center}
\hskip-0.1cm (a)\includegraphics[width=7cm,height=4cm]{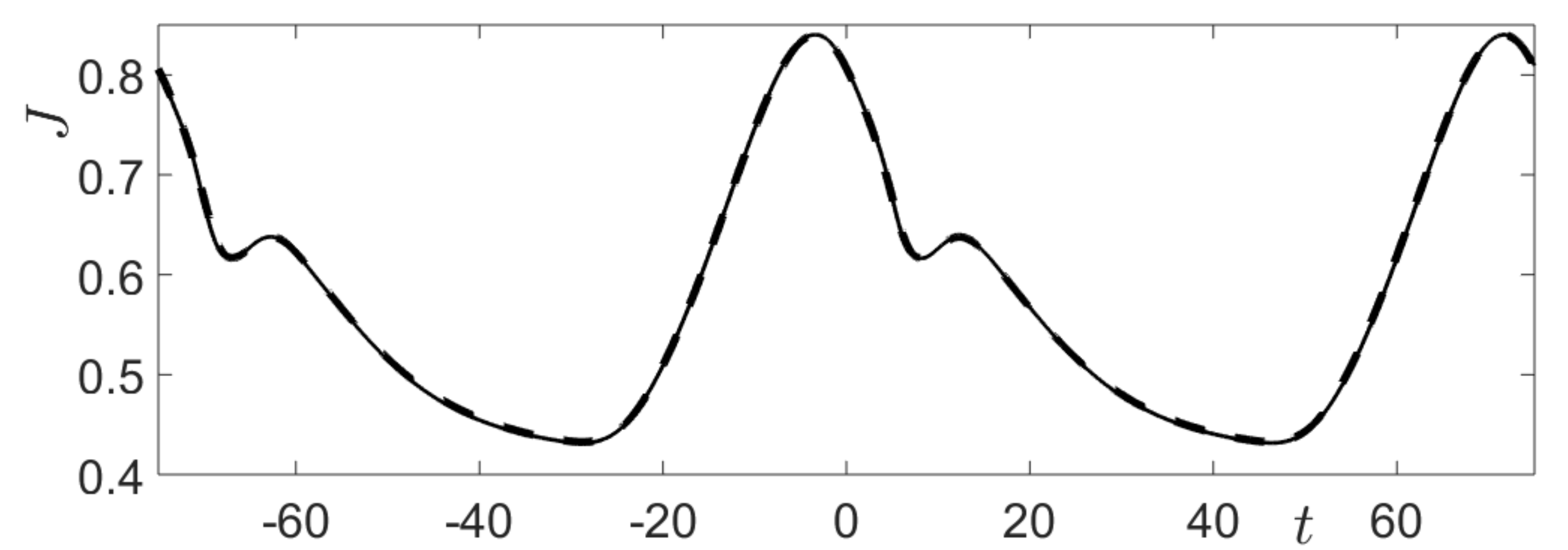}
\hskip0.5cm	(b)\includegraphics[width=6.9cm,height=4cm]{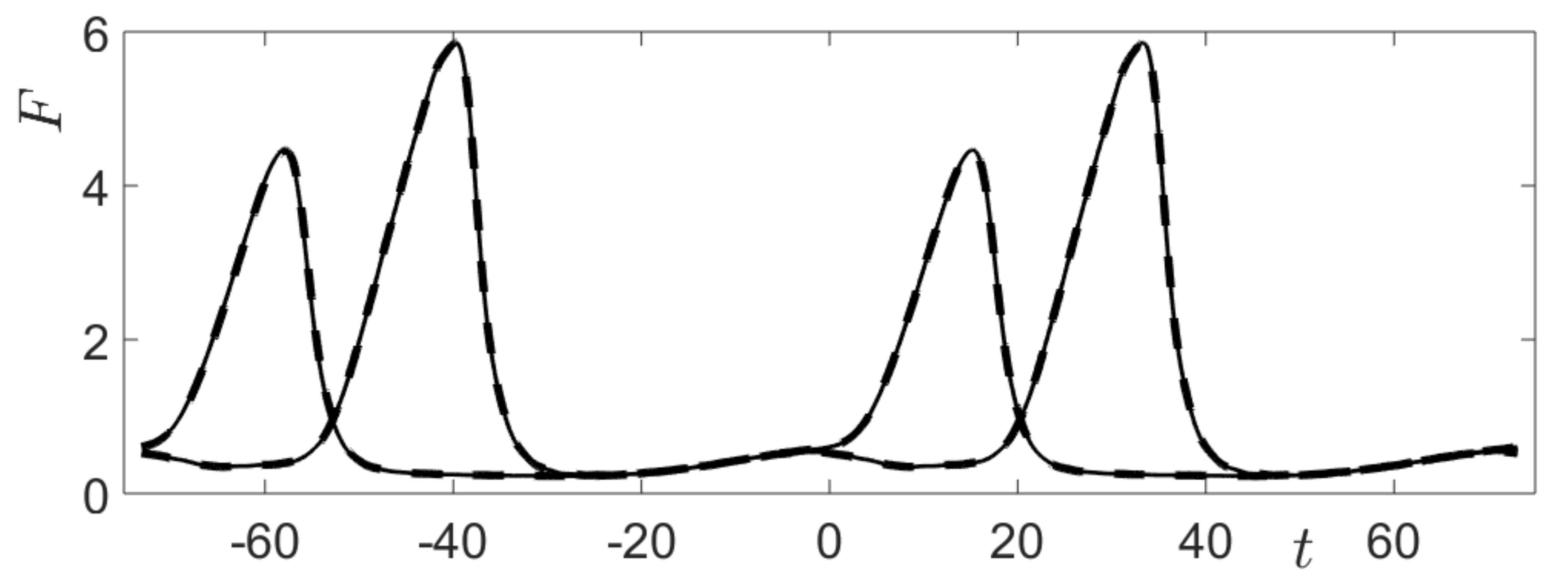}
\end{center}
\vskip-0.75cm
\caption{Counterpart of Fig.\ref{fig:PeriodicEvolutionROM-sigma065} for $\sigma^\text{new}=0.35$.
\label{fig:PeriodicEvolutionROM-sigma035}}
\end{figure}

In order to highlight the accuracy of the data-driven ROM, we consider two
very demanding  test cases corresponding to values of
 $\sigma$ near the end-points of the
parameter range (\ref{f0}). For $\sigma^\text{new}=0.65$, the outcome of the
ROM is compared in Fig.\ref{fig:PeriodicEvolutionROM-sigma065} with its
`exact' counterpart computed by the numerical solver.
Note that the  reference and approximated periodic attractors
are plot-indistinguishable.
For $\sigma^\text{new}=0.35$, the counterpart of Fig.\ref{fig:PeriodicEvolutionROM-sigma065}
is Fig.\ref{fig:PeriodicEvolutionROM-sigma035},
where it can be seen that the approximated and `exact' periodic orbits
are again plot-indistinguishable.
For  other values of the parameter $\sigma$ in
the range (\ref{f0}), the outcome of the HODMD-based
 data-driven ROM is either equally
good or even better.
Due to the large computational cost of the
numerical solver, the CPU time needed to generate the considered snapshots is
$\sim 2$ CPU hours. Instead, the cost of the remaining
HODMD tasks required in the offline stage of the ROM is much smaller, namely $\sim 30$ CPU seconds.
On the other hand, the online operation of the ROM, for any
new value of $\sigma$, is quite cheap, namely it takes $\sim0.5$ CPU seconds.


\section{Concluding remarks\label{sec:conclusion}}
We have studied self-oscillations of the current through a one
miniband semiconductor superlattice and the associated electric
field waves by  using the HODMD and STKD data processing methods.
The properties of the self-oscillations (i.e., their spectrum
comprising growth rate vs. frequency, wave shape, and propagation)
depend on the device configuration and, in particular,
 on the conductivity of the injecting contact.

Firstly, we have applied the HODMD tool to data of a complete
numerical simulation that includes both the transient stage and
the final time periodic attractor of the current self-oscillations.
As a result, we have identified the attractor and separated it
from the purely decaying dynamics of the transient stage.
 We have also performed additional
test cases in which we reconstruct the periodic attractor
 by using snapshots in various limited timespans.  In particular,
 we are able to approximate well the attractor
 using snapshots taken during the transient stage.
This last approach is important because it decreases
 the computational cost of calculating the snapshots,
 while preserving the physical properties within a satisfactory accuracy.

Secondly, we have used the STKD method to characterize
fairly well the electric field traveling pulse and
its average propagation velocity. This is a challenging endeavor
because the pulse wave is quite localized in both space and time, which implies
that   the STKD description involves a large number of spatio-temporal modes.

In addition, we have constructed a preliminary version of a novel HODMD-based data-driven reduced order model for the superlattice dynamics.
This ROM enables very fast online simulations of the device response
over a range of values of the contact conductivity.
For each test case, the online ROM operation requires
 only $\sim 0.5$ CPU seconds,
which is much smaller than the computational cost
 needed by the considered
standard numerical solver
(namely $\sim 20$ CPU minutes). Thus,  in this context, the
data-driven ROM divides the computational
 effort by a factor $\sim 2500$.
The drawback of our preliminary version is that the ROM offline preprocess takes
$\sim 2$ CPU hours. However,
most of this CPU time is due to the
required runs  to calculate the input snapshots.
Hence, it could be drastically
reduced by replacing the standard
numerical solver by a  low-dimensional model based on POD
and Galerkin projection of the governing equations.
This improvement is far beyond the scope of the
present work and will be pursued elsewhere.
Finally, it is worth remarking that an extension of the
developed data-driven approach
to analyses involving more than one parameter would be straightforward.

Our results show that appropriate data processing tools can be used
to $(i)$ uncover the dynamics underlying
a physical system, and $(ii)$ construct
purely data-driven ROMs for a parametric study of the involved mechanisms.  For semiconductor superlattices, these methods could be particularly valuable when a tilted external magnetic field is present \cite{fro04,ale12,sos19,bon17}.
In this case, the periodic motion of charge dipole waves is 2D and the numerical computations needed to explore parameter regions are much more costly \cite{bon17}. Hence, the techniques presented in this paper could be very helpful indeed.


\acknowledgments
 The authors are indebted to two anonymous referees for some useful comments
and suggestions on an earlier version of the manuscript.
This work has been supported by the FEDER / Ministerio de Ciencia, Innovaci\'on y Universidades -- Agencia Estatal de Investigaci\'on,
under grants TRA2016-75075-R,  MTM2017-84446-C2-2-R, and PID2020-112796RB-C22, and
by the Madrid Government (Comunidad de Madrid-Spain) under
the Multiannual Agreement with UC3M in the line of Excellence
of University Professors (EPUC3M23) and in the context
of the V PRICIT (Regional Programme of Research and Technological Innovation).


\appendix

\setcounter{equation}{0}
\renewcommand{\theequation}{A.\arabic{equation}}

\section{The HODMD method\label{sec:HODMD}}
The {\it higher order dynamic mode decomposition} (HODMD) and
the {\it spatio-temporal Koopman decomposition} (STKD) are
 summarized in this and the following appendices.
These tools have proven to give good results
in uncovering the nature of spatio-temporal patterns,
from either numerical or experimental data, in a variety
of dynamical systems of scientific and industrial interest
\cite{LeClaincheVegaComplexity18}, including,
e.g., PIV wind tunnel measurements
\cite{LeClaincheVegaSoria17,LeClaincheetalAIAA17},
wind turbine operation \cite{LeClaincheetalEnergies18,LeClaincheetalWE19},
aircraft flight flutter testing \cite{LeClaincheetalJA18,Mendezetal21},
basic fluid dynamics \cite{Beltranetal19}, and
pattern-forming systems \cite{SanchezNetVega2019}.
See also \cite{VegaLC2020} for a reader-friendly description of the HODMD and
STKD methods and some of their applications, as well as for
specific MATLAB implementations of their algorithms.

The HODMD method \cite{LeClaincheVegaSIADS17} is now briefly detailed.
The outcome of this technique
is a discrete expansion of the form
\beqn
\bq_k \equiv \bq(t_k) \simeq \sum_{n=1}^Na_n\bu_n
\er^{(\delta_n+\smallir\,\omega_n)\, t_k}, \label{ap1}
\eeqn
with $t_k = (k-1)\,\Delta t$, for $k = 1,\ldots,K$.
Here, $\bq_k$ are {\it snapshots} for the involved state variable
at $\Delta t$-spaced values of time,
$a_n > 0$ are real {\it amplitudes}, $\bu_n$ are conveniently normalized
(generally complex) {\it modes}, and $\delta_n$ and $\omega_n$ are the associated
{\it growth rates} and {\it frequencies}, respectively.
HODMD is an improvement of standard dynamic mode
decomposition (DMD) \cite{SchmidH,Schmid2010}.
By improving standard DMD we mean that, conveniently calibrated, HODMD
yields robust results  when standard DMD fails.
Indeed, the latter can only cope with cases in which the spectral
complexity coincides with the spatial complexity, while
HODMD gives good results
also when the spatial complexity is strictly smaller
than the spectral complexity (see below).
On the other hand, HODMD could be compared to FFT
and well-known improvements of FFT, such as the power spectral density
\cite{Pressetal98}
and the so-called Laskar method \cite{Laskar93,Laskar93a}.
However, compared to FFT and its improvements, HODMD exhibits
two main advantages, since
$(i)$ it provides not only the frequencies but also the associated
growth rates and
$(ii)$ it requires a smaller amount of data, in a shorter timespan.
Furthermore, once the expansion (\ref{ap1}) has been obtained, replacing
$t_k$ by $t$
readily yields the continuous expansion (\ref{a2}), which
can be seen as an analytical description of the dynamics
associated with the given data (i.e., those provided by the dynamical
system behind the data).

Without loss of generality, it is assumed that
the coefficients $\delta_n+\ir\,\omega_n$ appearing in the  exponential of eq.(\ref{ap1})
are different from each other. Indeed, if some of them coincide,
they can be collected in a single term.
The rank (or approximate rank) of the set of modes, $\bu_n$,
denoted  by $M$,
is known as the {\it spatial complexity},
while the number of terms appearing in  eq.(\ref{ap1}), $N$,
is the {\it spectral complexity}, which obviously verifies $N\geq M$.
Note that the amplitudes
could have been absorbed into the modes. However, isolating $a_n$ and
$\bu_n$ by
normalizing the latter such that, e.g., they exhibit unit
RMS norm, the amplitudes give  a quantitative measure of
the contribution of the various modes. This helps to identify
the {\it dominant modes}
as those exhibiting the largest amplitudes. Also,
retaining only those terms whose
amplitudes are larger than a desired threshold permits appropriate
{\it truncation}  of the expansion. In any event, the expansion (\ref{a2})
(the continuous counterpart of eq.(\ref{ap1}))
is only an approximation of $\bq(t)$ for various reasons, including  errors
in the given data (which can be significant in experimental data),
truncation, and computational errors
in the calculation of the amplitudes, modes, growth rates, and frequencies.
We  remark that HODMD gives the expansion
(\ref{a2}) in a purely data-driven fashion,
namely not relying on the governing equations, but  using only a
limited amount of associated data.

The derivation of eq.(\ref{ap1}) is now summarized for the case
used in this paper. See \cite{LeClaincheVegaSIADS17}
for further details and more general versions of the HODMD method
that allow for treating spatially multi-dimensional data and
filtering noisy artifacts in experimental data.
To begin with, the snapshots $\bq_k$ are assumed to be vectors of size $I$ and
are organized  as columns of a {\it snapshot matrix}, namely
\beqn
\bQ^K_1=[\bq_1, \bq_2, \cdots, \bq_K].\label{ap3}
\eeqn
Then, the snapshots are dimension-reduced by
applying truncated SVD \cite{GolubvL96}
to the $I\times K$-snapshot matrix $\bQ^K_1$, which yields
\beqn
\bQ_1^K\simeq\bU\,\bSigma\,\bV^\top\equiv \bU\,\widehat{\bQ}_1^K\quad
(\text{or }\bq_k=\bU\,\widehat\bq_k), \label{ap5}
\eeqn
where the $M\times K$-matrix $\widehat\bQ_1^K = \bSigma\,\bV^\top$
 is known as the
{\it dimension-reduced snapshot matrix} and its columns,
whose size is $M$, are
the {\it dimension-reduced snapshots}, $\widehat\bq_k$.
The number of  retained singular values, $M\leq I$,
is precisely the spatial complexity and
is determined by a (tunable) {\it dimension reduction threshold},
$\varepsilon_\text{SVD}$, requiring that
the  RRMS error of the approximation, as defined in eq.(\ref{d21}),
be smaller than  $\varepsilon_\text{SVD}$. Note that
this error is easily computed in terms of the singular
values using well-known SVD formulae \cite{GolubvL96}.

As a second step, the counterpart of the expansion (\ref{ap1})
for the reduced snapshots, namely
\beqn
\widehat \bq_{k}\simeq\sum_{n=1}^N  a_n
\widehat \bu_{n}\,\er^{(\delta_n+\smallir\,\omega_n)\, t_k}
\quad\text{ for }\,k=1,\ldots, K,\label{ap7}
\eeqn
is derived as follows. For the reduced snapshots,
standard DMD relies on the assumption
\beqn
\widehat\bq_{k+1}\simeq\widehat\bR\,\widehat\bq_k
\quad\text{for }\,k=1,\ldots,K,\label{ap9}
\eeqn
where the $M\times M$ matrix  $\widehat\bR$
(the reduced {\it Koopman matrix}) is computed from the
dimension-reduced snapshots via the pseudo-inverse \cite{LeClaincheVegaSIADS17}.
The non-zero eigenvalues of the reduced Koopman matrix,
$\mu_n$, give the  growth
rates and frequencies appearing
in eq.(\ref{ap7}) as
\beqn
\delta_n+\ir\,\omega_n=\frac{1}{\Delta t}
\log(\mu_n) \quad\text{for }\,n=1,\ldots,N,\label{ap11}
\eeqn
while the (conveniently normalized)
associated eigenvectors $\widehat\bu_n$ yield the
reduced modes. The mode amplitudes
$a_n$ are computed via least-squares fitting between the two sides
of eq.(\ref{ap7}). This computation
is similar to what is done in optimized DMD
\cite{Chenetal12}.
Using the mode amplitudes computed in this way,
the reduced expansion (\ref{ap7}) is finally truncated
retaining only those terms such that
\beqn
\frac{a_n}{\max\{a_n\}}>\varepsilon_\text{DMD},\label{ap13}
\eeqn
for some (small) tunable {\it mode truncation threshold}
$\varepsilon_\text{DMD}$.

Once eq.(\ref{ap7}) has been obtained, pre-multiplying
it by the matrix $\bU$ appearing in eq.(\ref{ap5}) gives
a first version of the expansion
(\ref{ap1}),  where
\beqn
\bu_n=\bU\,\widehat\bu_n.\label{ap15}
\eeqn
In fact, the amplitudes and modes computed in eqs.(\ref{ap13})-(\ref{ap15})
are jointly  rescaled
(namely, scaled again after the
previous implicit scalings performed above, to
scale the right SVD modes with the singular values in eq.(\ref{ap5}) and
to normalize the reduced modes $\widehat\bu_n$ appearing in  eq.(\ref{ap7})),
requiring that the modes
exhibit unit RMS norm while preserving the product
$a_n\bu_n$. It is precisely these rescaled amplitudes and modes
that are used in the right-hand side of eq.(\ref{ap1}), while the growth
rates and frequencies are as computed for the reduced
expansion (\ref{ap7}), according to eq.(\ref{ap11}).
Note that, since the growth rate/frequency pairs are all different
from each other,
invoking eq.(\ref{ap11}), the eigenvalues $\mu_n$ are also different
and the eigenvectors $\widehat\bu_n$ are linearly independent.
This means that, in the present case,
$M=N$. In other words, in standard DMD, the spatial and spectral complexities
coincide, as anticipated.

The general  case $N \geq M$ is dealt with via HODMD, in which
the assumption (\ref{ap9}) is replaced by
\beqn
\widehat\bq_{k+d}\simeq\widehat\bR_1\, \widehat\bq_{k+d-1}+
\widehat\bR_2\, \widehat\bq_{k+d-2}+\ldots+\widehat\bR_d\, \widehat\bq_{k},
\label{ap17}
\eeqn
which increases the spectral complexity, as seen  below. For convenience,
the assumption (\ref{ap17}) is rewritten as
\beqn
\widetilde\bq_{k+1}\simeq\widetilde\bR\,\widetilde\bq_k
\quad  \text{for }\,k=1,\ldots,K-d+1, \label{ap19}
\eeqn
where the {\it enlarged snapshots} $\widetilde\bq_{k}$ are defined in terms
of the reduced snapshots  $\widehat\bq_k$ as
\beqn
\widetilde\bq_{k}\equiv
\left[\begin{array}{c}
	\widehat\bq_k\\ \widehat\bq_{k+1}\\
	\vdots \\ \widehat\bq_{k+d-2}\\ \widehat\bq_{k+d-1}
\end{array}\right].\label{ap21}
\eeqn
The index $d\geq1$ appearing in eqs.(\ref{ap17}) and (\ref{ap21}) is tunable
in this method.
Comparing eq.(\ref{ap19}) with eq.(\ref{ap9})
suggests to apply standard DMD to the enlarged
snapshots, which gives
\beqn
\widetilde \bq_{k}\simeq\sum_{n=1}^N  a_n
\widetilde \bu_{n}\,\er^{(\delta_n+\smallir\,\omega_n) t_k}
\quad  \text{ for }\,k=1,\ldots,K-d+1.\label{ap23}
\eeqn
In this application of standard DMD, the dimension
reduction threshold, $\varepsilon_\text{SVD}$, coincides with its counterpart
in the first dimension reduction  of the original snapshots.
Once the expansion (\ref{ap23}) has been calculated, invoking eq.(\ref{ap21}),
the first $M$ components of the vectors appearing
in the left and right-hand sides of eq.(\ref{ap23})
lead to an expansion of type (\ref{ap7}) for the reduced snapshots $\widehat \bq_k$.
The mode amplitudes $a_n$ are recalculated via least-squares fitting between
the two sides of eq.(\ref{ap7}).
Truncating (using a tunable threshold
$\varepsilon_\text{DMD}$) the latter expansion and pre-multiplying it
by the matrix $\bU$ appearing in eq.(\ref{ap5}) lead to the expansion
(\ref{ap1})
for the original snapshots. Finally, as explained
for the standard DMD right after eq.(\ref{ap15}), the amplitudes and
modes in the last expansion are jointly rescaled
requiring that the modes
exhibit unit RMS norm  while preserving the product
$a_n\bu_n$.

The HODMD method described above is called {\it DMD-$d$ algorithm}.
Obviously, for $d=1$, the DMD-1 algorithm reduces to standard DMD.
Also, for appropriate $d>1$, the DMD-$d$ algorithm is able to cope
with arbitrary spatial and spectral complexities.
A MATLAB solver for the DMD-$d$ algorithm can be
found in \cite{linkHODMD,VegaLC2020}.
The algorithm depends
on some tunable parameters, whose selection is commented in the following
(see \cite{LeClaincheVegaSIADS17} for further details).
\begin{itemize}
	\item In order to avoid aliasing \cite{Meseguer2020},
	the temporal distance between snapshots, $\Delta t$,
	must be much smaller
	(say, five times smaller)
	than the smallest period involved in the expansion  (\ref{ap1}).
	Likewise, the total timespan where the snapshots are selected,
	$t_K-t_1$, must
	be  somewhat larger (say, 1.5 times as large) than
	the largest involved period. These values define the total number of
	considered snapshots, $K$.
	\item  For `clean' snapshots, the dimension reduction and
	mode truncation thresholds
	($\varepsilon_\text{SVD}$ and $\varepsilon_\text{DMD}$, respectively)
	can be comparable to each other and quite small. This occurs in the
	applications described in the present paper. Decreasing both  of them
	typically increases the
	accuracy of the obtained
	HODMD expansion, but it also increases the
	number of retained modes. In other words,
	a trade-off is needed to select these thresholds. Let us
	mention that, for noisy snapshots obtained from
	experimental data, $\varepsilon_\text{SVD}$
	should be taken as comparable to the noise level.
	This helps to filter noise due to well-known noise-filtering
	properties of the method \cite{LeClaincheVegaSoria17}.
	\item The index $d$  appearing in eqs.(\ref{ap17}) and (\ref{ap21})
	allows to deal with \textit{time-lagged} snapshots and
	can be chosen to somewhat minimize the
	RRMS error (as defined in eq.(\ref{d21})) of the approximation
	of the reduced snapshots, defined in (\ref{ap7}).
	It must be noted that $d$ scales with $K$, namely, when
	doubling $K$, $d$ must be doubled as well.
	\item The algorithm is quite robust in connection with the tunable parameters
	$\varepsilon_\text{SVD}$, $\varepsilon_\text{DMD}$, and
	$d$. In particular, the plot of the  approximation
	RRMS error vs. $d$ is fairly flat near the
	optimal value, which means that the selection of $d$ is not critical.
\end{itemize}
%


\setcounter{equation}{0}
\renewcommand{\theequation}{B.\arabic{equation}}

\section{The STKD method\label{sec:STKD}}
For spatially  1D dynamics, the
 outcome of the STKD method is a discrete expansion of the
form \cite{LeClaincheVegaJNLS18}
\beqn
q(x_i,t_k)\simeq\sum_{m=1}^M\sum_{n=1}^Na_{mn} u_{mn} \,
\er^{(\nu_m-\smallir\,\kappa_m)\, x_i +(\delta_n+\smallir\,\omega_n)\, t_k},
\label{ap30}
\eeqn
 where $t_k = (k-1)\,\Delta t$, for $k = 1,\ldots,K$,
and the $I$ discrete values of $x$, required to be equispaced, are defined as
\beqn
x_i = (i-1)\,\Delta x, \quad \text {for }\,i=1,\ldots,I.\label{ap32}
\eeqn
Here, $q(x_i,t_k)$ are the components of snapshots for the involved state variable,
$a_{mn} > 0$ are real \textit{amplitudes},
$u_{mn}$ are normalized (generally complex) \textit{modes},
$\nu_m$ and $\delta_n$ are
{\it spatial} and {\it temporal growth rates}, respectively,
$\kappa_m$ are {\it wavenumbers}, and $\omega_n$ are {\it frequencies}.
The derivation of the discrete expansion (\ref{ap30})
is now  summarized for the case considered in this paper,
in which a scalar state variable and a single longitudinal
 coordinate $x$ are involved.
Full details can be found in \cite{LeClaincheVegaJNLS18},
 where more general
cases are considered, namely involving a vector state variable,
more than one longitudinal coordinates, or
additional transverse coordinates.

As a first step, the snapshots appearing in the left-hand side of
the expansion (\ref{ap30}) are organized in a $I\times K$-snapshot matrix, as
defined in  eq.(\ref{ap3}). This matrix is dimension-reduced
 via truncated  SVD
according to a
{\it dimension reduction threshold}, $\varepsilon_\text{SVD}$.
After truncation, the snapshot matrix is approximated as
\beqn
\bQ_1^K\simeq\bU\,\bSigma\,\bV^\top\equiv
\left(\widehat\bQ^x\right)^\top  \widehat\bQ^t,\label{ap34}
\eeqn
where
\beqn
\widehat \bQ^x=\sqrt{\bSigma}\,\bU^\top \text{ and  }
\,\widehat \bQ^t=\sqrt{\bSigma}\,\bV^\top.\label{ap36}
\eeqn
These matrices are called the {\it reduced spatial} and {\it temporal
snapshot matrices}, respectively, and their columns,
$\widehat q^x_i$ and $\widehat q^t_k$,
are called the {\it reduced spatial} and {\it temporal snapshots},
respectively. Comparison of eqs.(\ref{ap36}) and (\ref{ap5}) shows that, while the SVD
singular values  are used in the HODMD method to scale the
reduced temporal snapshots, here they are equidistributed between
the reduced spatial and temporal snapshots,
which are thus scaled by the square root of the SVD singular values.

As a second step,
the DMD-$d$ algorithm described in Appendix
\ref{sec:HODMD} is applied to both the reduced spatial snapshots and
the reduced temporal snapshots, using appropriate indices $d^x$ and $d^t$
(which do not necessarily coincide), respectively.
For simplicity, we  can use a
dimension reduction threshold, $\varepsilon_\text{SVD}$,
equal to its counterpart
in the first dimension reduction step leading to eq.(\ref{ap36}).
In addition, we can consider a common
mode truncation threshold, $\varepsilon_\text{DMD}$,
when applying HODMD to the reduced spatial and temporal snapshots.
It follows that
\begin{alignat}{1}
&\widehat q^x_{i}\simeq\sum_{m=1}^{M}\widehat a_{m}^x
\widehat u_{m}^x\er^{(\nu_m - \smallir\, \kappa_m)\, x_i}
\quad\text{for }\,i=1,\ldots,I, \label{ap38}\\
&\widehat q^t_{k}\simeq\sum_{n=1}^{N}\widehat a_{n}^t
\widehat u_{n}^t\er^{(\delta_{n}+\smallir\,\omega_{n})\, t_k}
\quad\text{for }\,k=1,\ldots,K.\label{ap40}
\end{alignat}
Substituting these expansions into the columns
of the matrices
$\widehat\bQ^x$ and $\widehat\bQ^t$ appearing in eq.(\ref{ap34})
yields the STKD expansion (\ref{ap30}), with
\begin{alignat}{1}
& a_{mn} = |\widehat a_{m}^x\,\widehat a_{n}^t\,\widehat u_{m}^x\,\widehat u_{n}^t|, \,\,\,\,
u_{mn} = \widehat a_{m}^x\,\widehat a_{n}^t\,
\widehat u_{m}^x\,\widehat u_{n}^t/a_{mn}.
\label{ap45}
\end{alignat}
Note that the (complex) scalar modes $u_{mn}$ exhibit unit
absolute value, as required.
Finally, truncation is performed in the expansion (\ref{ap30})
by retaining only those terms such that
\beqn
\frac{a_{mn}}{\max\{a_{mn}\}}>\varepsilon_\text{DMD},\label{ap50}
\eeqn
where the threshold $\varepsilon_\text{DMD}$ coincides with
the mode truncation threshold  previously set to truncate the expansions
(\ref{ap38})-(\ref{ap40}).


\end{document}